\documentclass[preprint,12pt]{elsarticle}

\usepackage{url}
\usepackage[bottom]{footmisc}
\usepackage[caption=false,font=footnotesize]{subfig}
\usepackage{amsmath}
\usepackage{amssymb}
\usepackage{amsfonts}
\usepackage{breakcites}
\usepackage{newtxmath}
\usepackage{natbib}
\usepackage{hyperref}
\usepackage{comment}
\usepackage{soul}
\usepackage{booktabs}
\usepackage[T1]{fontenc}
\usepackage{siunitx}
\usepackage{caption}

\graphicspath{{./}}

\journal{International Journal of Multiphase Flow}

\begin{document}
\begin{frontmatter}

\title{Evaporation of a deformable droplet under convection}


\author[label1]{Faraz Salimnezhad}
\author[label1]{Metin Muradoglu}
\affiliation[label1]{organization={Koc University},
            addressline={Rumelifeneri Yolu, Sariyer}, 
            city={Istanbul},
            postcode={34450}, 
            country={Turkey}}
\begin{abstract}
Evaporation of a deformable droplet under convection is investigated and performance of the classical and Abramzon-Sirignano (A--S) models is evaluated. Using the Immersed Boundary/Front-Tracking (IB/FT) method, interface-resolved simulations are performed to examine droplet evaporation dynamics over a wide range of Reynolds ($20 \leq Re \leq 200$), Weber ($0.65 \leq We \leq 9$), and mass transfer ($1 \leq B_M \leq 15$) numbers. It is shown that flow in the wake region is greatly influenced by the Stefan flow as higher evaporation rates leads to an earlier flow separation and a larger recirculation zone behind the droplet. 
Under strong convection, the  models fail to capture the evaporation rate especially in the wake region, which leads to significant discrepancies compared to interface-resolved simulations. Droplet deformation greatly influences the flow field around the droplet and generally enhances evaporation but the evaporation rate remains well correlated with the surface area. The A--S model exhibits a reasonably good performance for a nearly spherical droplet but its performance deteriorates significantly and generally underpredicts evaporation rate as droplet deformation increases. The A--S model is overall found to outperform the classical model in the presence of significant convection. 
\end{abstract}

\begin{graphicalabstract}
\begin{figure}[hbt!]
\centering
\includegraphics[scale=0.8]{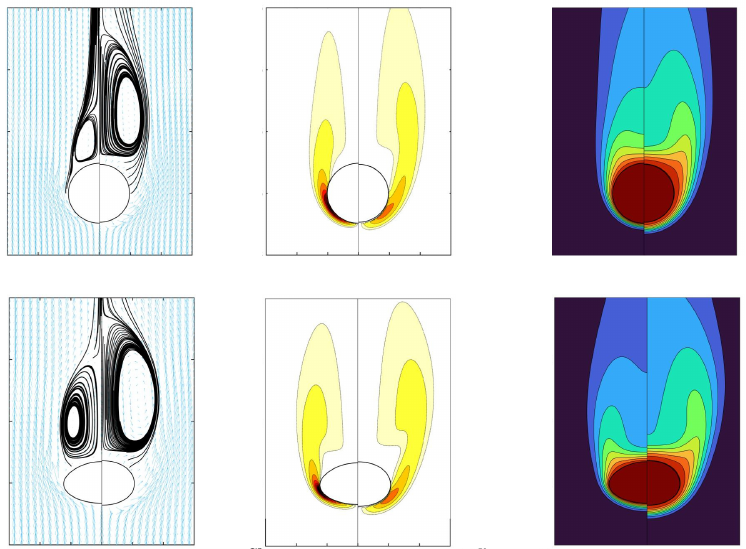}
\caption*{Velocity vectors and streamlines, contours of vorticity  and vapor mass-fraction in the absence and in presence  of Stefan flow for a nearly spherical and moderately deforming droplet.}
\end{figure}
\end{graphicalabstract}

\begin{highlights}
\item Droplet evaporation under convection is studied using IB/FT method.
\item Performance of evaporation models is examined over wide ranges of Reynolds, Weber and mass transfer  numbers.
\item Stefan flow thickens boundary layer and results in early flow separation with a larger recirculation zone.
\item Evaporation models generally fail to predict local Sherwood number in wake region and underpredict evaporation rate of deforming droplets.
\item Droplet deformation enhances evaporation rate but evaporation rate remains strongly correlated with droplet surface-area.

\end{highlights}
\end{frontmatter}


\section{Introduction}
Evaporation of liquid fuel droplets can significantly influence formation and homogeneity of reactive gaseous mixture in liquid-fueled combustion chambers \cite{LAW1982171,FAETH1987293}. Therefore efficiency of combustion process and pollutant formation largely depend on the vaporization process. Evaporation of atomized fuel droplets is governed by a number of factors including size of droplets, temperature of gas phase, pressure in the combustion chamber, and the characteristics of the flow field  \cite{SIRIGNANO1983291}. The dynamics of droplets in a spray environment is inherently complex, involving a range of interdependent processes such as primary atomization, secondary breakup, and droplet deformation. Spray combustion typically takes place in a turbulent environment, making the investigation of this already complex problem even more challenging \cite{GOKALP1992286}.
In addition to combustion applications, droplet evaporation is of fundamental significance across a diverse range of natural phenomena and industrial applications such as spreading of 
 respiratory diseases, cloud formation, and weather forecasting \cite{PhysRevFluids.6.020501,mittal_ni_seo_2020,sazhin2006advanced,BIROUK2006408,PhysRevE.93.032801}. 

In large-scale simulations of evaporating sprays and droplet clouds, use of interface-resolved techniques is practically not  feasible. Instead the Lagrangian point-particle methods are the primary approach widely used in modeling the evaporation of sub-grid droplets~\cite{JENNY2012846}. Here, the droplets are modeled
as point-particles, and low-order models are employed to account for the energy, mass, and momentum transfer between the dispersed and carrier phases. The earliest evaporation model was introduced by Maxwell~\cite{fuchs2013evaporation}. In this model, evaporation rate is assumed to be solely governed by diffusion process~\cite{sazhin2006advanced}. This model was later advanced~\cite{SPALDING1953847,GODSAVE1953818,fuchs2013evaporation} to account for the effects of Stefan flow in addition to the diffusion process for a spherical droplet, to be known as the classical model. The evaporation rate of small droplets in the creeping flow limit is well-predicted by the classical model \citep{RENKSIZBULUT1988189,WILLIAMS19731,FAETH1977191,FAETH1987293,LAW1982171}. On the other hand, vaporization of larger droplets with intermediate Reynolds numbers  ($5<Re<500$) \cite{RENKSIZBULUT1988189} is significantly influenced by the convective transport processes, where neither creeping flow approximations nor the classical boundary-layer analyses might be rigorously applicable. Our focus in this study is to better understand the evaporation process in this regime. 

One of the key physical phenomena in droplet evaporation under convection is the interaction between Stefan flow and the free stream flow. Stefan flow arises due to the substantial density variation of the liquid as it transitions into the vapor phase and it generally leads to thickening of the transport boundary layer. This phenomena reduces heat and mass transfer exchange between the carrier and droplet fluids, thereby influencing the evaporation dynamics \citep{abramzon1989droplet,sazhin2006advanced,sirignano2010fluid}. Abramzon and Sirignano \cite{abramzon1989droplet} modeled this phenomena by treating the droplet as a collection of evaporating wedges and employed a range of Falkner-Skan solutions to quantify the reduction in Sherwood and Nusselt numbers due to thickening of the boundary layer. Their analysis demonstrated that the variation in these parameters is primarily governed by the mass and heat transfer numbers ($B_M, B_T$), and proposed a correction to account for the reduction in Sherwood and Nusselt numbers due to boundary layer inflation.
Readers are referred to the review papers by \citet{FAETH1977191}, \citet{SIRIGNANO1983291}, \citet{LAW1982171} and \citet{sazhin2006advanced,SAZHIN201769} which provide extensive insights into the theoretical foundations and applications of low-order droplet evaporation models.

Interface-resolved simulations can provide unprecedented insight and reveal physics of droplet evaporation, which can be useful in improving/evaluating the low-order models to be used in engineering applications. So far, various numerical techniques have been employed to study droplet evaporation under convection. 
\citet{10.1115/1.3245591} employed a finite-difference method to investigate evaporation of a spherical droplet in a high-temperature gas stream for the Reynolds numbers in the range of $ 10 < Re < 100$. Their study demonstrated that outward flow of vapor from the droplet interface leads to boundary layer thickening, which in turn reduces heat transfer to the droplet. \citet{RENKSIZBULUT1988189} studied the transient evaporation of a spherical n-heptane droplet at intermediate Reynolds numbers using a finite-volume-based numerical method. The results showed that, particularly at elevated pressures, heating in the liquid phase plays an important role in the overall behavior of the droplet. In a later study \citet{haywood1989detailed} showed that receding of liquid/gas interface and transient processes in the gas phase do not play a determining role in droplet evaporation at lower pressures. 

Haywood et al.~\cite{R.J.Haywood,HAYWOOD19941401} conducted one of the earliest numerical investigations on the evaporation of deforming droplets under convection. They examined the transient behavior of a deformed droplet evaporation in low and moderate Reynolds numbers and investigated effects of internal recirculation as well as variable material properties. They observed that the droplet exhibited oblate/prolate oscillatory behavior in the initial stages of evaporation. However, their results indicated that deformation had a small impact on mass transfer rates and the Sherwood number could still be predicted within 8\% accuracy using the quasi-steady correlations, provided that the equivalent droplet radius is used in its definition. Additionally, they emphasized that the accurate incorporation of droplet projected area is crucial for precisely estimating the drag coefficient. \citet{MASHAYEK20011517} employed a finite-element method to investigate the influence of free oscillations on the evaporation rate of droplets under stagnant conditions. In this work, the deformation was introduced by applying spherical harmonics perturbations to the interface. The droplet was assumed to be at its boiling temperature, neglecting the heat-up period, while the far-filed temperature was maintained at a constant value. The results revealed that local mass flux over an evaporating droplet could be related to  local curvature of droplet interface. 

\citet{CHIANG19921307}  presented a detailed investigation of transient evaporation of a droplet injected in convective environments, focusing on variation of droplet drag coefficient and transfer parameters with respect to surface blowing and droplet deceleration. However droplet deformation was not incorporated in their model.  \citet{SCHLOTTKE20085215} presented a volume of fluid (VOF)-based numerical model for simulation of strongly deforming evaporating droplets. They observed complex, transient, and three dimensional flow field around a deformable droplet in convective environments.   \citet{PALMORE2019108954} employed a VOF-based method for interface-resolved computation of 3D liquid-gas flows with evaporation. It was shown that  interface properties play a crucial role in accuracy of vaporization simulations. In addition,  evaporation of a deformable droplet was studied at Reynolds number of $Re=400$  and it was found that droplet wake becomes asymmetrical and chaotic as time progresses.  \citet{NI2021120736} developed a high-fidelity interface capturing method based on a coupled level-set VOF method and studied evaporation of droplet in high temperature forced convective conditions. The surface-averaged Sherwood and Nusselt numbers were compared to the experimental correlations \cite{ranz1952evaporation,clift2005bubbles}.

\citet{DODD2021121157} simulated evaporation of a deforming droplet in homogeneous isotropic turbulence (HIT) using a geometrical VOF approach and validated their numerical method against the experimental results obtained by \citet{VERWEY201833}. Their study revealed that the conventional correlations underpredict the Sherwood number of evaporating droplets in HIT conditions, highlighting the limitations of evaporation models used in point particle methods. \citet{scapin2022finite} investigated evaporation of droplets in weakly compressible homogeneous shear turbulence using a new VOF method. They conducted a parametric study on role of gas phase temperature, Weber number, and ratio of droplet diameter to the Kolmogorov length scale on evaporation rate and evaluated the performance of empirical models considering three different methods for approximation of thermpophysical properties. They observed that the Ranz-Marshall correlation underestimates the Sherwood number at high temperatures and more deformable droplets evaporates faster due to the increased surface area. 

\citet{SETIYA2023104455} have recently investigated the quasi-steady evaporation of a deforming n-decane droplet at Reynolds numbers of $Re=25$ and $Re=120$ for a range of Weber numbers ($1 \le We \le 12$) in a high pressure environment. The results showed that droplet deformation significantly enhances evaporation rate at $Re=120$, while at a lower Reynolds number ($Re=25$),  evaporation rate is weakly affected by the deformation process. The enhancement in the evaporation rate is attributed to the increase in droplet surface area.

The previous works have predominantly focused on investigating fuel droplet evaporation under specific conditions considered in experimental studies. However, a parametric study focusing on critical assessment of widely used evaporation models against the interface-resolved simulations remains scarce. In the present work, the recently developed IB/FT method \cite{salimnezhad2024hybrid} is employed for a comprehensive examination of droplet evaporation in convective environments and evaluation of performance of the low-order models. For this purpose, simulations are carried out for low and moderate Reynolds numbers in the range of $20 \leq Re \leq 200$, for which flow around the droplet remains axisymmetric, and for a wide range of mass transfer numbers ($1 \leq B_M \leq 15$). Local mass flux and Sherwood number distribution over surface of an evaporating droplet as well as  overall evaporation rate and averaged Sherwood number are quantified for each case and compared to the Abramzon–Sirignano~\cite{abramzon1989droplet} and the classical~\cite{SPALDING1953847,sazhin2006advanced} models. Effects of the Stefan flow on flow field around the droplet  and wake dynamics are also studied in detail. Four cases of nearly spherical, weakly deforming, moderately deforming, and highly deforming droplets are considered to examine the effects of deformation on evaporation rate. The coupled interactions of droplet deformability and the Stefan flow are explored and the relevance of the analytical models in case of deforming droplets is evaluated. 

The rest of the paper is organized as follows. The mathematical formulation and the numerical methods are briefly  discussed in Section~\ref{Mathform}. The computational setup is described in Section~\ref{Comp}. The results are presented and discussed in Section~\ref{results}. Concluding remarks are outlined in Section ~\ref{conlusions}.
\label{Int}
\section{Mathematical formulation and numerical method}
\label{Mathform}

The governing equations are described in the context of the hybrid Immersed-Boundary/Front-Tracking (IB/FT) method of \citet{salimnezhad2024hybrid}. In this method, a one-field formulation is used to solve the flow equations in the entire computational domain and the jump conditions at the interface are taken into account appropriately \cite{salimnezhad2024hybrid}. Considering an incompressible multiphase system and using a moving reference frame (MRF) to keep the center of the droplet fixed in the computational domain \cite{rusche2003computational}, the momentum equation can be written as \citep{sato2013sharp,BHUVANKAR2020115919}
\begin{eqnarray} 
\rho \frac{\partial { \boldsymbol{u}_{rel}}}{\partial {t}} &+&\rho \left[\nabla \cdot \left(\boldsymbol{u}_{rel}\boldsymbol{u}_{rel}\right) -\boldsymbol{u}_{rel} \left( \nabla \cdot\boldsymbol{u}_{rel}\right) \right]= - \nabla p+\rho \boldsymbol{a}_{MRF} \nonumber \\ 
&+& \nabla \cdot \mu \left(\nabla \boldsymbol{u}_{rel}+\nabla {\boldsymbol{u}}^T_{rel}\right)+ \int_A{\sigma \kappa \boldsymbol{n}\delta \left(\boldsymbol{x}\mathrm{-}{\boldsymbol{x}}_{\mathit{\Gamma}}\right)dA},
\label{eqn:momentum}
\end{eqnarray}
where $\boldsymbol{u_{rel}}$ is the relative velocity, $p$ is the pressure, and $\boldsymbol{a}_{MRF}$ is the acceleration introduced to fix the droplet's center of mass with respect to the domain. The discontinuous density and viscosity fields are denoted by $\rho$ and $\mu$, respectively. The last term on the right-hand side of Eq.~\eqref{eqn:momentum} denotes the body force due to the surface tension. Here, $\sigma$, $\kappa$ and $\boldsymbol{n}$ denote the surface tension coefficient, twice the mean curvature and the outward normal vector at the interface, respectively. 

The non-divergence-free velocity field at the interface due to the phase change requires the continuity equation to be modified as \citep{tryggvason2001front,esmaeeli2004front,esmaeeli2004computations,irfan2017front,irfan2018front},
\begin{equation} 
\nabla \cdot \boldsymbol{u}_{rel}=\frac{1}{h_{lg}}\left(\frac{1}{{\rho }_g}-\frac{{1}}{{\rho }_l}\right)\int_A{\delta \left(\boldsymbol{x}-{\boldsymbol{x}}_{\mathit{\Gamma}}\right)\dot{q}_{\Gamma} \ dA}, 
\label{eqn:continuity}
\end{equation}
where the divergence of velocity field is made non-zero at the interface to account for volume expansion due to evaporation. Here, $h_{lg}$ and $\dot{q}_{\Gamma}$ are the latent heat of vaporization and  heat flux per unit time at the interface, respectively. The subscripts $l,g,$ and $\Gamma$, respectively, indicate the liquid phase, gas phase, and interface. The species mass fraction equation solved in the gas phase only is given by
\begin{equation} 
\frac{\partial Y}{\partial t}+\boldsymbol{u_{rel}} \cdot \nabla Y= \nabla \cdot D_{vg}\nabla Y,
\label{eqn:specieseq}
\end{equation}
where $Y$ is the mass fraction and $D_{vg}$ is the binary diffusion coefficient.  

Considering a species-driven evaporation of a single component droplet, the evaporative mass flux at the interface can be computed as \cite{irfan2017front}:
\begin{equation} 
{\dot{m}^{\prime\prime}}=\frac{{\rho }_g\ D_{vg}\ {\left(\frac{\partial Y}{\partial n}\right)}_{\Gamma}}{1-Y^{\Gamma}},
\label{eqn:speciessingle}
\end{equation}
where ${\left(\frac{\partial Y}{\partial n}\right)}_{\Gamma}$ denotes the vapor mass fraction gradient at the interface in the normal direction. Following \citet{salimnezhad2024hybrid}, the interface mass fraction, $Y^{\Gamma}$, is set as a Dirichlet boundary condition on the interface, corresponding to a specific mass transfer number, $B_M$, using the ghost cell and image point methodology adapted from \citet{mittal2008versatile}. 

We assume that the material properties following a fluid particle remain constant, i.e., $\frac{D\rho }{Dt}=\frac{D\mu }{Dt}=\frac{DD_{vg} }{Dt}=0$.  The material properties  are set throughout the computational domain using an indicator function defined to distinguish between the liquid and gas phases \cite{salimnezhad2024hybrid}.  

The coupled form of the flow equations is solved on a uniform staggered grid using a finite-difference front-tracking method \citep{irfan2018front,tryggvason2001front,unverdi1992front,juric1998computations,esmaeeli2004computations}. All scalar fields including  material properties, pressure, and species concentrations are stored at the cell center while the velocities are located on the cell faces on a staggered Cartesian grid~\cite{salimnezhad2024hybrid}. The convective terms in the species equation are approximated using a fifth-order WENO-Z scheme \citep{borges2008improved} while the convective terms in the momentum equation are approximated employing a third-order QUICK method \cite{LEONARD197959}. The other spatial derivatives are discretized using central differences.  The flow equations are integrated in time using Chorin's explicit projection method \cite{chorin1968numerical}, while the species transport equation is advanced using an explicit first-order Euler scheme. The numerical method is explicit and first order in time but a second-order time accuracy can be easily recovered using a predictor–corrector scheme as described by \citet{tryggvason2001front}. The first-order method is employed in the present study because the temporal discretization error is generally negligible compared to the spatial error, owing to the small time step required for numerical stability \citep{irfan2018front,salimnezhad2024hybrid}.

\begin{figure}[!h]
\centering
\begin{tabular}{cc}
\includegraphics[scale=1.25]{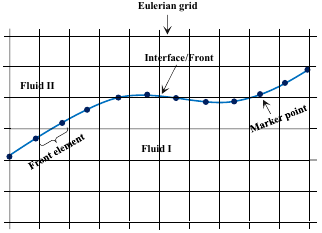} 
\end{tabular}
\caption{A schematic illustration of the Lagrangian grid cast on the stationary Eulerian grid.}
\label{Fig:FTillust}
\end{figure}

A separate Lagrangian grid (or front) is used to track the interface. As sketched in Fig.~\ref{Fig:FTillust}, the Lagrangian grid consists of connected marker points and a segment of the interface between two neighboring markers points is called as a front element. The marker points are advected by the local flow velocity interpolated from the Eulerian grid and the interface receding velocity due to the evaporation, i.e.,
\begin{equation} 
\frac{d{\boldsymbol{x}}_{\Gamma }}{dt}=u_n{\boldsymbol{n}}_{\Gamma },
\label{eqn:frontadv}
\end{equation}
where ${\boldsymbol{x}}_{\Gamma }$ is the position of the marker point and ${\boldsymbol{n}}_{\Gamma }$ is the outward normal vector. The normal component of velocity at the interface, $u_n$, is computed as
\begin{equation}
u_{n}=\frac{1}{2}\left({\boldsymbol{u}}_l+{\boldsymbol{u}}_g\right) \cdot \boldsymbol{n}-\frac{\dot{q_{\Gamma}}}{2 h_{lg }}\left(\frac{1}{{\rho }_l}+\frac{1}{{\rho }_g}\right),
\label{eqn:normalv}
\end{equation}
where ${\boldsymbol{u}}_l$ and ${\boldsymbol{u}}_g$ are the velocities evaluated on the liquid and gas side of the interface, respectively \cite{salimnezhad2024hybrid}. The last term on the right hand side of Eq.~\ref{eqn:normalv} is the evaporation-induced interface receding velocity.  

A complete description of the numerical method can be found in our recent paper \cite{salimnezhad2024hybrid}.

\section{Computational setup}
\label{Comp}
\begin{figure}[h!]
\centering
\includegraphics[scale=0.8]{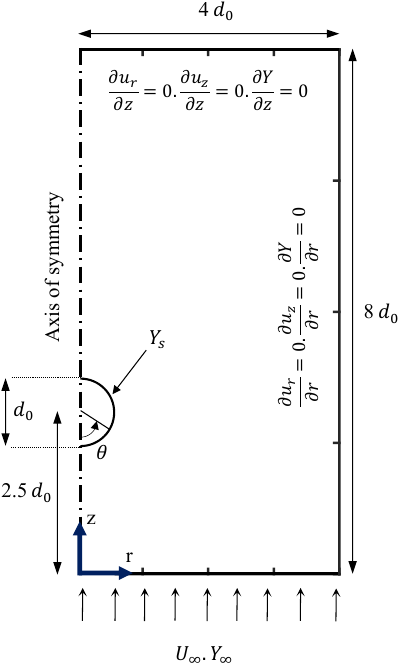}
\caption{Sketch of the computational domain and the boundary conditions used in simulations of droplet evaporation in a convective environment.}
\label{Fig:Domain}
\end{figure}

The computational setup used in the present study follows the framework established in our previous work \citep{salimnezhad2024hybrid} and it is schematically shown in Fig.~\ref{Fig:Domain}. Center of mass of a deformable droplet is fixed in space using a moving reference frame methodology. The ambient flow has a velocity of $U_{\infty}$. An initially spherical droplet with a diameter of $d_0$ is located at an axial distance of $2.5d_0$ from the inlet. The computational domain spans $4d_0 \times 8d_0$ and is discretized using a uniform Cartesian grid with $512 \times 1024$ cells in the radial and axial directions, respectively. \citet{salimnezhad2024hybrid} showed that this grid resolution is sufficient to reduce the spatial error to below 0.5\% for all the flow quantities.

In all the simulations, a uniform velocity $U_{\infty}$ is prescribed at the south boundary, corresponding to a specified Reynolds number. Symmetry and full-slip boundary conditions are imposed at the left (centerline) and right (far-field) boundaries, respectively. On the droplet surface, a Dirichlet boundary condition is applied for the vapor mass fraction using the ghost cell and image point methodology \cite{salimnezhad2024hybrid} corresponding to a specified mass transfer number ($B_M$). At the south boundary, the vapor mass fraction is set to zero, corresponding to a dry gas. The gravitational forces are neglected to facilitate a direct comparison with the evaporation models.

The relevant non-dimensional parameters are the Reynolds number ($Re$), Weber number ($We$), Schmidt number ($Sc$), average Sherwood number ($Sh$), local Sherwood number ($Sh_l$) and mass transfer number ($B_M$), defined as

\begin{eqnarray} 
Re &=& \frac{ \rho_g U_{\infty} d_0 }{ \mu_g }, \;
We = \frac{ \rho_g U_{\infty}^2 d_0 }{ \sigma }, \;
Sc = \frac{ \mu_g }{ \rho_g D_{vg} }, \nonumber \\ 
Sh &=& \frac{ \dot{m} }{ \pi d \rho_g D_{vg} B_M }, \; 
Sh_l = \frac{ -d }{ Y_\Gamma - Y_\infty } 
\left( \frac{\partial Y}{\partial n} \right)_{\Gamma}, \;
B_M = \frac{ Y_\Gamma - Y_\infty }{ 1 - Y_\Gamma },
\label{eqn:nodimensional}
\end{eqnarray}
where $U_\infty$ is the free-stream velocity, $d_0$ is the initial droplet diameter, $d$ is the instantaneous droplet diameter, $\rho_g$ and $\mu_g$ are the gas density and viscosity, $\sigma$ is the surface tension, $\dot{m}$ is the evaporation rate, $Y_\Gamma$ and $Y_\infty$ are, respectively, the mass fraction on the droplet surface and at the far field, and $D_{vg}$ is the vapor diffusivity in the gas phase. The governing equations are solved in their dimensional forms but the results are non-dimensionalized using the length, velocity, and time scales of $d_0$, $U_{\infty}$ and $ d_0 / U_{\infty}$, respectively. Note that, for a spherical droplet, the surface-averaged local Sherwood number is equivalent to average Sherwood number ($Sh$) defined in Eq.~\eqref{eqn:nodimensional}. But, for a deforming droplet, these two quantities may differ due to the increased surface area associated with deformation. Nevertheless, following the common approach used in point-particle simulations, we use the average Sherwood number definition given in Eq.~\eqref{eqn:nodimensional} for both nearly spherical and deforming droplets.
\section{Results and discussion}
\label{results}
Simulations are performed to examine evaporation dynamics of a volatile droplet in a convective environment particularly focusing on effects of convection (Reynolds number), evaporation intensity (mass-transfer number) and droplet deformability (Weber number) in ranges relevant to practical applications. Effect of Stefan flow on droplet evaporation is also examined in detail.

The Schmidt number is fixed at $Sc = 0.7$, assuming the thermophysical properties of air for the gas phase. Simulations are carried out for a range of convective conditions corresponding to  $20 \leq Re \leq 200$. It is important to note that increasing the Reynolds number beyond $Re=200$ would result in three dimensional flow structures, i.e., vortex shedding, which cannot be captured by the axisymmetric setting employed in the present study. To investigate influence of droplet deformability on evaporation, the Weber number is varied in the range of $1 \leq We \leq 9$,  which is discussed in greater detail in Section~\ref{DefOnEvrate}. The viscosity ratios is fixed at $\mu_l / \mu_g = 15.34$. The density ratio in the momentum equation (see Eq.~\eqref{eqn:momentum}) is set to $\rho_l / \rho_g = 25.75$  following \citet{salimnezhad2024hybrid}. Whereas the physical density ratio of $\rho_l / \rho_g = 530.36$, corresponding to a light hydrocarbons–air multiphase system, is used in Eqs.~\eqref{eqn:continuity} and \eqref{eqn:normalv}. Here the subscripts $l$ and $g$ denote the liquid and gas phases, respectively. Note that, in Eq.~\eqref{eqn:momentum} the density and viscosity ratios are kept relatively small compared to the values in typical evaporating multiphase systems in order to enhance the numerical stability, relax the time-step restrictions and thus facilitate extensive simulations~\cite{salimnezhad2024hybrid,tryggvason2011direct}. The physical density ratio employed in Eqs.~\eqref{eqn:continuity} and ~\eqref{eqn:normalv} also allows for the investigation of droplet behavior without unphysical and fast extinction of droplet under higher convective flow or at high mass transfer numbers. It is emphasized here that a further increase in the property ratios does not affect the results significantly as demonstrated by \citet{irfan2018front} and \citet{salimnezhad2024hybrid}. In spray combustion applications, the slip Reynolds number is typically on the order of 100 or lower~\cite{sirignano2010fluid,RENKSIZBULUT1988189}. Therefore, investigating the physics of droplet evaporation and assessing the accuracy of the low-order models at low to moderate Reynolds numbers are of practical importance.  

\begin{figure}[hbt!]
\centering
\begin{tabular}{cc}
\includegraphics[scale=0.45]{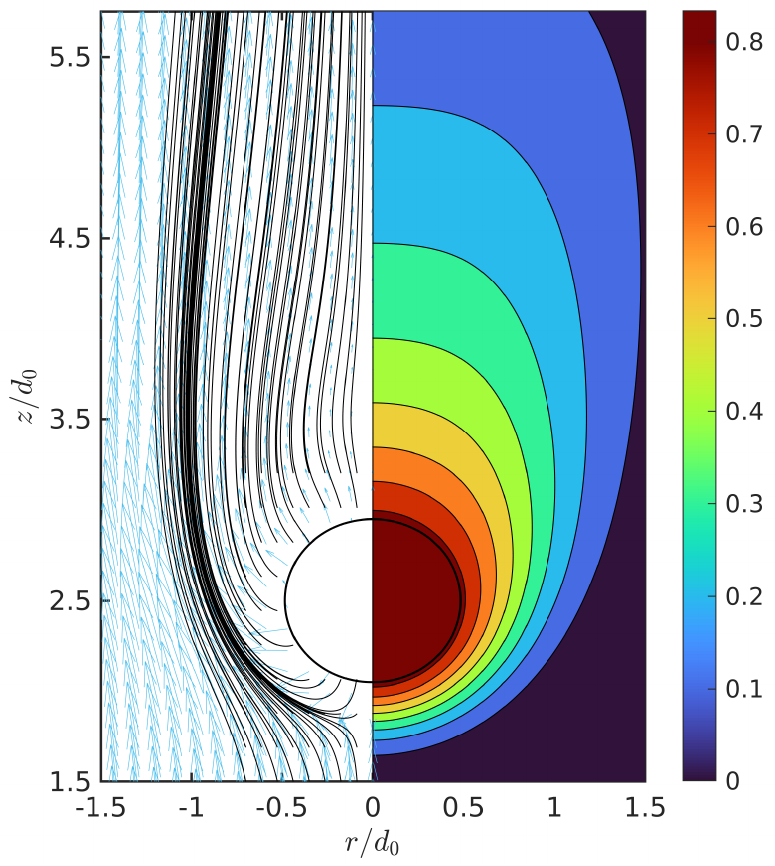}
&
\includegraphics[scale=0.45]{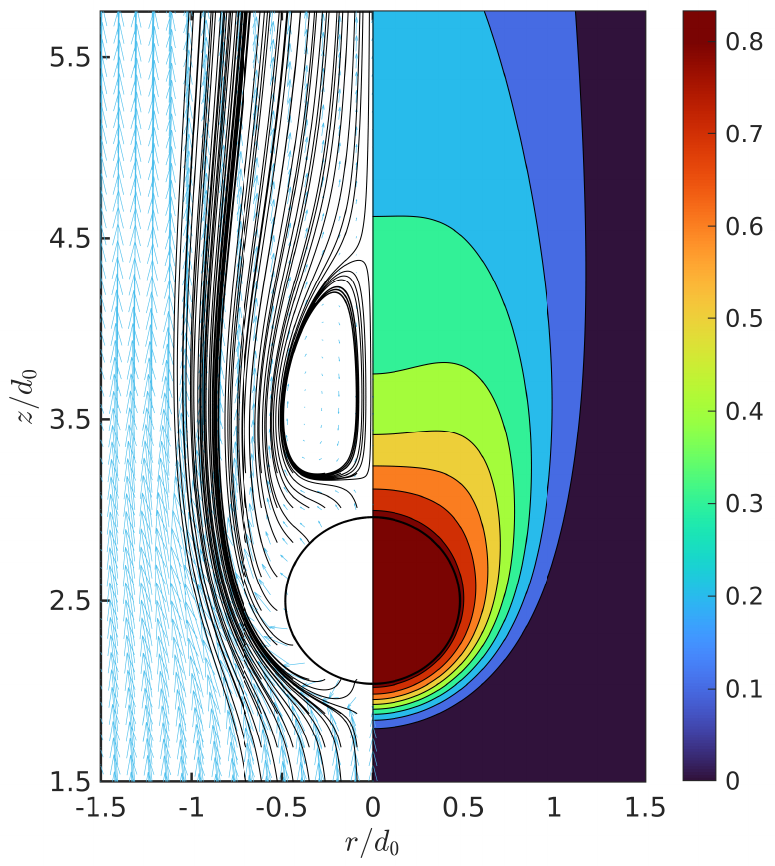}
\\
\scriptsize{(a) $Re=20,t^*=5$} & \scriptsize{(b) $Re=50,t^*=10$} 
\\
\includegraphics[scale=0.45]{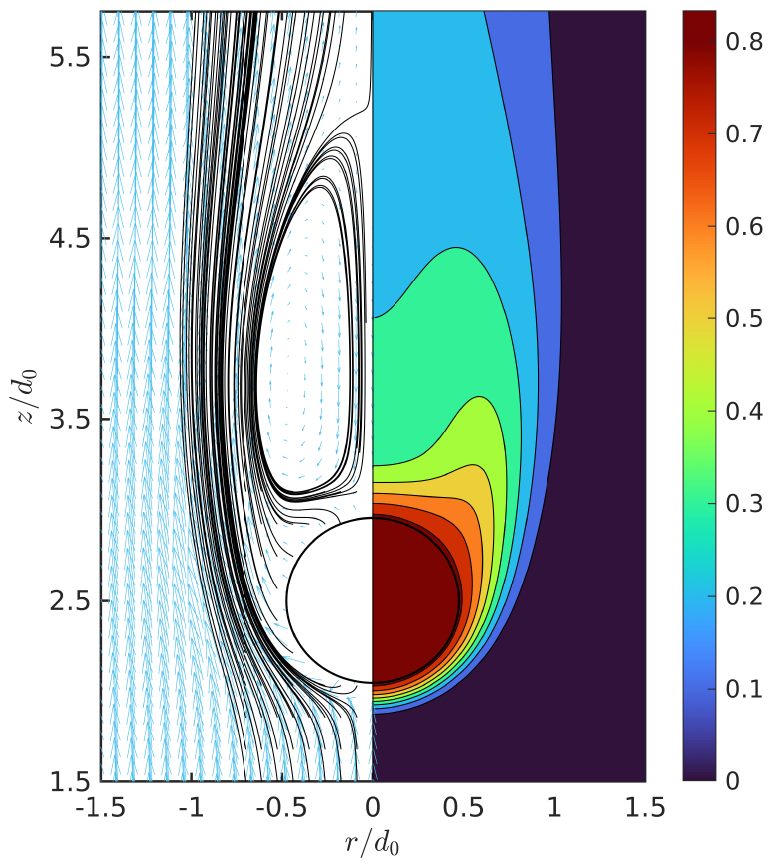}
&
\includegraphics[scale=0.45]{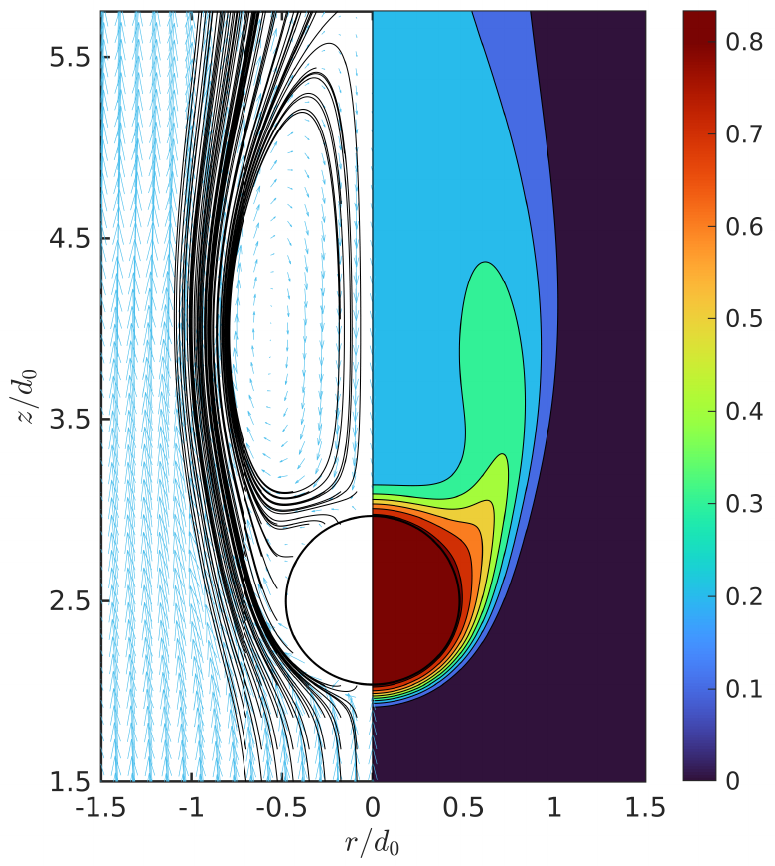}
\\
 \scriptsize{(c) $Re=100,t^*=20$} & \scriptsize{(d) $Re=200,t^*=30$} 
\\
\end{tabular}
\caption{Velocity vectors and streamlines (left portion),  and mass fraction field (right portion) around a nearly spherical droplet ($We=0.65$) for Reynolds numbers of $Re\in[20,\;50,\; 100,\;200]$ at the mass transfer number of $B_M=5$. Domain: $4d_0 \times 8d_0$; Grid: $512 \times 1024$.}
\label{Fig:Re2050100200}
\end{figure}

Simulations are first carried out for a nearly spherical droplet by fixing the Weber number at $We=0.65$ to facilitate direct comparison with the low-order models in which droplet is assumed to be spherical.  Velocity vectors  and  mass fraction field are plotted in Fig.~\ref{Fig:Re2050100200} for $Re\in [20,\;50,\;100,\; 200]$ in a moderately evaporating regime of $B_M = 5$ when the droplet attains a quasi-steady state to show the overall flow around an evaporating droplet. At a low Reynolds number ($Re=20$), the mass fraction field indicates that convection due to Stefan flow as well as diffusion play a dominant role in the evaporation process and shaping the flow field around the droplet. The velocity vectors also indicate that intensity of Stefan flow is significant compared to mean flow. In this regime, the boundary layer assumption over the droplet might be questionable as its thickness is comparable to the droplet radius. As Reynolds number increase to $Re=50$, the flow field undergoes significant changes: a thinner boundary layer develops in the upstream of the droplet and separates in the presence of a strong Stefan flow, resulting in a closed wake region downstream of the droplet. Size of the wake increases further at $Re=100$ which significantly affects distribution of mass fraction field in this region. At $Re = 200$, the wake region becomes significantly elongated and a thin boundary layer forms in the leading edge of the droplet. Comparing Figs.~\ref{Fig:Re2050100200}d and \ref{Fig:Re2050100200}b reveals that mass fraction close to the rear stagnation point is lower in the former, highlighting the increasing role of recirculation zone downstream of droplet as Reynolds number increases.

\begin{figure}[tbp!]
\centering
\begin{tabular}{ccc}
\includegraphics[scale=0.28]{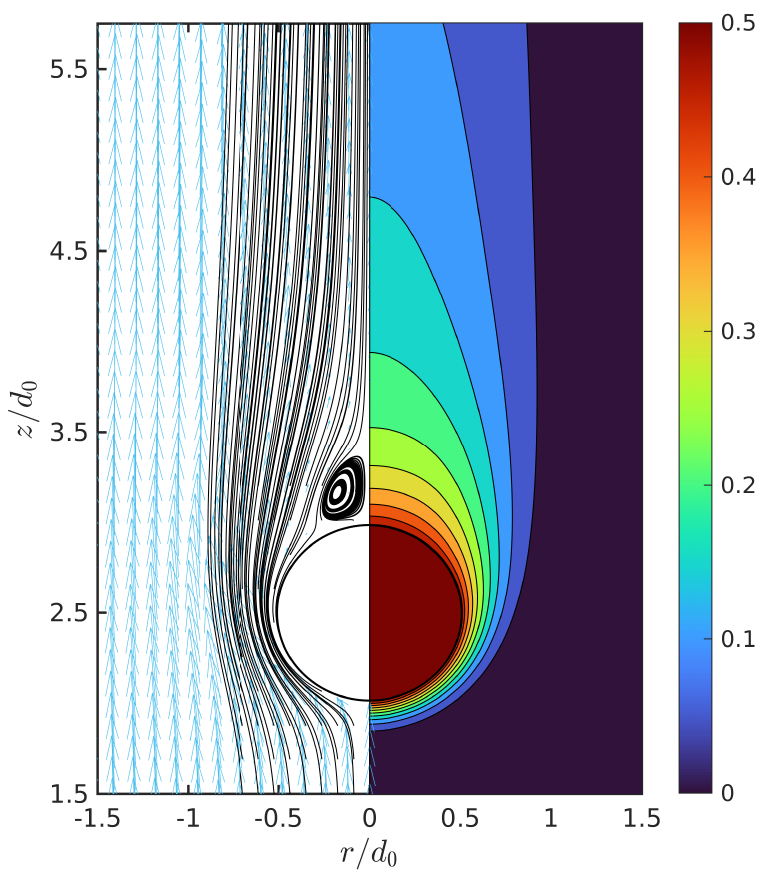}
&
\includegraphics[scale=0.28]{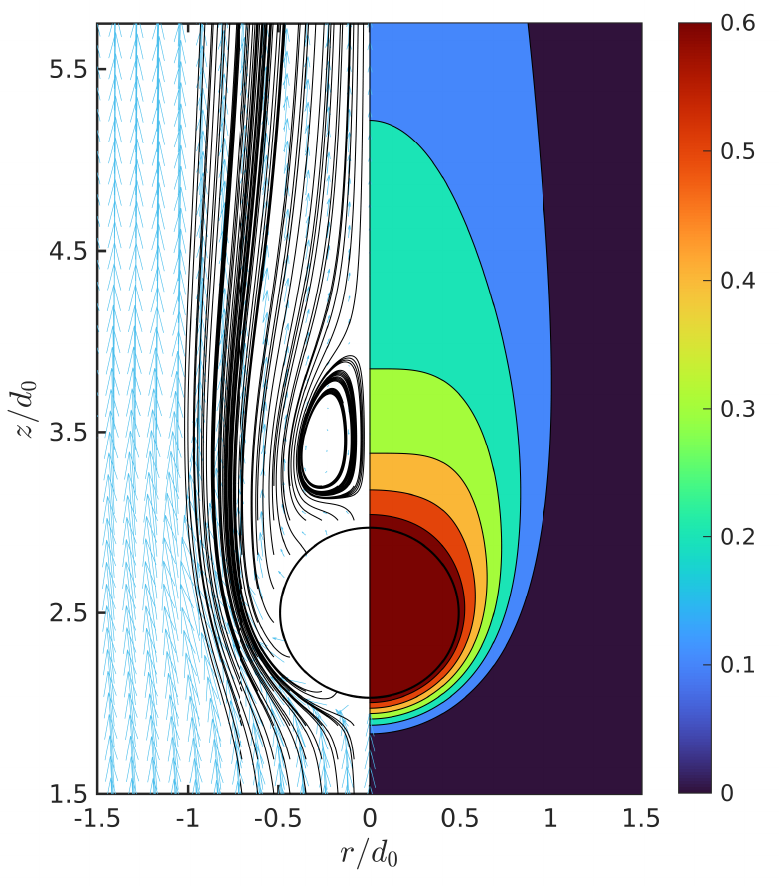}
&
\includegraphics[scale=0.28]{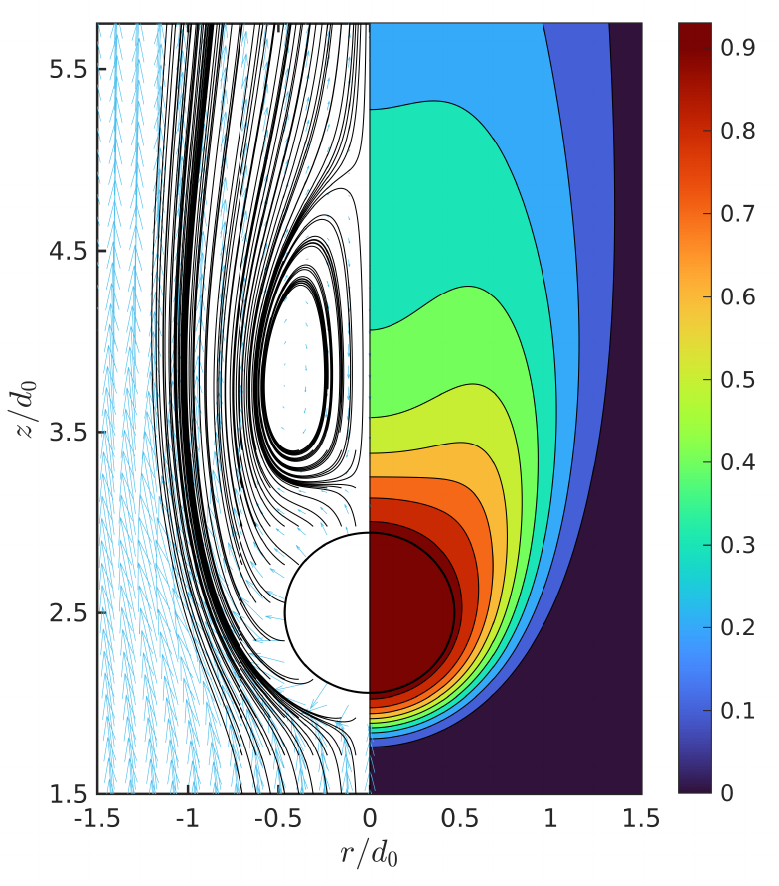}
\\
\scriptsize{ NSF, $Re=50, t^*=10$} & \scriptsize{ $B_M=2, Re=50,t^*=10$} & \scriptsize{$B_M=15, Re=50, t^*=10$} 
\\
\includegraphics[scale=0.28]{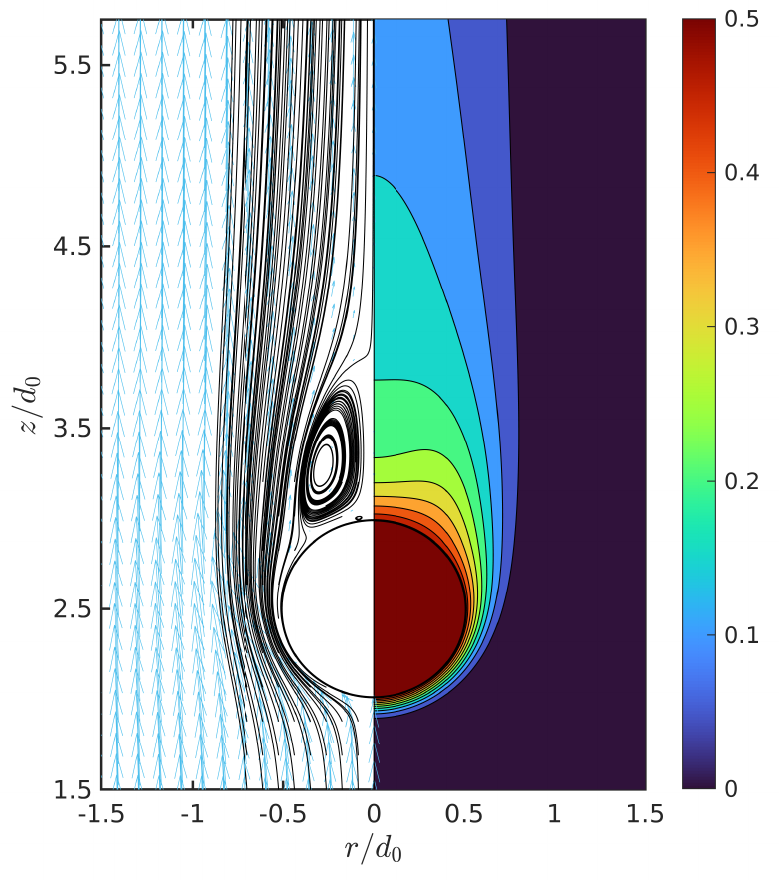}
&
\includegraphics[scale=0.28]{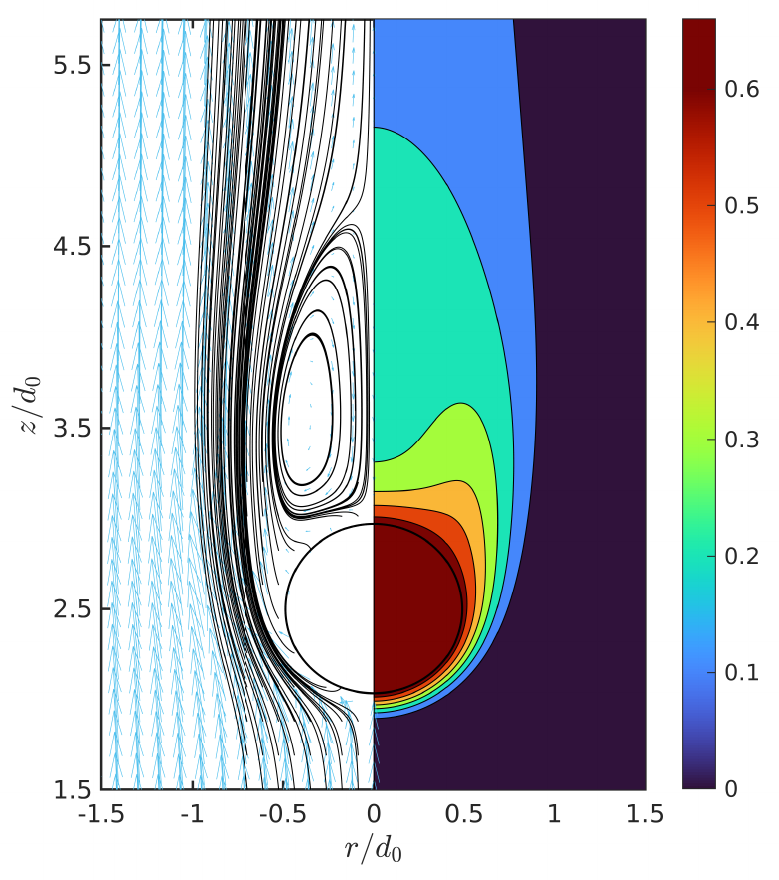}
&
\includegraphics[scale=0.28]{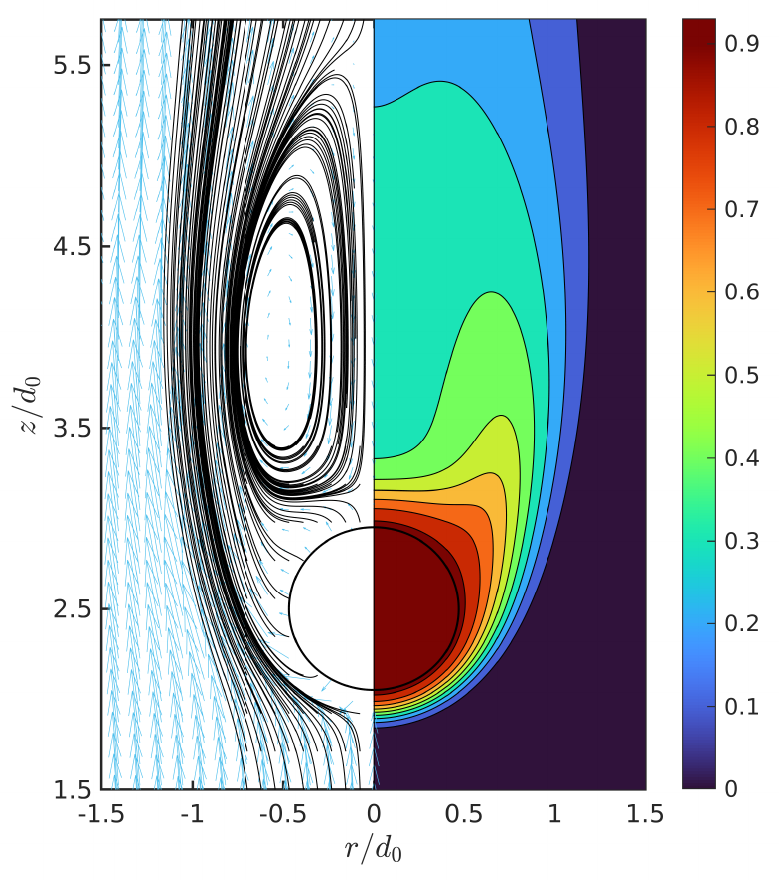}
\\
 \scriptsize{NSF, $Re=100, t^*=20$} & \scriptsize{ $B_M=2, Re=100,t^*=20$} & \scriptsize{$B_M=15, Re=200, t^*=20$}
\\
\includegraphics[scale=0.33]{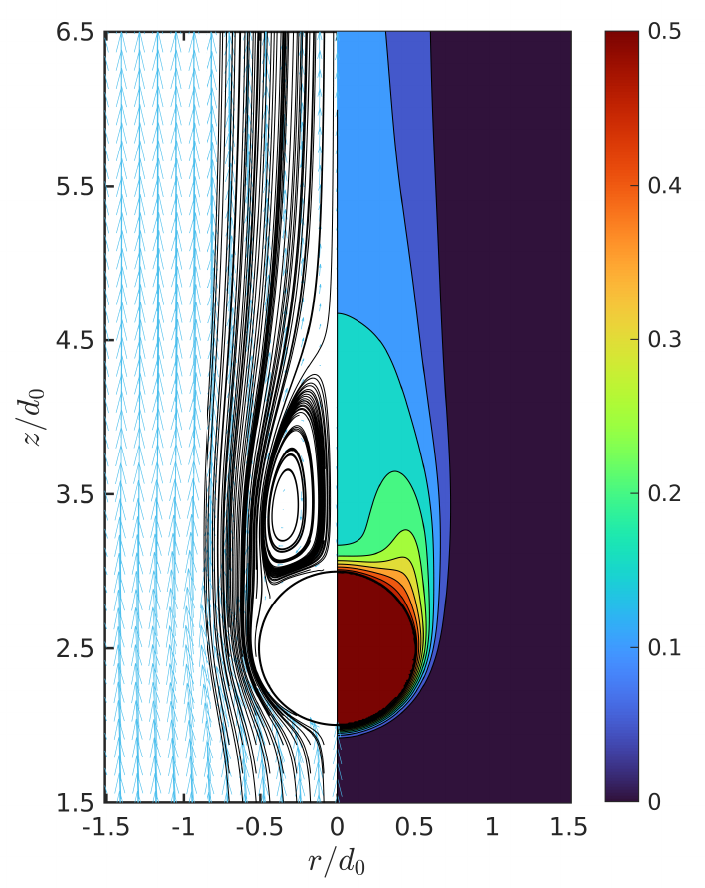}
&
\includegraphics[scale=0.33]{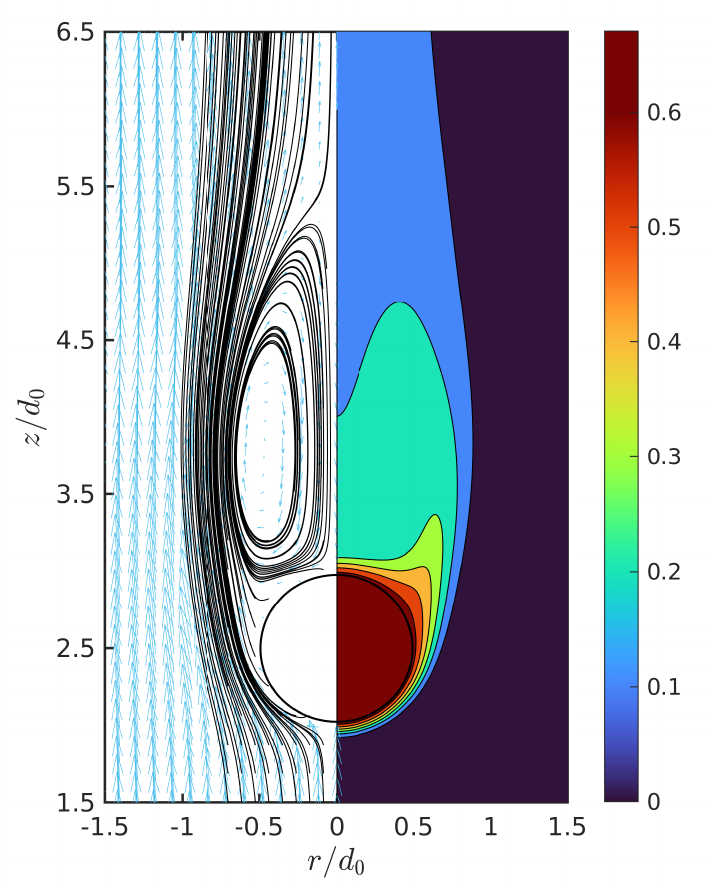}
&
\includegraphics[scale=0.33]{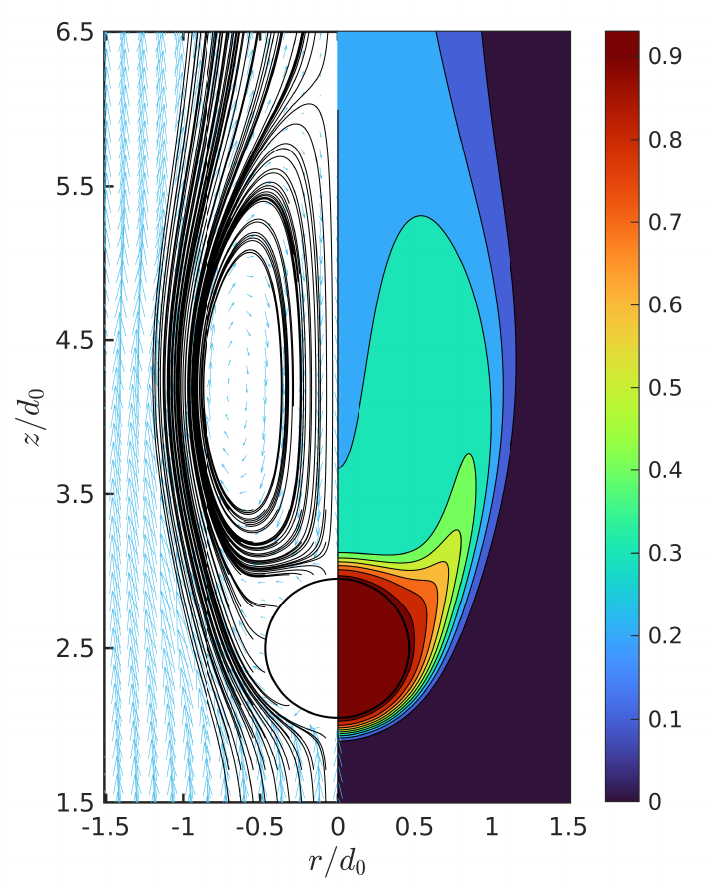}
\\
 \scriptsize{NSF, $Re=200, t^*=30$} & \scriptsize{ $B_M=2, Re=200, t^*=30$} & \scriptsize{$B_M=15, Re=200, t^*=30$}
\\
\end{tabular}
\caption{Velocity vectors (left portion) and mass fraction (right portion) fields around a nearly spherical droplet ($We=0.65$) for the no-Stefan flow (NSF, top row), the moderately evaporating ($B_M=5$, middle row) and the strongly evaporating ($B_M=15$, bottom row) cases at $Re= 50$ (left column), $Re=100$ (middle column) and $Re=200$ (right column). Color bars indicate values of mass fraction. Domain: $4d_0\times 8d_0$; Grid: $512\times 1024$.}
\label{Fig:Re50100200_We065_BM215}
\end{figure}

Effects of evaporation intensity are qualitatively shown in Fig.~\ref{Fig:Re50100200_We065_BM215} where velocity and mass fraction fields are plotted around a nearly spherical droplet for $B_M=0$ (non-evaporating), $B_M=2$ (moderately evaporating) and $B_M=15$ (strongly evaporating) cases at $Re=[50,\;100,\;200]$. Outward  velocity vectors at droplet surface show  evaporation-induced Stefan flow that intensifies and pushes streamlines further away from droplet surface as $B_M$ increases. As a result, the front and back stagnation points detach from surface of the droplet. Blowing effect of Stefan flow thickens the boundary layer and leads to an early separation with a larger wake region behind the droplet as seen in Fig.~\ref{Fig:Re50100200_We065_BM215}. Note that, for a fixed Reynolds number, an increase in mass transfer number leads to formation of a larger recirculation zone. For example, in the case of $Re=200$ and $B_M = 15$, the wake length becomes as large as $3.1d_0$, which is much longer than that of wake behind a solid sphere which can reach up to $2.5d_0$ at $Re = 400$ \cite{clift2005bubbles}. It is likely that this large recirculation zone is an artifact of imposing axisymmetry in the present simulations. Resolution of this issue requires interface-resolved full 3D simulations. 

\begin{figure}[hbt!]
\centering
\includegraphics[scale=0.5]{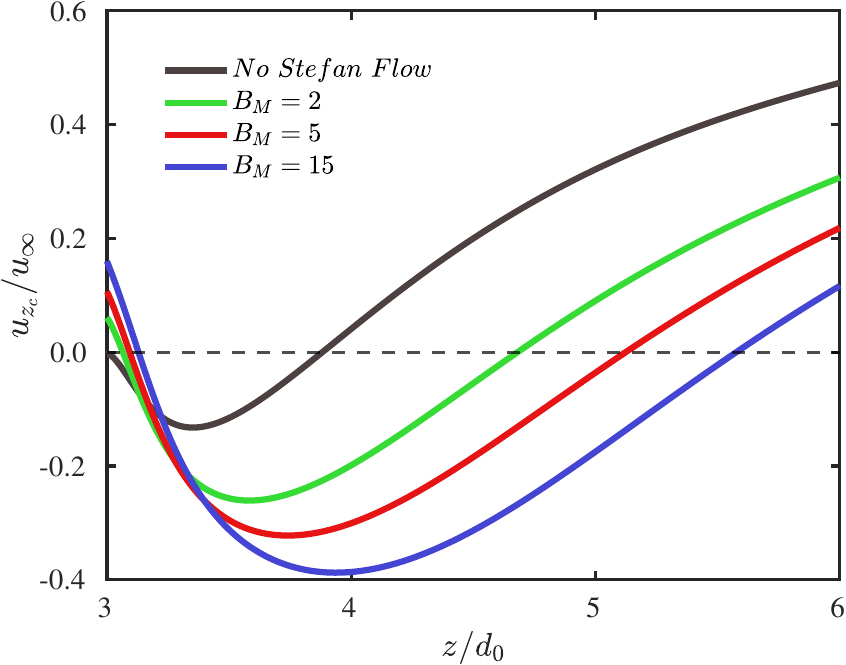}
\caption{Axial velocity profiles along the symmetry axis behind the droplet at $Re = 100$ and mass transfer numbers of $ B_M\in[2,\; 5,\; 15]$. Domain: $4d_0\times 8d_0$; Grid:$512\times 1024$.}
\label{Fig:FV_wake_Re100_BM_0_2_5_15_updated-cropped}
\end{figure}

The axial velocity profiles along the centerline behind the droplet are plotted in Fig.~\ref{Fig:FV_wake_Re100_BM_0_2_5_15_updated-cropped} for the $B_M\in [2,\;5,\; 15]$ cases at $Re=100$. The velocity profile for the `no Stefan flow` case is also plotted as a reference. Non-zero velocity on the droplet surface (i.e., at $z/d_0=3$) indicates the magnitude of the Stefan flow. Note that the intersection point between  velocity profile and horizontal dashed line indicates axial length of  recirculation zone.  As seen, recirculation zone enlarges as $B_M$ increases as also seen visually in Fig.~\ref{Fig:Re50100200_We065_BM215}. 
\subsection{Comparison with the evaporation models}
\label{low-orderComp}
After the qualitative description, the evaporation rate and total mass transfer are quantified and the results are used to evaluate performance of the classical and Abramzon–Sirignano (A–-S) models.  The evaporation models involve gross simplifying assumptions such as a well-behaved boundary layer (BL) theory~\cite{abramzon1989droplet} on a non-deformable spherical droplet~\cite{sazhin2006advanced}. Thus, their performance requires a critical assessment particularly under the conditions that these assumptions are not valid, e.g., separated flow with a significant droplet deformation. 

We consider the classical~\citep{sazhin2006advanced}
and Abramzon and Sirignano (A–-S) \citep{abramzon1989droplet, sirignano2010fluid} models as they are widely used in engineering applications. In the classical model, average Sherwood number ($Sh$) is given by
\begin{equation} 
Sh = Sh_0 \frac{\ln\left(1+B_M \right)}{B_M},   
\label{eqn:Classicalmodel}
\end{equation}
where $Sh_0$ represents the Sherwood number in the absence of Stefan flow and it is usually evaluated using the Frossling's correlation~\cite{frossling1963evaporation}
\begin{equation} 
Sh_0 = 2 + 0.552 {Re}^{\frac{1}{2}} {Sc}^{\frac{1}{3}}.
\label{eqn:Frossling}
\end{equation}
Abramzon and Sirignano~\citep{abramzon1989droplet, sirignano2010fluid} approximated the droplet as a collection of evaporating wedges and utilized the Falkner–Skan boundary layer solution. They also took into account Stefan flow-induced thickening of boundary layer. The Abramzon and Sirignano model introduces a correction to the classical model and can be written as
\cite{abramzon1989droplet}
\begin{eqnarray}
   Sh &=& (2+\frac{Sh_0-2}{F_M}) \frac{\ln\left(1+B_M \right)}{B_M}, \nonumber \\ 
   F_M &=& {\left(1+B_M \right)}^{0.7} \frac{\ln\left(1+B_M \right)}{B_M},
\label{eqn:A-S}
\end{eqnarray}
where $F_M$ is the correction factor. Note that the classical model is recovered for $F_M = 1$.

\begin{figure}[hbt!]
\centering
\includegraphics[width=\textwidth]{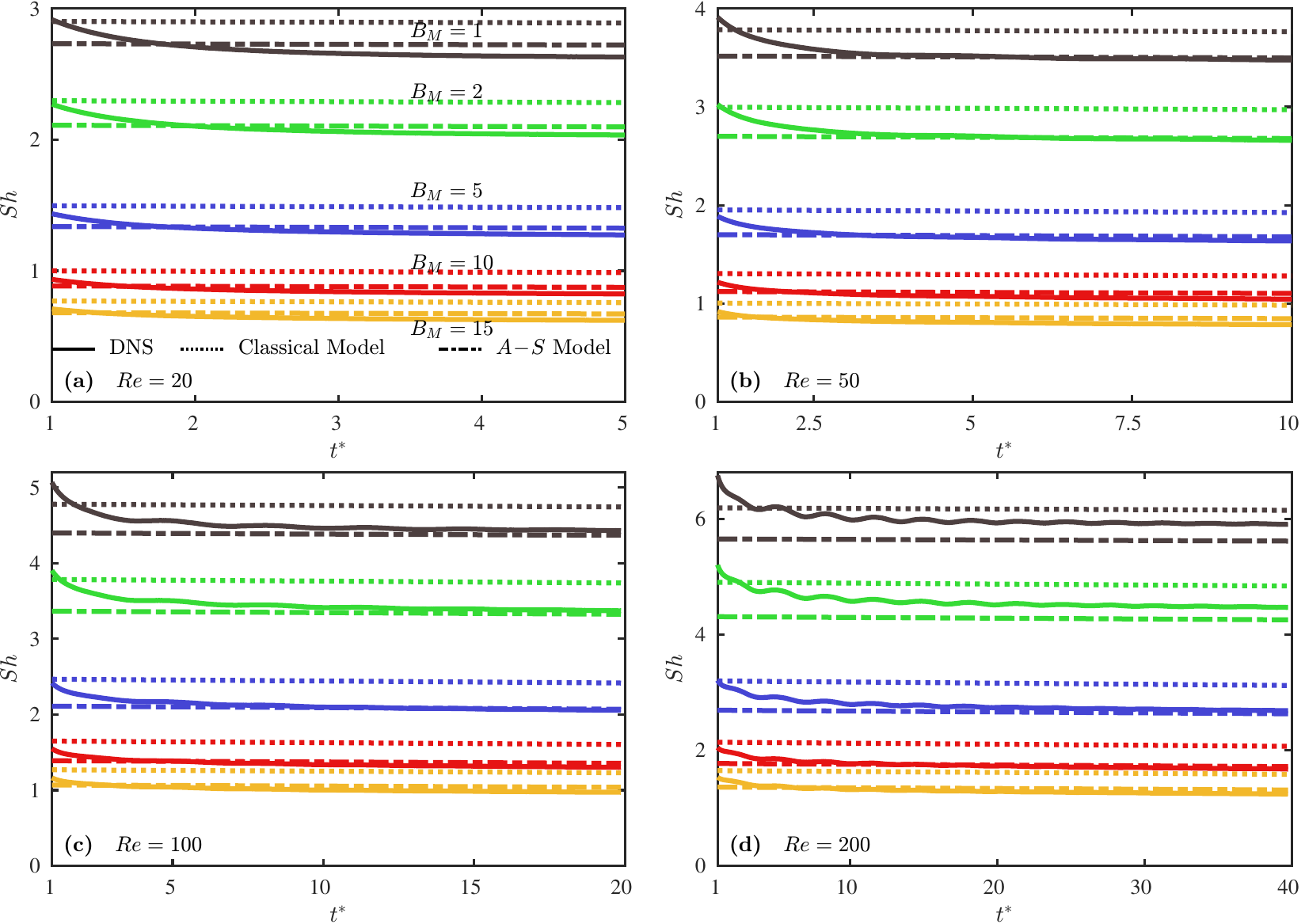}
\caption{Interface-resolved results of surface-averaged Sherwood number over a nearly spherical droplet at $Re\in[20,\; 50,\; 100,\; 200]$ in comparison with the low-order models.}
\label{Fig:ShRe2050100200}
\end{figure}

Figure~\ref{Fig:ShRe2050100200} shows temporal evolution of average Sherwood number for an evaporating droplet at $Re\in[20,\; 50,\; 100,\; 200]$ and $B_M\in[1,\; 2,\; 5,\;10,\; 15]$. The numerical results are compared with the low-order models. In all the cases, $Sh$ is initially high but decays over time as a quasi-steady evaporation regime is reached. Note that, for the low-order models, $Sh_0$ is evaluated using the Frossling's correlation (Eq.~\eqref{eqn:Frossling}) where the Reynolds number is evaluated using the instantaneous droplet diameter obtained using the interface-resolved simulations. We also note that, hereafter, direct numerical simulation (DNS) is interchangeably used to denote the interface-resolved simulation results. As seen, the classical model consistently over-predicts $Sh$ compared to the interface-resolved simulation results mainly because it neglects the boundary layer thickening by Stefan flow and its adverse effect on convective mass transfer from droplet. On the other hand, the A--S model agrees more closely with the numerical results across the parameter space.  

In practical applications, the main goal is to predict overall mass flux and mass transfer rate from a droplet. Thus, following \citet{SETIYA2023104455}, we introduce dimensionless mass flux ($\dot{m}^{\prime\prime}/\dot{m}_{d^2}^{\prime\prime}$) and mass transfer rate ($\dot{m}/\dot{m}_{d^2}$) to quantify local and overall evaporation characteristics of a droplet. Here, $\dot{m}_{d^2}^{\prime\prime}$ and $\dot{m}_{d^2}$, respectively, denote the mass flux and mass transfer rate obtained from the $d^2-$law law, i.e.,
\begin{eqnarray}
\dot{m}_{d^2} &=&  2 \pi\rho_gD_{vg} d \ln(1+B_M), \nonumber\\ 
\dot{m}_{d^2}^{\prime\prime} &=& \frac{\dot{m}_{d^2}}{\pi d^2}=\frac{2}{d}\rho_g D_{vg} \ln(1+B_M), 
\end{eqnarray}
where $d$ is the instantaneous droplet diameter.  

\begin{figure}[hbt!]
\centering
\includegraphics[width=\textwidth]{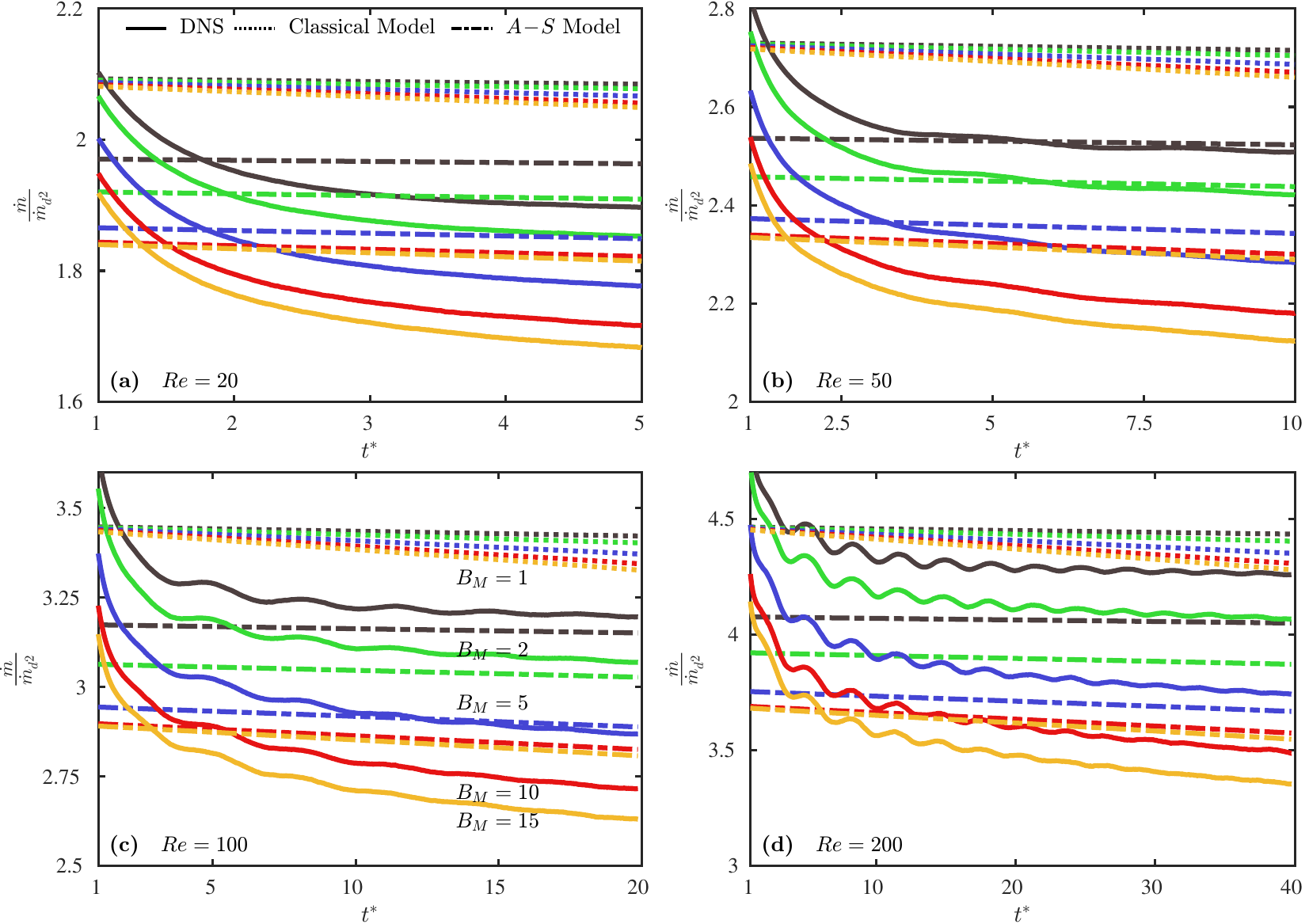}
\caption{Interface-resolved results of normalized evaporation rate of a nearly spherical droplet at $Re\in[20,\;50,\;100,\;200]$ in comparison with the low-order models. }
\label{Fig:ShRe2050100200mdot}
\end{figure}

Time evolution of interface-resolved results and evaporation model predictions of normalized evaporation rate are plotted in Fig.~\ref{Fig:ShRe2050100200mdot} for the same cases as in Fig.~\ref{Fig:ShRe2050100200}. The results show that the classical model predicts nearly constant normalized evaporation rates regardless of mass transfer number ($B_M$). A slight decrease observed in values predicted by the classical model with increasing $B_M$ is solely caused by evaporative reduction in droplet size, as a droplet with a higher mass transfer number ($B_M$) evaporates faster. A decaying trend in the DNS results is also attributed to shrinking of droplet size as Sherwood and Reynolds numbers are proportional to droplet radius (see Eq.~(\eqref{eqn:nodimensional})). The fluctuations seen in $Sh$ for the cases of $Re=100$ and $Re=200$ are attributed to small oscillations in droplet shape. Figure~\ref{Fig:ShRe2050100200mdot} reveals that the normalized evaporation rate decreases as $B_M$ increases. This is attributed primarily to boundary layer inflation caused by Stefan flow, which is neglected by the classical model as well as in the $d^2$-law. The A--S model partially captures this phenomenon but its predictions are not consistently accurate across all Reynolds and mass transfer numbers. For instance, the A--S model tends to overpredict the evaporation rate at higher mass transfer number ($B_M\in[10,\;15]$) for all the Reynolds numbers. Conversely, at lower mass transfer numbers ($B_M\in[1,\;2]$), the model predictions agree better with the DNS results although slight deviations remain visible at higher Reynolds numbers, particularly for $Re=200$. The discrepancy between the A--S model and the DNS results is pronounced as $B_M$ and Reynolds numbers increase. For instance, at $Re=200$ and $B_M=15$, the A--S model overpredicts the normalized evaporation rate by as much as 15\%.

\begin{figure}[hbt!]
\centering
\begin{tabular}{cc}
\includegraphics[width=\textwidth]{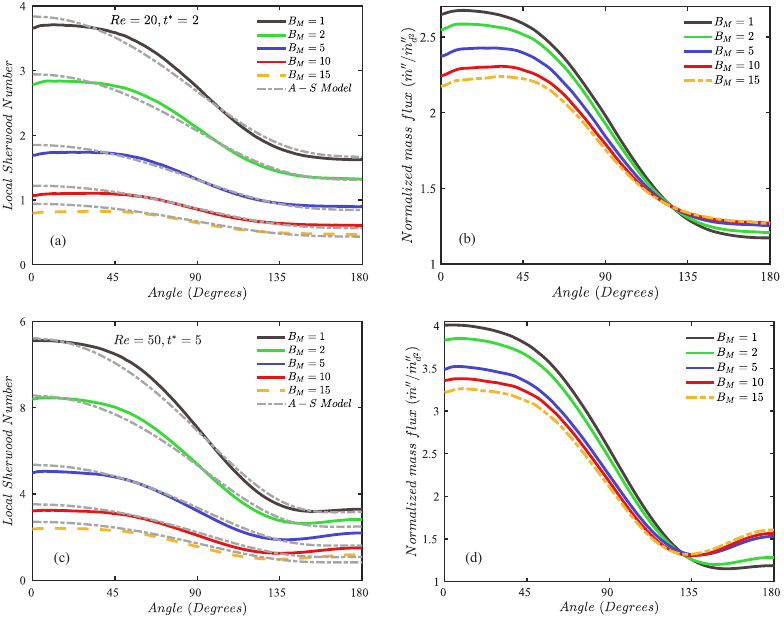}
\end{tabular}
\caption{Local Sherwood number and normalized mass flux over a nearly spherical droplet at $Re=20$, $t^*=2$ (top row) and $Re=50$, $t^*=5$ (bottom row). }
\label{Fig:Re2050mdotSh}
\end{figure}

\begin{figure}[hbt!]
\centering
\begin{tabular}{cc}
\includegraphics[width=\textwidth]{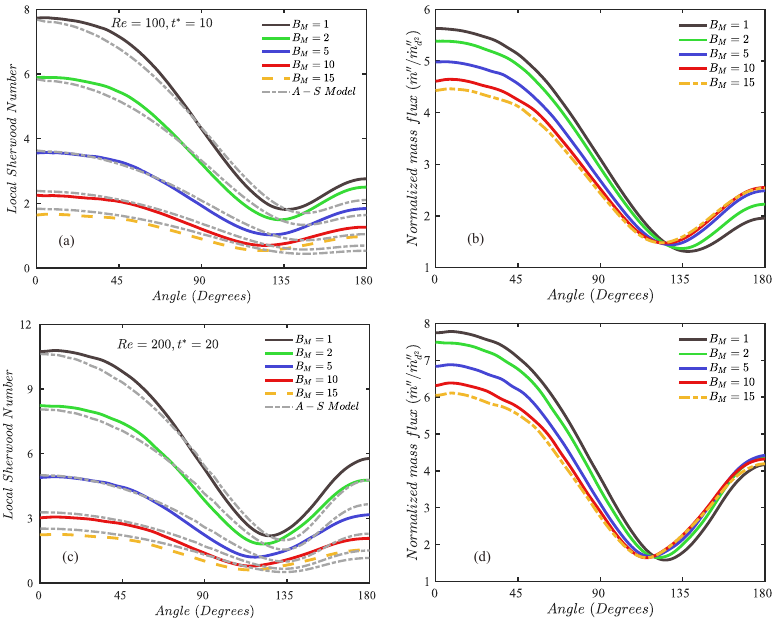}
\end{tabular}
\caption{Local Sherwood number and normalized mass flux over a nearly spherical droplet at $Re=100$, $t^*=10$ (top row) and $Re=200$, $t^*=20$ (bottom row).}
\label{Fig:Re1002000mdotSh}
\end{figure}

We next show the distribution of local Sherwood number ($Sh_l$) and normalized evaporative mass flux for these cases to elucidate potential sources of discrepancy between interface-resolved simulations and low-order models. We consider a quasi-steady evaporation regime where flow field is fully developed around droplet. Figures~\ref{Fig:Re2050mdotSh} and \ref{Fig:Re1002000mdotSh} show distribution of local Sherwood number and normalized mass flux over an evaporating droplet for $Re\in[20,\;50,\;100,\;200]$ and mass transfer numbers of $B_M \in [1,\;2,\;5,\;10,\;15]$. Here, $Sh_0$ is computed using a separate sharp-interface immersed-boundary simulation for mass transfer from a solid sphere with the same initial droplet radius and external flow conditions as in the interface-resolved simulations \cite{salimnezhad2024hybrid}. Then,  this $Sh_0$ is  used in the A--S model to compute the distribution of $Sh$ over the droplet since the Frossling correlation gives only a surface-averaged value for $Sh_0$. As shown in Fig.~\ref{Fig:Re2050mdotSh}a, the A--S model agrees reasonably well with the interface-resolved simulations for all the mass transfer numbers.  However, the discrepancy increases in the wake region especially after $135^o$ as shown in Fig.~\ref{Fig:Re2050mdotSh}c. The discrepancy is also intensified in the leeward side of droplet as Reynolds number increases as shown in Figs.~\ref{Fig:Re1002000mdotSh}a \& \ref{Fig:Re1002000mdotSh}c. We later elaborate on this phenomena in a greater detail. 


An interesting and somewhat counterintuitive trend is seen from comparison of Figs.~\ref{Fig:Re2050mdotSh}b and \ref{Fig:Re2050mdotSh}c as well as Figs.~\ref{Fig:Re1002000mdotSh}b and \ref{Fig:Re1002000mdotSh}c: the droplets with a higher mass transfer number exhibit lower normalized evaporative mass flux in upstream region and yet, surprisingly, this trend reverses in the rear portion of the droplet. This behavior can be explained by considering the thickening of the boundary layer in windward side of droplet caused by a stronger Stefan flow at a higher $B_M$, which reduces convective mass transfer rate. In contrast, on leeward side where boundary layer assumption is not valid any more, the same phenomena leads to a higher normalized evaporative flux for the cases with higher $B_M$ values. This reversed trend arises since, at a fixed Reynolds number, a larger value of mass transfer number leads to formation a larger recirculation zone behind droplet as shown in Fig.~\ref{Fig:Re50100200_We065_BM215}. This, in turn, leads to a stronger convective flow within the wake region. As shown in Fig.~\ref{Fig:FV_wake_Re100_BM_0_2_5_15_updated-cropped}, the velocity magnitude along wake centerline increases as recirculation zone enlarges. Consequently, vapor is carried from leeward side of droplet by both Stefan flow and the wake flow toward the outer periphery of the closed wake and further downstream of the droplet, thereby enhancing evaporation on the rear side of the droplet. Enlargement of the recirculation zone poses a challenge to accuracy of the A--S model, which inherently relies on the boundary-layer assumption. As demonstrated in Figs.~\ref{Fig:Re2050mdotSh}b,d and \ref{Fig:Re1002000mdotSh}b,d, an increasing portion of droplet experiences a reversed trend in normalized evaporative flux distribution as Reynolds number increases.  For example, at $Re=200$, a noticeable discrepancy between the A--S model and DNS predictions emerges starting from approximately $120^o$. This observation indicates that, as Reynolds number increases, fidelity of boundary-layer-based low-order models such as the A–S model may start to deteriorate, particularly in predicting evaporation rates in regions dominated by wake flows. 

\subsection{Effects of Stefan flow}
\label{CF}
 
\begin{figure}[t]
\centering
\begin{tabular}{ccc}
\includegraphics[scale=0.32]{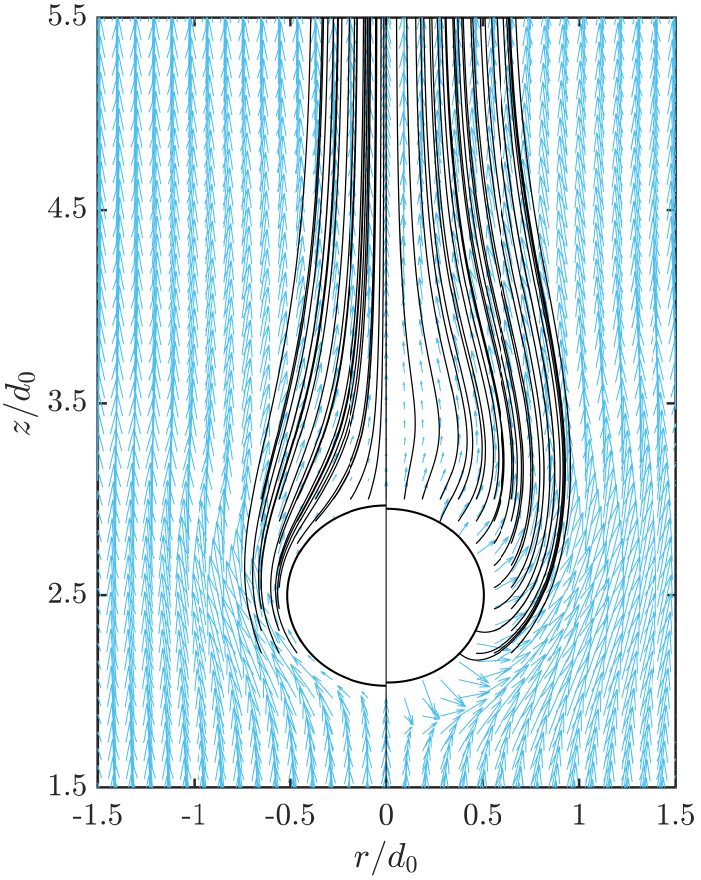}
&
\includegraphics[scale=0.32]{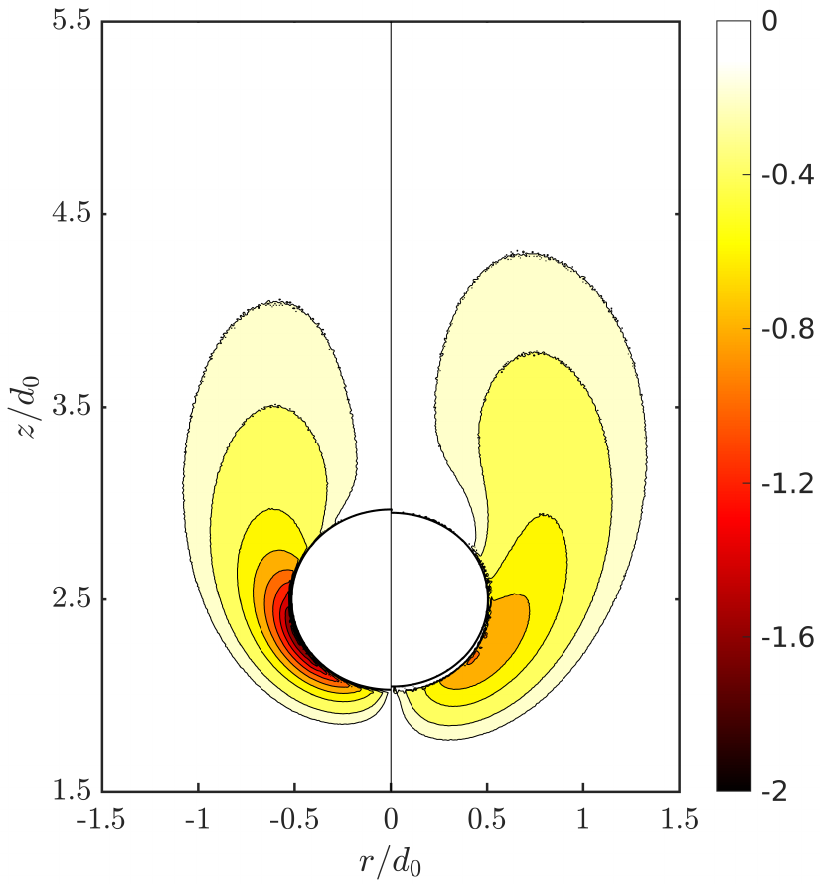}
&
\includegraphics[scale=0.32]{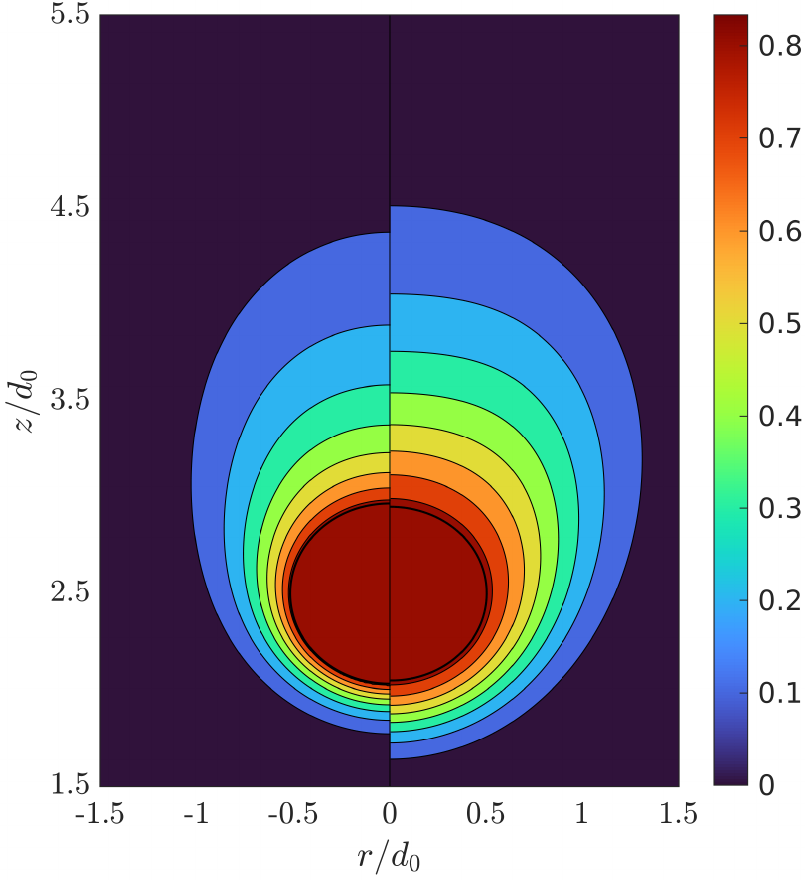} 
\\
\includegraphics[scale=0.32]{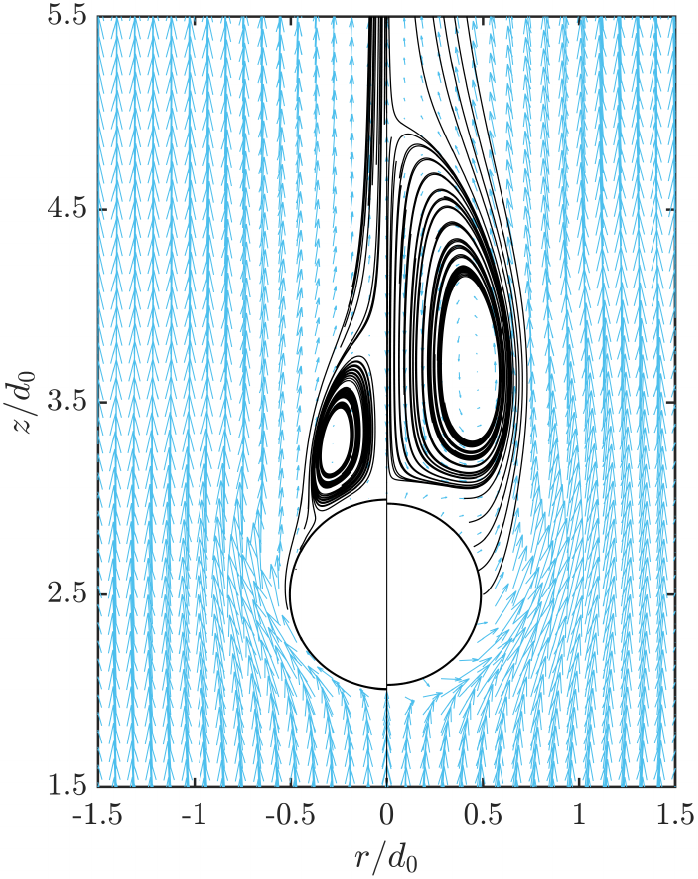}
&
\includegraphics[scale=0.32]{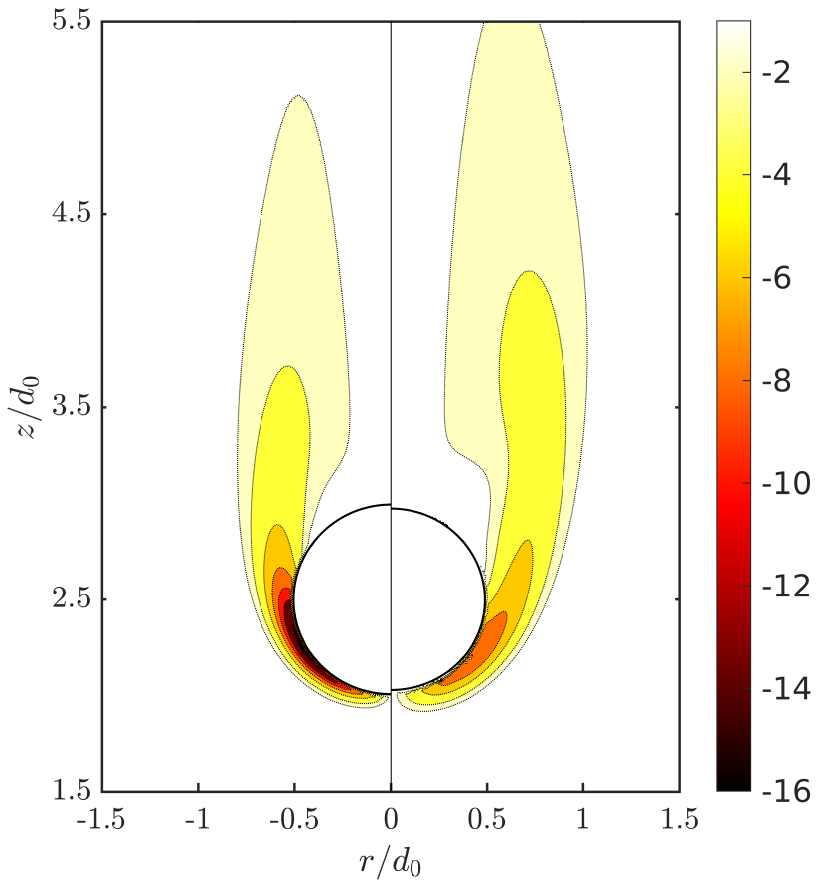}
&
\includegraphics[scale=0.32]{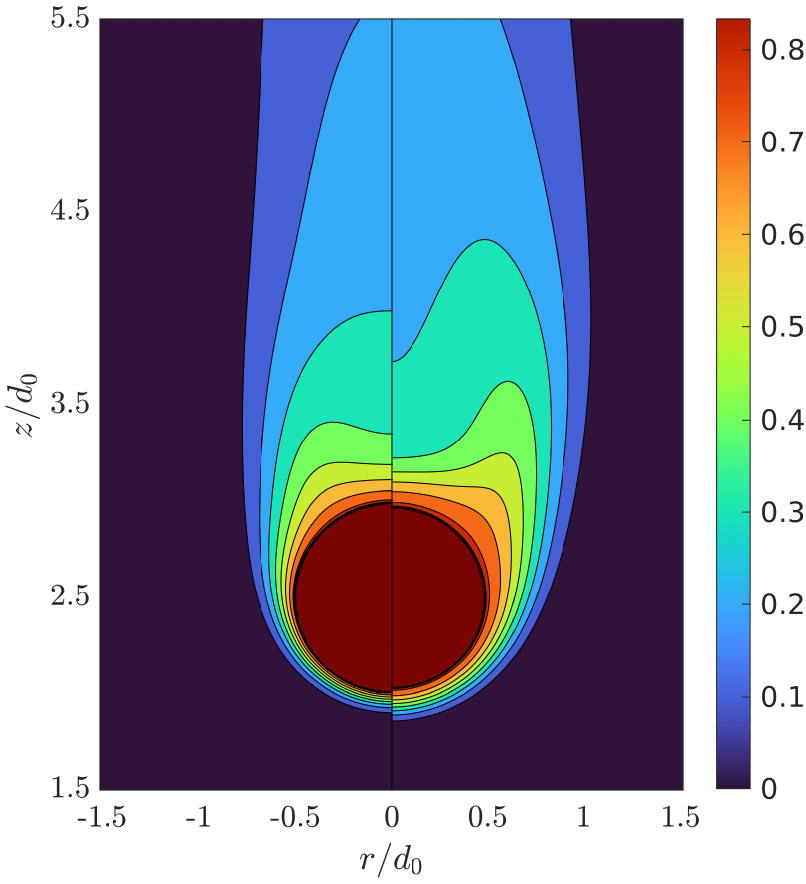}
 \\
\scriptsize{(a)} & \scriptsize{(b)} & \scriptsize{(c)}
\end{tabular}
\caption{(a) Velocity vectors, (b) vorticity field,  and (c) vapor mass-fraction field in the absence (left portion) and in presence of Stefan flow (right portion) for a nearly spherical droplet ($We=0.65$) at $Re=20$ \& $t^*=2$ (top row) and $Re=100$ \& $t^*=10$ (bottom row). (Domain:$4d_0 \times 8d_0$, Grid:$512 \times 1024$, $B_M=5$, $We=0.65$).}
\label{Fig:VelVorY-t9}
\end{figure}

In this section, we leverage interface-resolved simulations to investigate the adverse effect of Stefan flow on droplet evaporation under various convective flow conditions and compare the findings with  results obtained using the low-order models. For this purpose, simulations are performed for two distinct scenarios: in the presence and absence of Stefan flow. In the later case, Stefan flow is artificially suppressed. Figure~\ref{Fig:VelVorY-t9} qualitatively shows  effects of Stefan flow on flow and mass-fraction fields for $Re=20$ and $Re=100$ cases at $We=0.65$ and $B_M=5$. As seen, Stefan flow has a profound effect on droplet evaporation and associated flow and mass fraction fields. Also the presence of Stefan flow significantly changes  vorticity field as seen in Fig.~\ref{Fig:VelVorY-t9}b.

To quantify effects of Stefan flow, we utilize the Sherwood number ratio ($SR$) defined as
\begin{equation}
 SR = Sh/Sh_0.
\label{eqn:CF}
\end{equation}
This parameter signifies reduction in surface-averaged Sherwood number $(Sh)$ due to the blowing effect of  Stefan flow. Note that $Sh_0$ is evaluated by suppressing Stefan flow in interface-resolved simulations. This is achieved by simply setting the right hand side of the continuity equation (Eq.~\ref{eqn:continuity}) to zero. For the low-order models, $Sh_0$ is obtained from the Frossling correlation (Eq.~\eqref{eqn:Frossling}).  

\begin{figure}[t!]
\centering
\includegraphics[width=\textwidth]{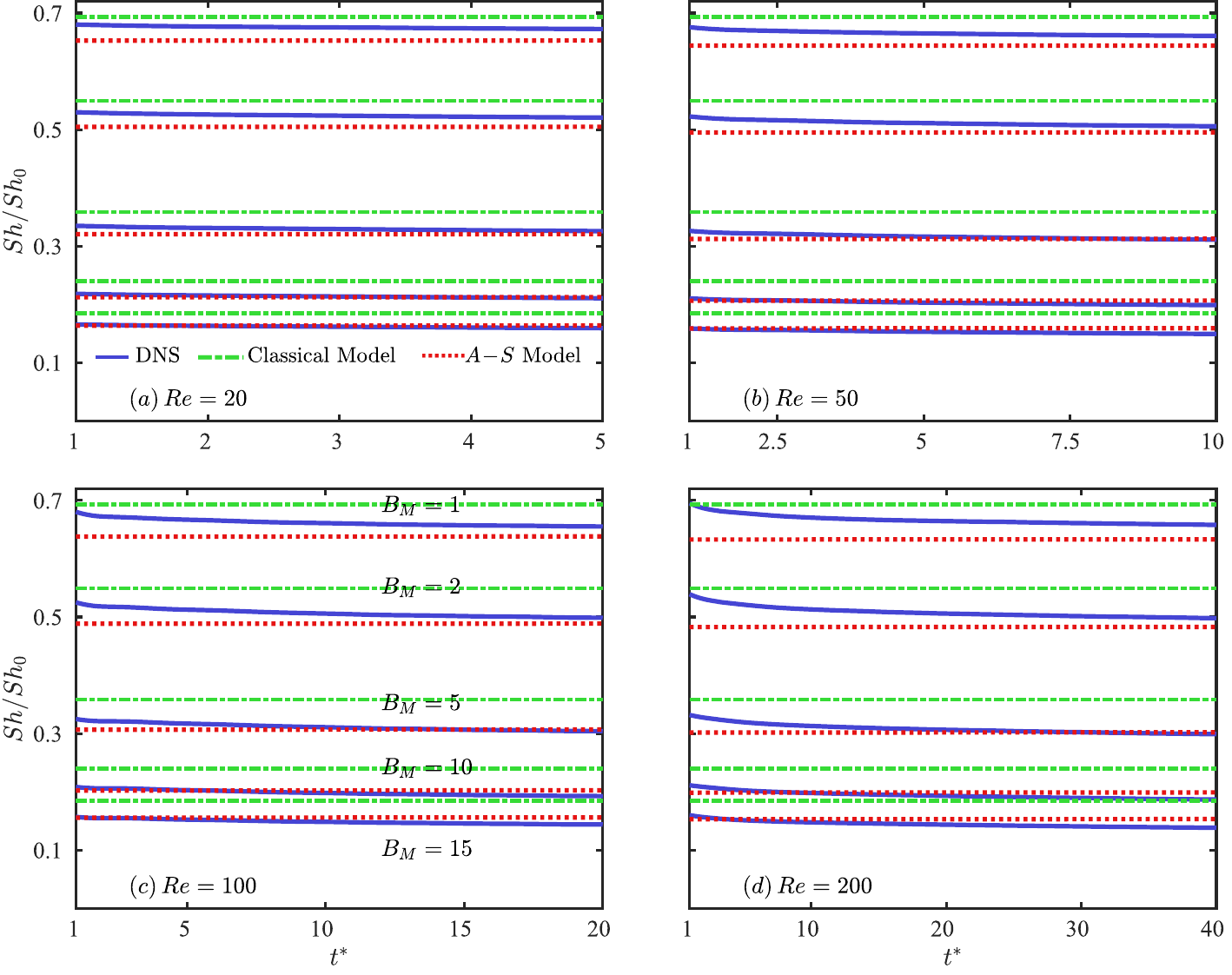}
\caption{Time evolution of the averaged Sherwood number ratio ($Sh/Sh_0$) for the case of a nearly spherical droplet ($We=0.65$) in the ranges of $Re\in [20\; 50\; 100\; 200]$ and $B_M \in [1\; 2\; 5\; 10\; 15]$.) }
\label{Fig:CFAve}
\end{figure}

Simulations are performed for a nearly spherical droplet in the ranges of $Re\in [20,50,100,200]$ and $B_M\in [1,2,5,10,15]$. The numerical results are plotted in Fig.~\ref{Fig:CFAve} together with the evaporation model predictions. For the case of $Re=20$, agreement between the A--S model and interface-resolved simulations improves  as  mass transfer number increases while the classical model generally overpredicts $Sh/Sh_0$ especially at higher mass transfer numbers. To examine effects of evaporation-induced reduction in droplet volume, additional simulations are conducted, where droplet volume is held constant by injecting liquid into droplet to compensate the evaporative mass loss in every time step. The effect of volume reduction on $Sh/Sh_0$ is found to be less than 3.5\% for the case of $B_M=15$ and $Re=200$, demonstrating that the influence of volume reduction is generally small.

\begin{figure}[t]
\centering
\includegraphics[width=\textwidth]{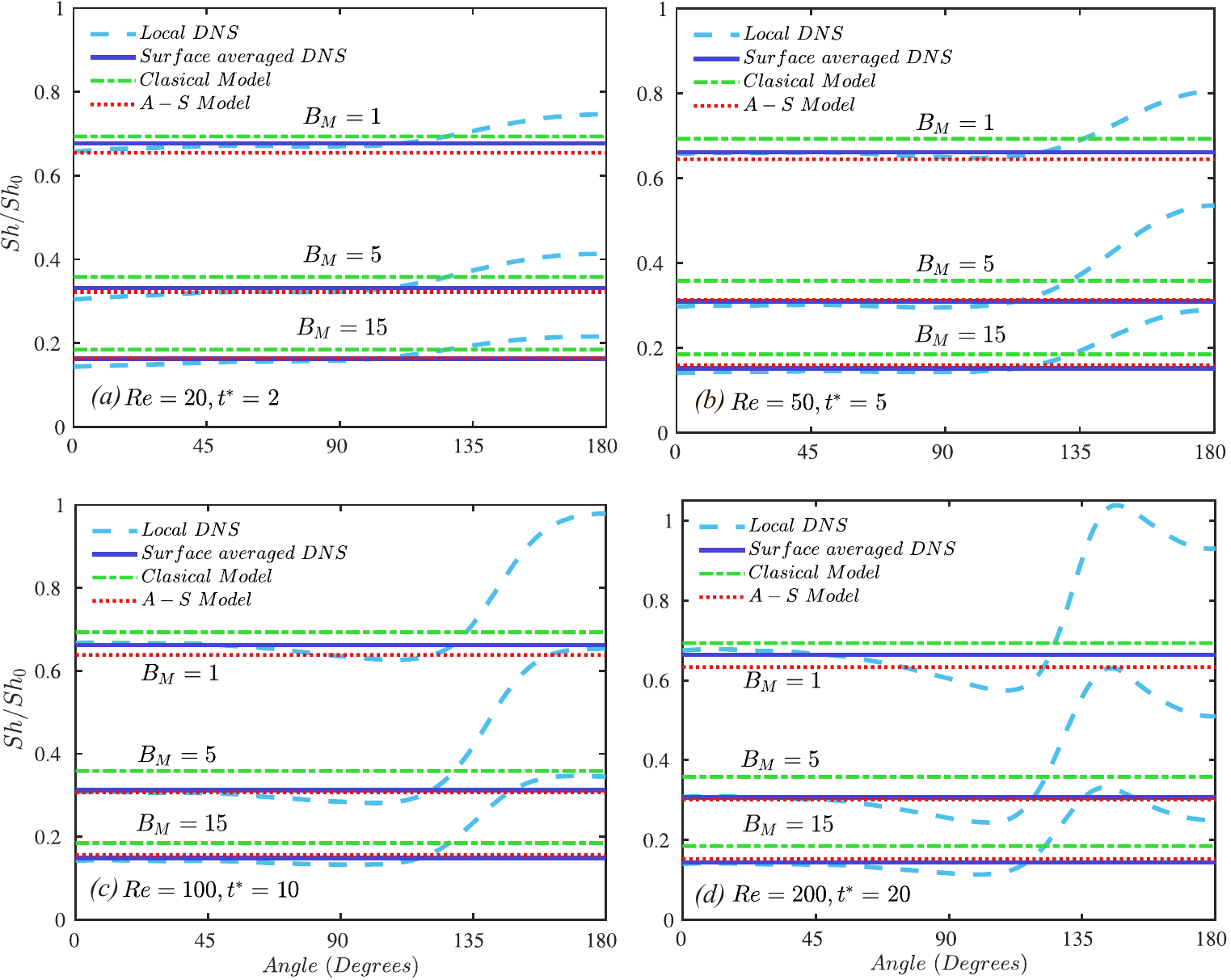}
\caption{Variation of the local Sherwood number ratio ($LSR$) over a nearly spherical droplet for $Re\in [20,\; 50,\; 100,\; 200]$ and $B_M \in [1,\; 5,\; 15]$. }
\label{Fig:CFRe20Re50Re100Re200}
\end{figure}

While $SR$ provides an overall picture for the impact of Stefan flow on Sherwood number, examining its local variation can provide deeper insights into the nuanced effects of Stefan flow under different convective flow conditions. We thus introduce the local Sherwood number ratio defined as
\begin{equation}
    LSR(\theta) = Sh(\theta) / Sh_0(\theta),
\label{eqn:CF2}
\end{equation}
where $LSR$ is a function of the angle ($\theta$) and is evaluated using the interface-resolved simulations. Figure~\ref{Fig:CFRe20Re50Re100Re200}  shows quasi-steady variations of  Sherwood number ratio for a nearly spherical droplet  in the ranges of $Re\in [20\; 50\; 100\; 200]$ and $B_M\in[1\;5\;15]$.  As seen, both $Re$ and $B_M$ have profound effects on local evaporation rate especially in the wake region. $LSR$ remains nearly uniform over droplet before the boundary layer separation. In this region, the numerical results are relatively in good agreement with the model predictions. However, as Reynolds number increases, the discrepancy between the numerical results and the model predictions becomes more pronounced especially in the wake region. An examination of Figs.~\ref{Fig:CFRe20Re50Re100Re200}a and \ref{Fig:CFRe20Re50Re100Re200}d reveals even qualitative differences between results obtained for the $Re=20$ and $Re=200$ cases. For $Re \le 100$, $LSR$ becomes maximum at the rear stagnation point ($\theta=180^o$) but, for the $Re=200$ case, it attains a maximum value at about $\theta = 145^o$ and then starts decreasing until reaching nearly a zero slope at the rear stagnation point. As $B_M$ increases, a qualitatively similar trend is observed but the discrepancy is amplified significantly due to the boundary layer thickening and early flow separation owing to intensified Stefan flow. It is interesting to observe that the discrepancy is negative until about $\theta\approx 120^o$ and then it becomes positive. Since the positive and negative discrepancies partially cancel each other, the low-order models predict the average evaporation rates significantly better than the local distributions as seen in Fig.~\ref{Fig:CFAve}.

\begin{figure}[t]
\centering
\begin{tabular}{cc}
   
\includegraphics[scale=0.61]{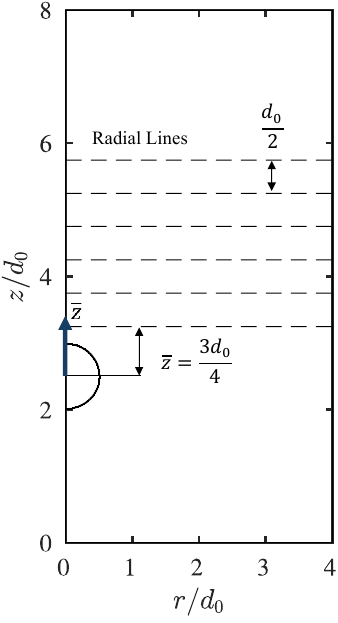}  &  
\includegraphics[scale=0.61]{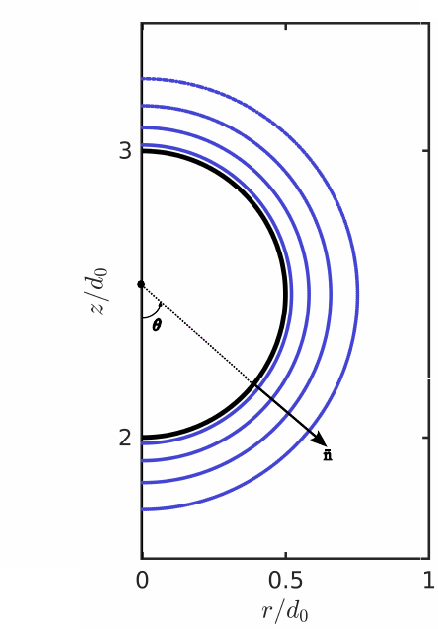}
\\
\scriptsize{(a)} &  \scriptsize{(b)}
\end{tabular}
\caption{(a) Radial lines along which mass fraction and axial velocity are examined in the wake of the droplet. (b) Probe locations around the droplet corresponding to normal distances of $0.02d_0,\; 0.08d_0,\; 0.15d_0$, and $0.25d_0$ from the droplet surface.}
\label{Fig:Probes}
\end{figure}

We next examine the wake region in the presence and absence of Stefan flow. For this purpose, we consider six equally-spaced radial probe lines in downstream of an evaporating droplet as shown in Fig.~\ref{Fig:Probes}a. Distance between the radial lines is $d_0 / 2$ with the closest line positioned at $3d_0 / 4$ from the droplet center. Lines are selected to encompass a significant portion of recirculation zone for the range of Reynolds numbers considered here. 

\begin{figure}[t]
\centering
\includegraphics[width=\textwidth]{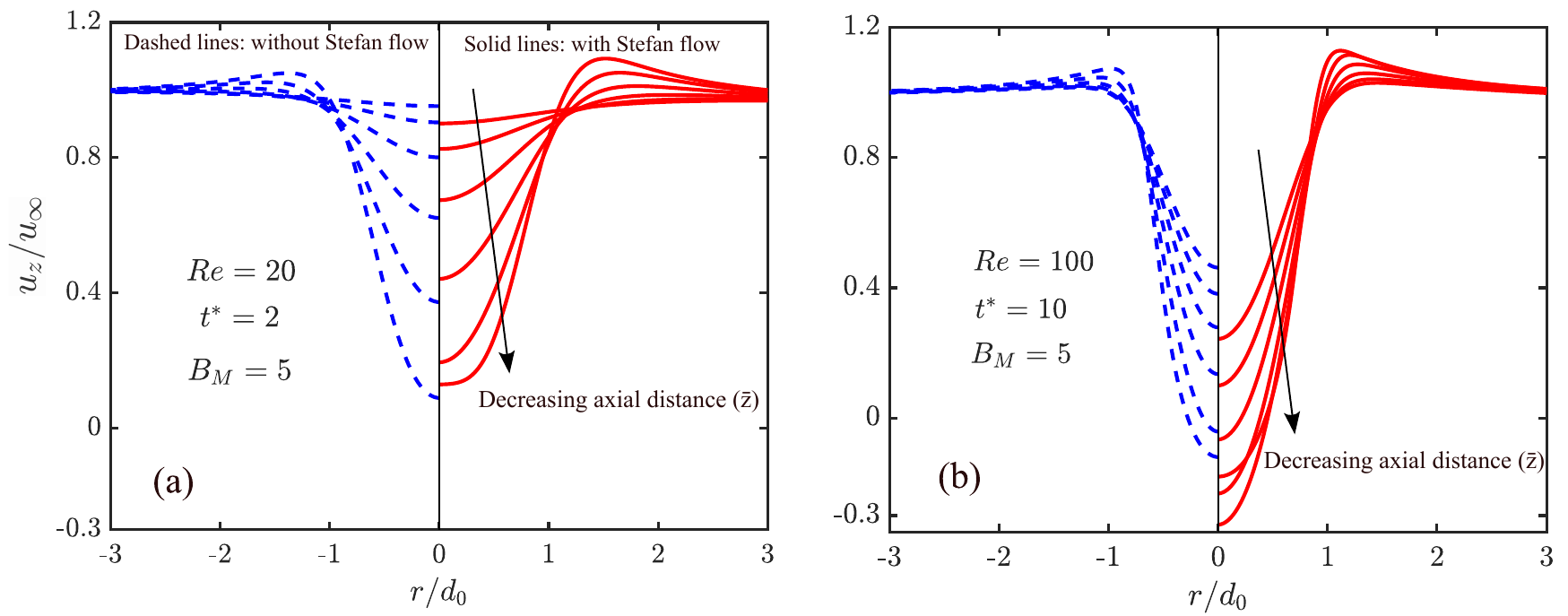}
\caption{Normalized axial velocity profiles in the wake of a nearly spherical ($We=0.65$) evaporating droplet at $B_M=5$ for (a) $Re=20$  and  (b) $Re=100$. Red solid and blue dashed lines indicate the cases in the presence (right) and  absence (left) of Stefan flow, respectively.}
\label{Fig:highBMVaxial}
\end{figure}

\begin{figure}[t]
\centering
\includegraphics[width=\textwidth]{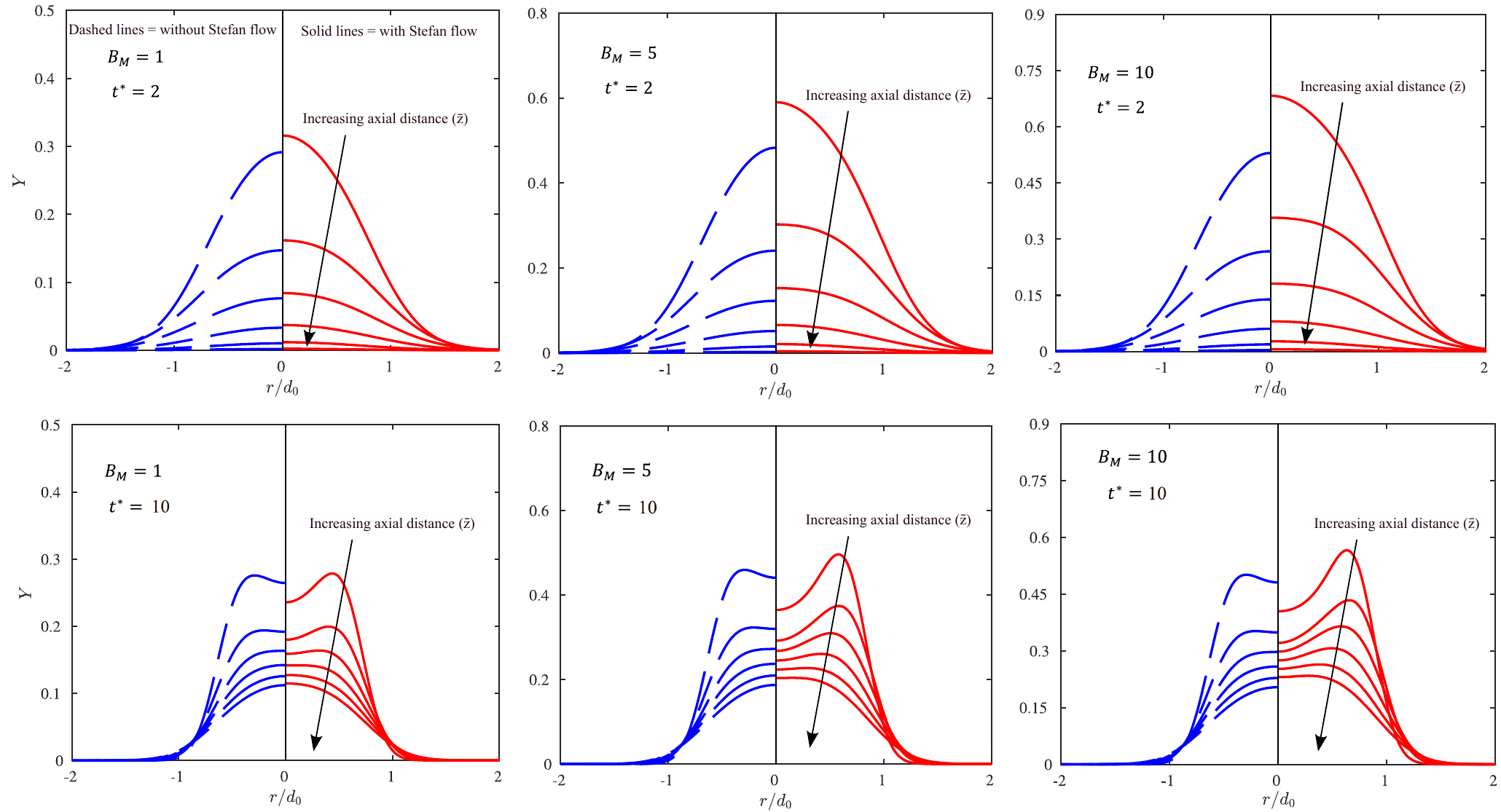}
\caption{Mass fraction profiles in the wake of an evaporating droplet for $Re=20$ at $t^*=2$ (top row) and $Re=100$ at $t^*=10$ (bottom row). Red solid and blue dashed lines indicate the cases in the presence (right) and  absence (left) of Stefan flow, respectively.}
\label{Fig:MassfracRe_20_100}
\end{figure}

Figure~\ref{Fig:highBMVaxial} shows the impact of Stefan flow on normalized axial velocity profiles for the $Re=20$ and $Re=100$ cases at $B_M=5$. As seen, there is no backflow for the case of $Re=20$ both in  presence and absence of Stefan flow whereas, at $Re=100$, there is a significant backflow indicating a recirculating wake region behind the droplet and the backflow is amplified by Stefan flow. Figure~\ref{Fig:MassfracRe_20_100} shows quasi-steady vapor mass fraction profiles in the wake region for the $Re=20$ and $Re=100$ cases at mass transfer numbers of $B_M\in [1,\; 5,\; 10]$. For the $Re=20$ case, the mass fraction is the maximum at the centerline and decreases monotonically toward the edge of the wake eventually reaching the freestream value. For all the mass transfer numbers, mass-fraction profiles are similar but the magnitude is reduced in the absence of Stefan flow. For the case of $Re=100$,  mass-fraction profiles are qualitatively different than those in the $Re=20$ case, i.e.,  mass-fraction increases from a local minimum at the centerline and attains a maximum before relaxing toward the free stream value at the edge of the wake. Stefan flow considerably alters flow field in the recirculation zone and impacts distribution of vapor concentration. As seen, in contrast with the $Re=20$ case, the presence of Stefan flow results in a reduction in mass fraction field close to the droplet surface in the wake region for this high Reynolds number case.

\begin{figure}[t]
\centering
\includegraphics[width=\textwidth]{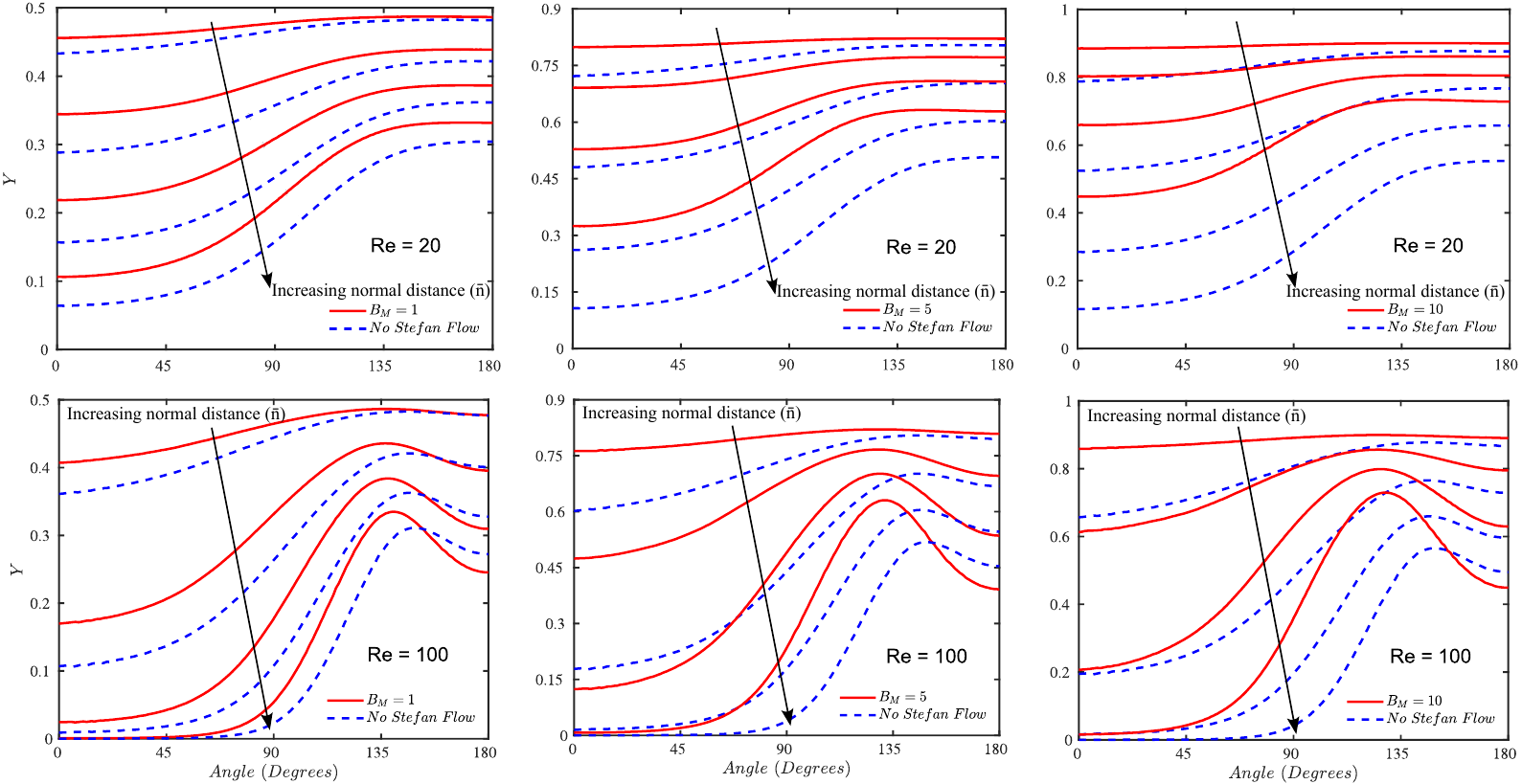}
\caption{Mass fraction field around a nearly spherical droplet ($We=0.65$) at $B_M=1\;, B_M=5$ and $ B_M=10$ (from left to right) and various normal distances from the droplet surface, as sketched in Fig.~\ref{Fig:Probes}b, for $Re=20$ at $t^* = 2$ (top row) and for $Re=100$ at $t^* = 10$ (bottom row). Red solid and blue dashed lines indicate the cases in the presence and absence of Stefan flow, respectively.}
\label{Fig:highBMt9}
\end{figure}

Next we evaluate influence of Stefan flow on distribution of mass fraction in the vicinity of the droplet. Figure~\ref{Fig:Probes}b shows the probe locations around the droplet at normal distances of $0.02 d_0, \ 0.08 d_0, \ 0.15 d_0$, and $0.25 d_0$ from the droplet surface where mass fraction is plotted. These locations are selected based on the film theory suggested by \citet{abramzon1989droplet} and \citet{bird1960stewart}. The film thickness is defined by $\delta_{t_0} = d_0 / (Sh_0 - 2)$ for the case of mass transfer from droplet without taking into account the blowing effect of Stefan flow. Here, $Sh_0$ is  Sherwood number in the absence of Stefan flow, evaluated using Eq.~\eqref{eqn:Frossling}. For the $Re=100$ case, the film thickness is approximately $\delta_{t_0}\approx 0.3d_0$; thus, the probes shown in Fig.~\ref{Fig:Probes}b are positioned inside the boundary layer (BL) for all the Reynolds numbers considered in this section. Figure~\ref{Fig:highBMt9} shows effects of Stefan flow on distribution of vapor mass fraction around an evaporating droplet at the probe locations. The analysis is presented for mass transfer numbers of $B_M \in [1,\; 5,\; 10]$), and Reynolds numbers of $Re\in [20,\; 100]$. For each Reynolds number, the results are shown at $t^* = 2$ and $t^*=10$, respectively, corresponding to a quasi-steady state of evaporation process. The results indicate that, in all the cases, Stefan flow significantly enhances vapor mass fraction field around the droplet. At $Re=20$, mass fraction is less sensitive to the convective effects as evidenced by smooth variations along the probe locations.  In contrast, at $Re=100$, influence of convection is clear in shaping mass fraction field around the droplet, such that the boundary layer region and recirculation zone can be differentiated  examining $Y$ values at the farthermost probes. The increase in mass transfer number saturates near droplet with more amount of vapor, giving rise to a significant difference compared to that of the no Stefan flow case. Note that, at $B_M=10$, mass fraction at the closest probe location to the droplet is approximately uniform for all the Reynolds numbers, and does not exhibit considerable change with respect to the angular position. This is consistent with the results shown in Fig.~\ref{Fig:Re1002000mdotSh}a, where local Sherwood number is considerably lower for the $B_M=10$ case compared to other cases.  This behavior might be attributed to a stronger influence of Stefan flow compared to the convective effects in vicinity of an evaporating droplet, particularly at higher mass transfer numbers. Note that the difference in mass fraction distribution between cases with and without Stefan flow is more pronounced in the upstream of the droplet compared to that in the recirculation zone, particularly for $Re=100$. This observation shows that vapor mass fraction distribution in the boundary-layer region might be more significantly influenced by  Stefan flow, while the recirculation zone experiences only a modest variation in $Y$ values, particularly at higher Reynolds numbers.

\subsection{Effects of droplet deformation}
\label{DefOnEvrate}

In the low-order evaporation models used in large-scale spray simulations, impact of droplet deformation is usually ignored and droplets are assumed to remain spherical. This is obviously not the case especially for large droplets that undergo a significant deformation. Droplet deformation is expected to increase evaporation rate as it increases surface area. In addition, deformation also alters flow field around droplet, which, in turn, influences mass transfer rate. At high evaporation rates, Stefan flow further complicates these interactions as it can greatly affect flow field around droplet as well as its deformation~\cite{salimnezhad2024hybrid}. Further simulations are performed to shed light on these highly non-linear interactions and resulting effect on droplet evaporation in this section. 

\begin{figure}[t]
\centering
\begin{tabular}{cc}
\includegraphics[scale=0.32]{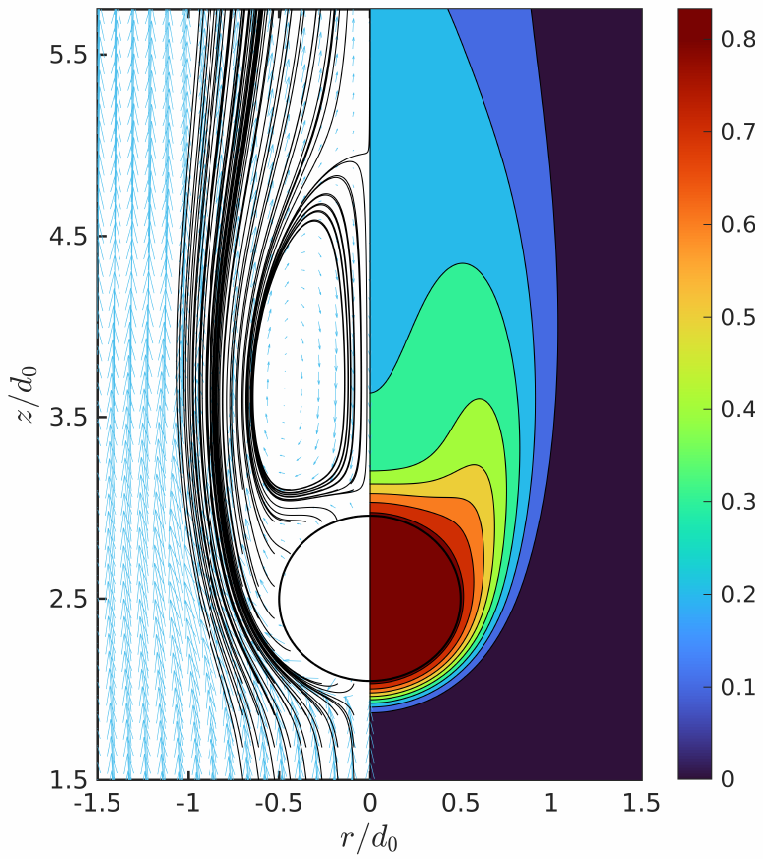}
&
\includegraphics[scale=0.32]{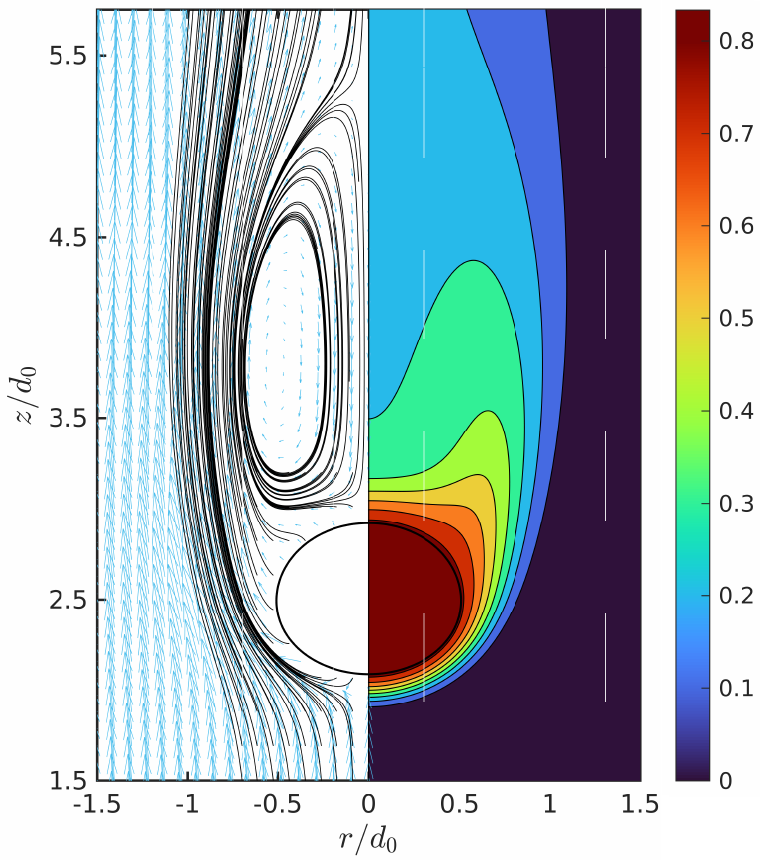}
\\
\scriptsize{(a) $We = 1$} & \scriptsize{(b) $We = 3$} 
\\
\includegraphics[scale=0.32]{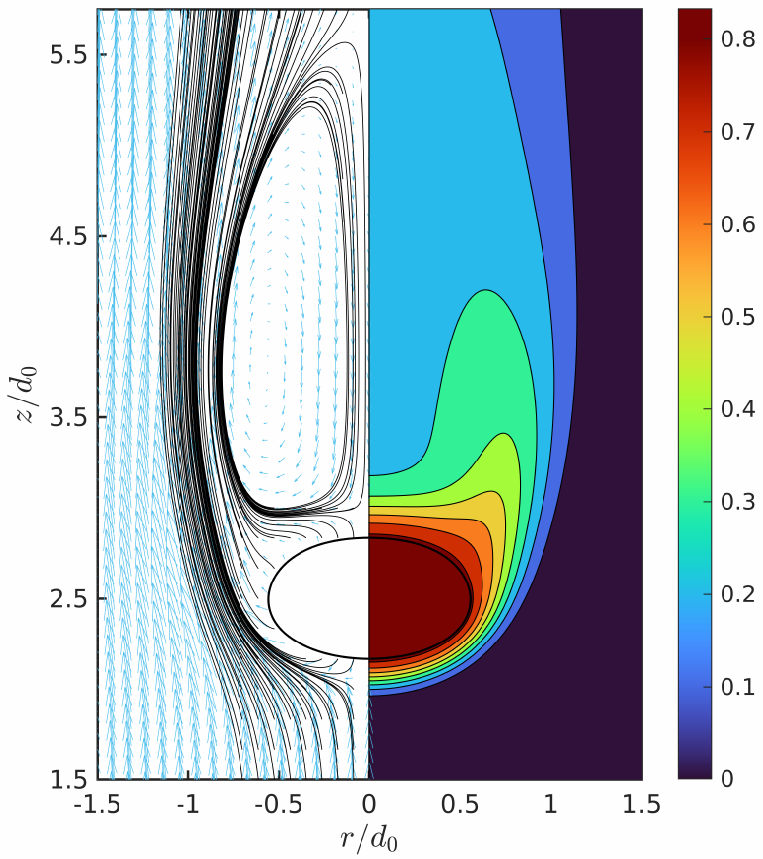}
&
\includegraphics[scale=0.32]{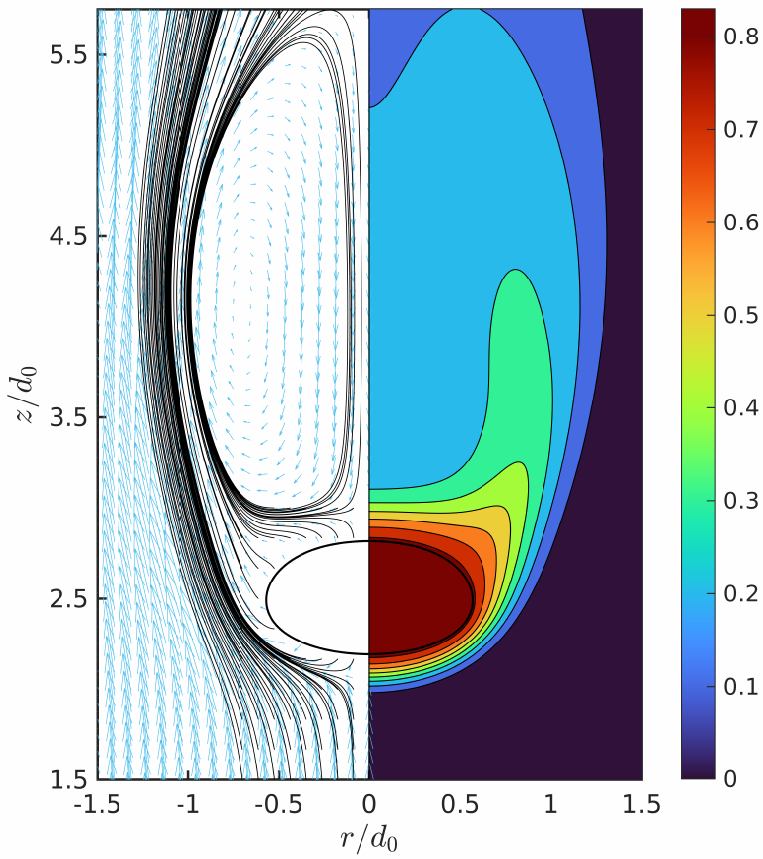}
\\
 \scriptsize{(c) $We = 6$} & \scriptsize{(d) $We = 9$} 
\\
\end{tabular}
\caption{Velocity vectors and streamlines (left portion)  and mass fraction field (right portion) in the vicinity of (a) a nearly spherical ($We=1$), (b) a weakly deforming ($We=3$), (c) a moderately deforming ($We=6$) and (d) a highly deforming droplet ($We=9$) at $t^*=15$. ($Re=100$, $B_M =5$, Domain: $4d_0\times 8d_0$, Grid: $512 \times 1024$)}
\label{Fig:We1369}
\end{figure}

Droplet deformation is primarily controlled by the Weber number. Thus, simulations are first performed for Weber numbers in the range of $We\in [1,\; 3,\; 6,\; 9]$ at $Re=100$ and $B_M=5$. The results are plotted in Fig.~\ref{Fig:We1369} that shows flow field  and vapor mass fraction distribution around a droplet at $t^*=15$. The droplet remains nearly spherical at a low Weber number of $We=1$ as seen in Fig.~\ref{Fig:We1369}a. As Weber number is increased to $We=3$,  the droplet gradually transitions to a more oblate shape as illustrated in Fig.~\ref{Fig:We1369}b and the recirculation zone gets enlarged slightly. By further increasing Weber number to $We=6$ and $We=9$, the droplet deforms significantly and elongates in the radial direction as seen in Figs.~\ref{Fig:We1369}c \& d. Flow field is also modified significantly by droplet deformation. 

\begin{figure}[hbt!]
\centering
\begin{tabular}{ccc}
\includegraphics[scale=0.28]{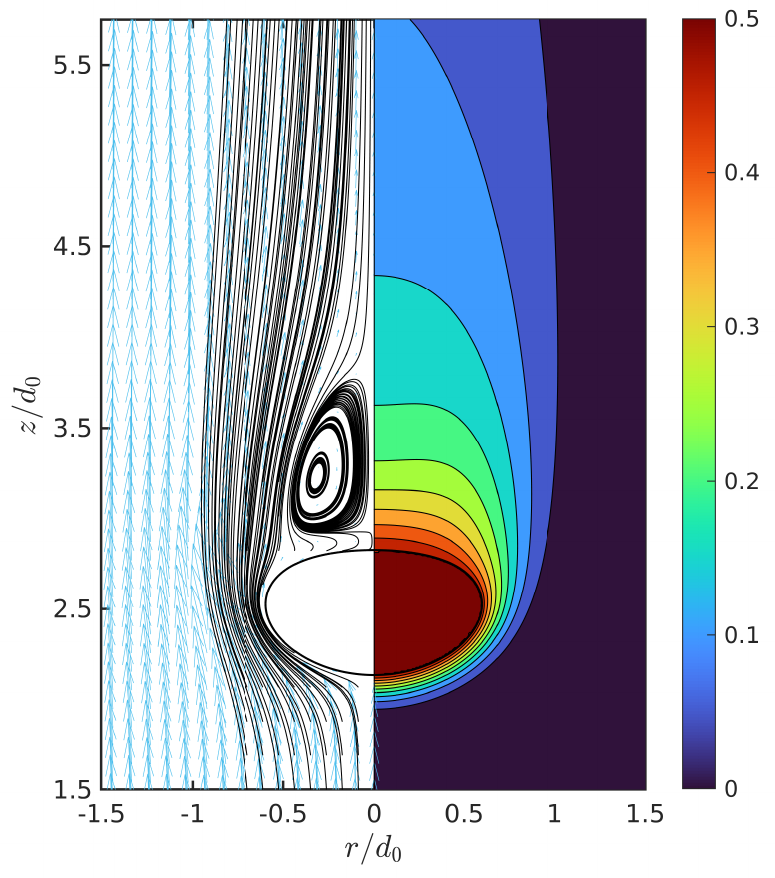}
&
\includegraphics[scale=0.28]{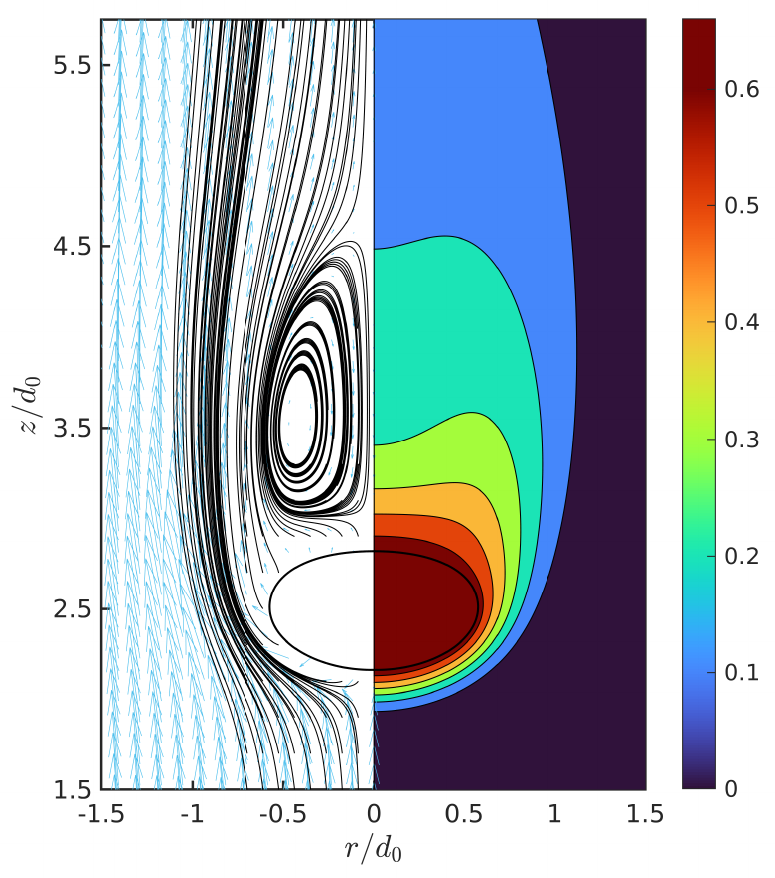}
&
\includegraphics[scale=0.28]{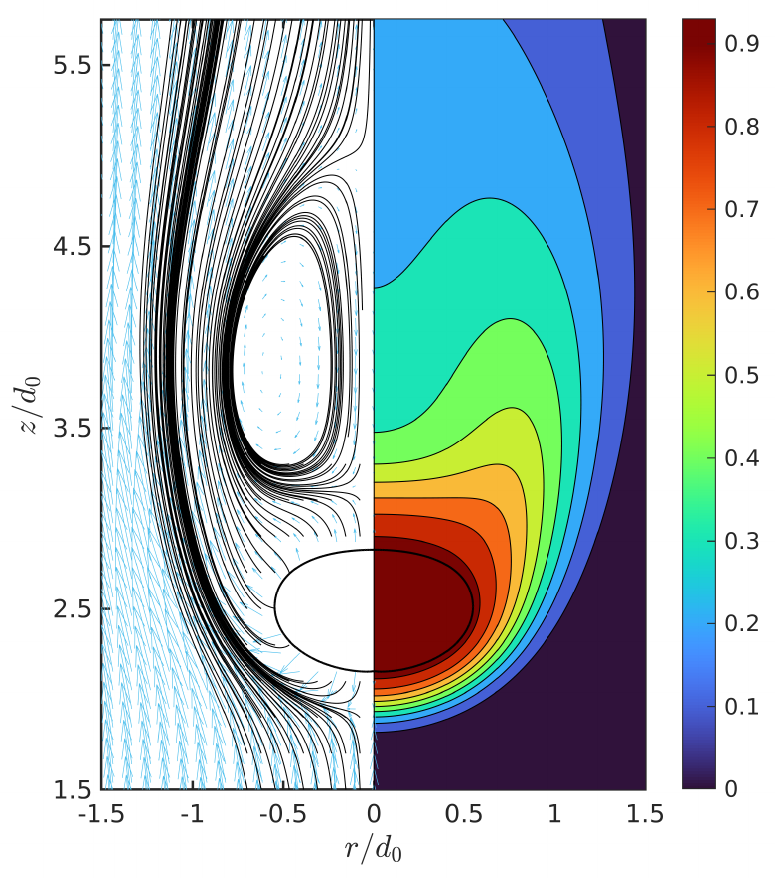}
\\
\scriptsize{(a) $Re=50,t^*=10,$ NSF} & \scriptsize{(b) $Re=50,t^*=10,B_M=2$} & \scriptsize{(c) $Re=50,t^*=10,B_M=15$} 
\\
\includegraphics[scale=0.28]{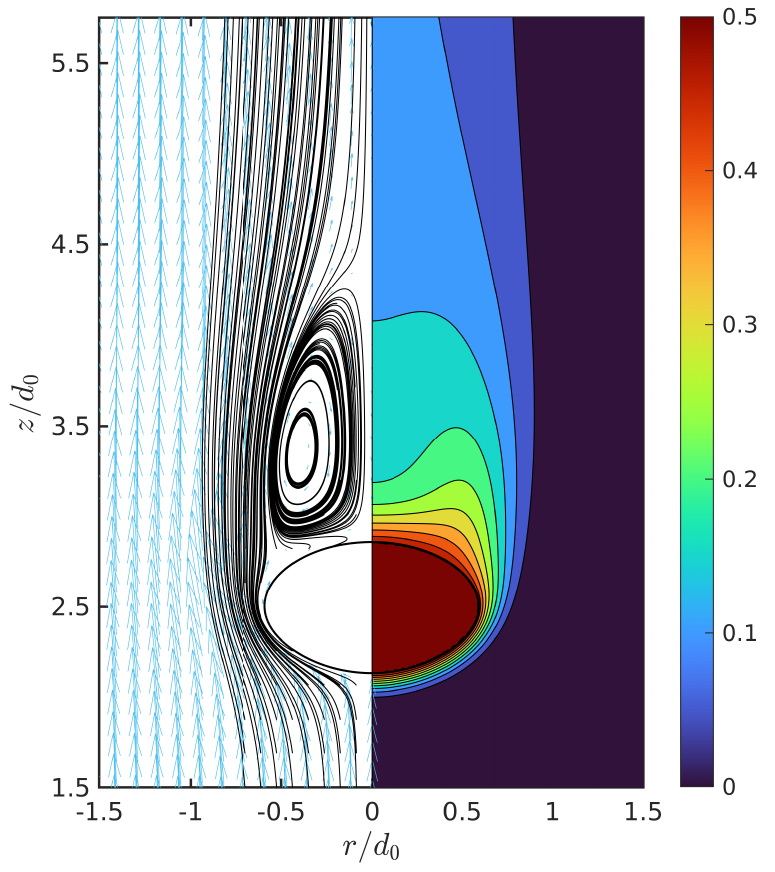}
&
\includegraphics[scale=0.28]{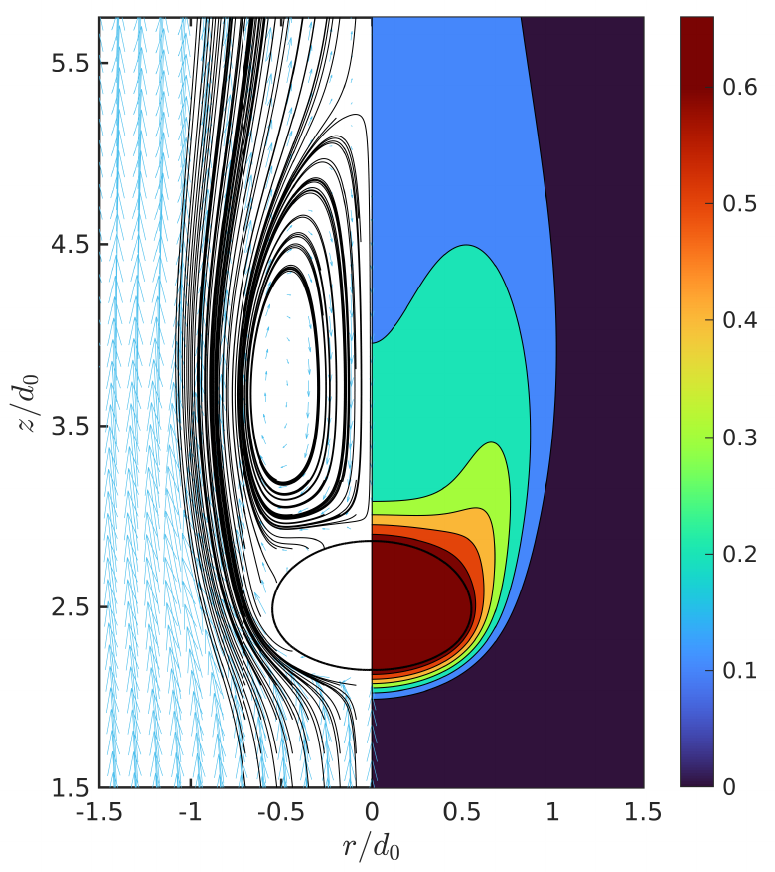}
&
\includegraphics[scale=0.28]{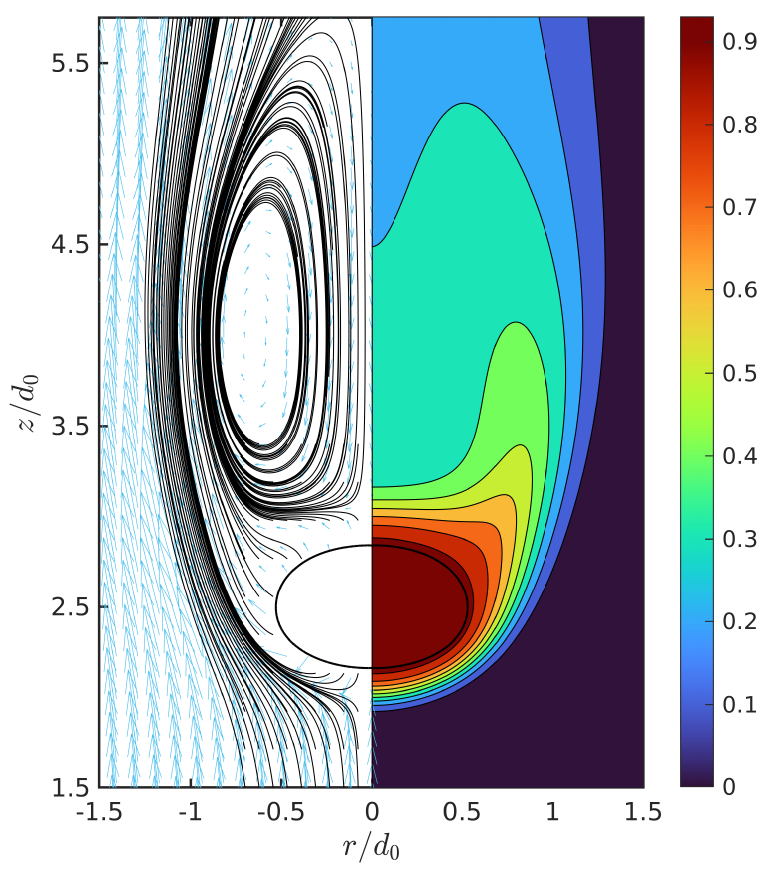}
\\
 \scriptsize{(d) $Re=100,t^*=20,$ NSF} & \scriptsize{(e) $Re=100,t^*=20,B_M=2$} & \scriptsize{(f) $Re=200,t^*=20,B_M=15$}
\\
\includegraphics[scale=0.33]{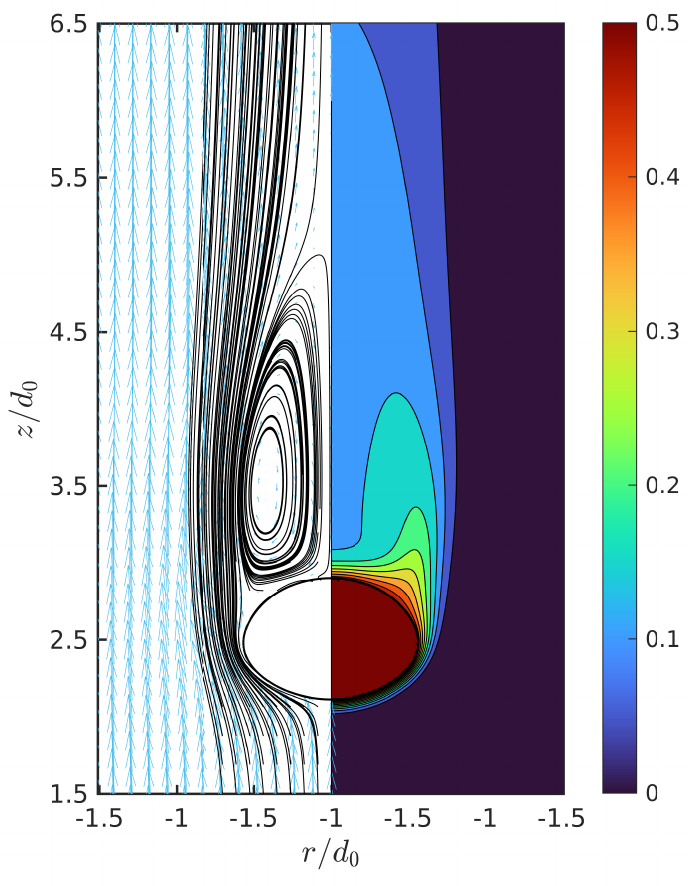}
&
\includegraphics[scale=0.33]{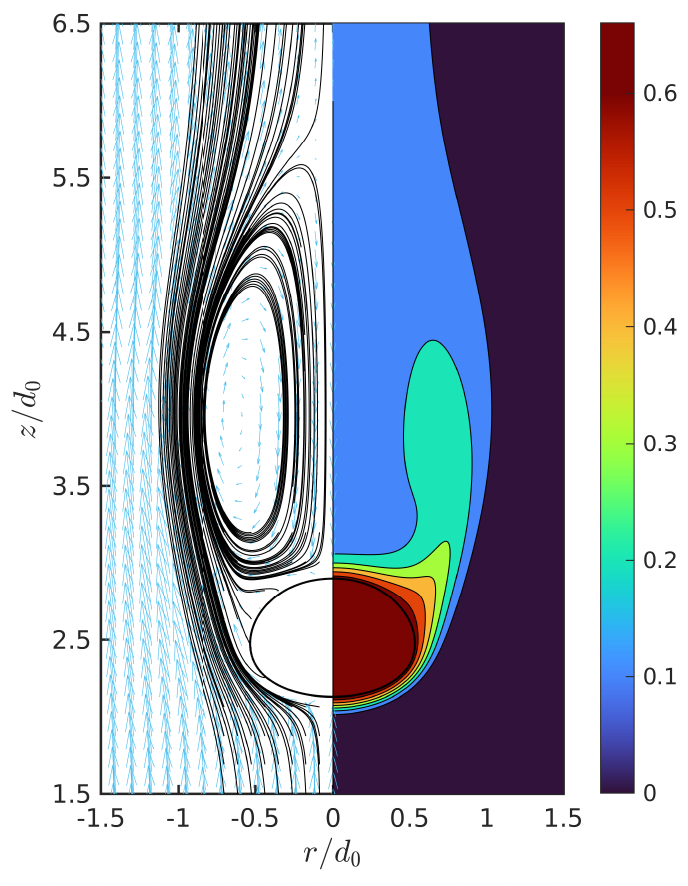}
&
\includegraphics[scale=0.33]{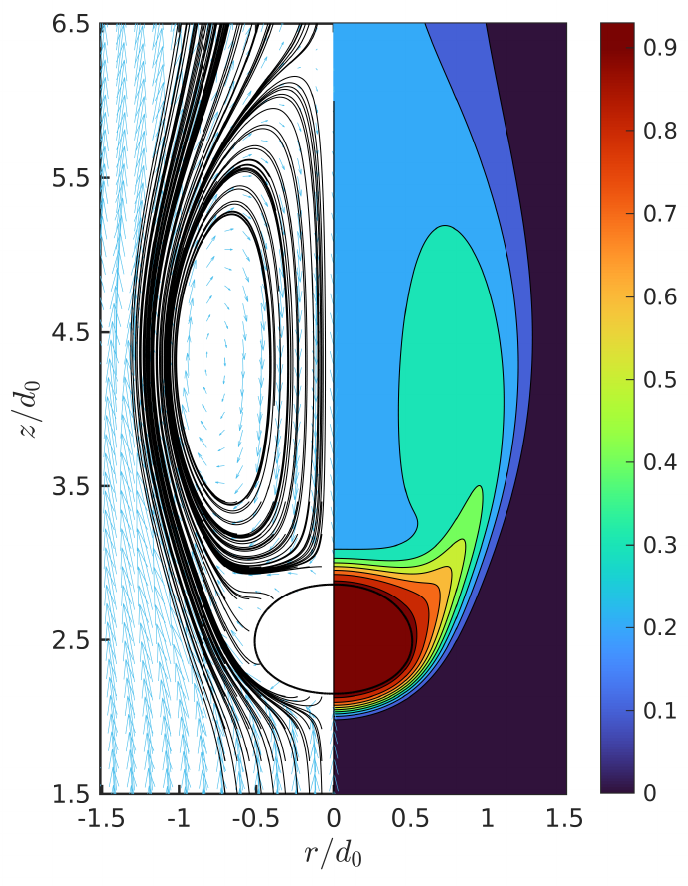}
\\
 \scriptsize{(g) $Re=200,t^*=30,$ NSF} & \scriptsize{(h) $Re=200,t^*=30,B_M=2$} & \scriptsize{(i) $Re=200,t^*=30,B_M=15$}
\\
\end{tabular}
\caption{Quasi-steady velocity vectors and streamlines (left portion)  and mass fraction field (right portion) around a moderately deforming droplet ($We=6$) at the Reynolds numbers of $Re\in[20,\;50,\; 100,\;200]$ for moderately ($B_M=2$) and strongly ($B_M=15$) evaporating cases. The no Stefan flow (NSF) results are also plotted as a reference. Color bars indicate the mass fraction values. Domain:$4d_0 \times 8d_0$; Grid:$512 \times 1024$.}
\label{Fig:Re50100200_We6_BM215}
\end{figure}

Figure~\ref{Fig:Re50100200_We6_BM215} shows flow and mass-fraction fields around a moderately deforming droplet ($We=6$) for the moderately ($B_M=2$) and strongly ($B_M=15$) evaporating cases at $Re\in [50,\; 100,\; 200]$. Results for the no Stefan flow (NSF) cases are also shown for comparison. As seen, Stefan flow has a profound effect on flow and mass fraction fields. Wake region gets enlarged as either $Re$ or $B_M$ increases similar to the nearly spherical droplet case, as shown in Fig. \ref{Fig:Re50100200_We065_BM215}. 

\begin{figure}[t]
\centering
\includegraphics[width=\textwidth]{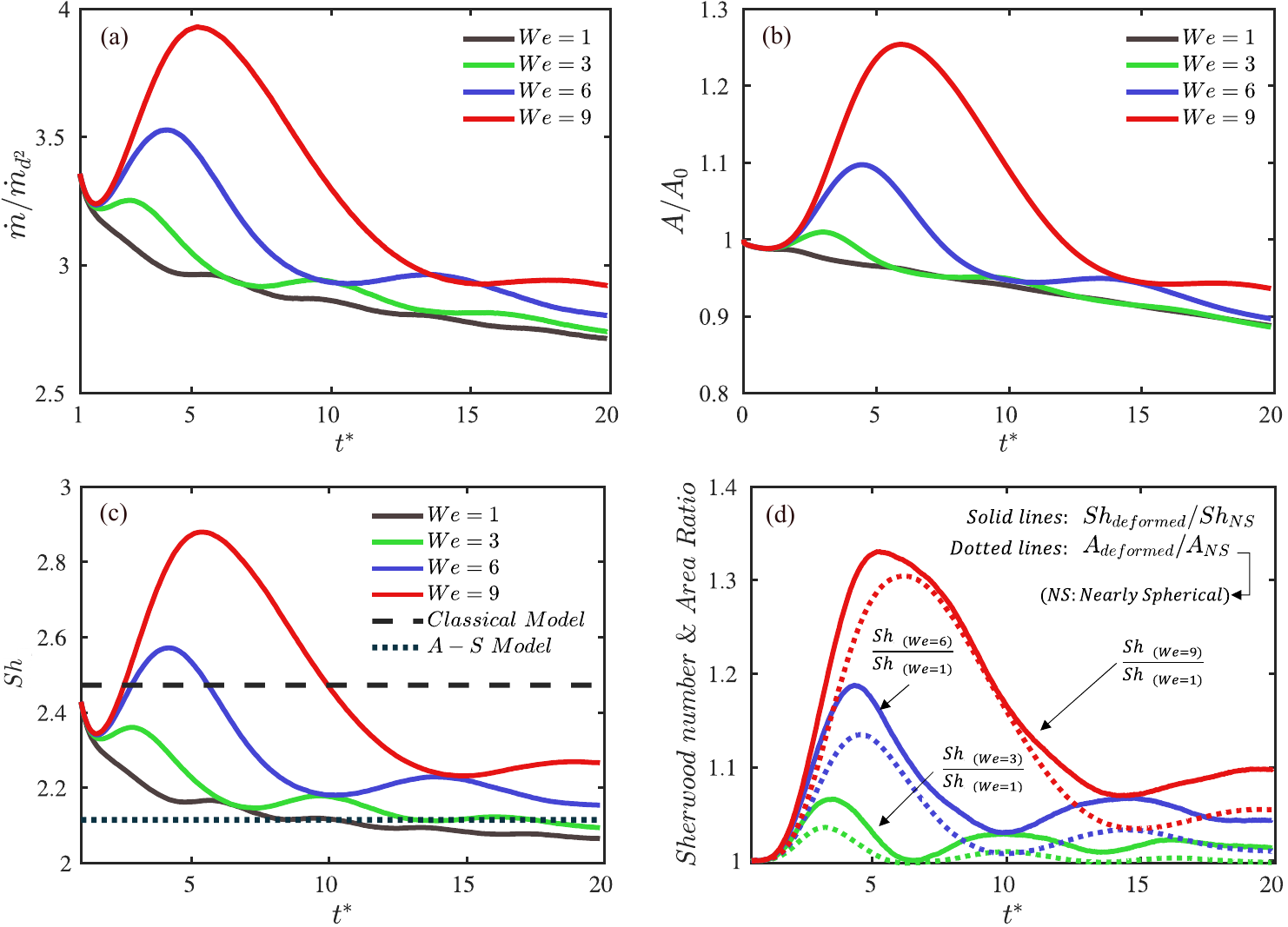}
\caption{(a) Temporal evolution of (a) the normalized mass transfer rate and (b) ratio of the surface area of deforming droplets (c) Temporal variation of the average Sherwood number, classical model \cite{sazhin2006advanced}, and Abramzon-Sirignano model \cite{abramzon1989droplet}. (d) Ratio of the Sherwood number and surface area of the deformed droplets to that of the nearly spherical droplet ($Re=100$ and $B_M=5$.) Domain and grid sizes are $4d_0 \times 8d_0$ and $512 \times 1024$. }
\label{Fig19_combined}
\end{figure}

Figure~\ref{Fig19_combined}a \& b, respectively, show time evolution of normalized evaporation rate and surface area for the deforming droplets at $Re=100$ and $B_M=5$. It is seen that evaporation rate correlates well with surface area.  In the initial phase, there is a considerable surge in evaporation rate of the most deformable droplets ($We=6$ and $We=9$) due to initial oscillations in droplet shape \cite{hase2004transient}. Thereafter, evaporation rate decreases and reaches a quasi-steady state at about $t^*=15$. Although the most deforming droplets exhibit significantly larger evaporation rates with the difference reaching as much as $30\%$ during the transient period, the differences decrease considerably and the maximum difference becomes less $7.5 \%$ once quasi-steady conditions are reached. 

This study aims to shed light on accuracy and relevance of the evaporation models commonly used in point-particle simulations. Therefore, it is particularly valuable to develop a strategy for estimating evaporation rate of deformed droplets. Such an approach could be incorporated into single-droplet evaporation models.
Following a species-driven evaporation regime, the single droplet evaporation models give the evaporation rate of a droplet by \cite{sazhin2006advanced,abramzon1989droplet}, 
\begin{equation} 
\dot{m} = 2\pi\rho_g D_{vg} R Sh B_M,
\label{eqn:singledroplet}
\end{equation}
where $Sh$ is the surface averaged Sherwood number of an evaporating droplet. However, it is known that Eq.~\eqref{eqn:singledroplet} is obtained assuming a perfectly spherical droplet~\cite{sazhin2006advanced}, and effects of increased surface area of a deformed droplet are inherently ignored. Thus, we evaluate average Sherwood number that is directly calculated based on Eq.~\eqref{eqn:singledroplet}. Note that, in Eq.~\eqref{eqn:singledroplet}, equivalent radius of a deformed droplet is readily calculated from interfaced-resolved simulations. 
Evolution of averaged Sherwood number, $Sh$, is plotted in Fig.~\ref{Fig19_combined}c together with the classical~\cite{sazhin2006advanced} and Abramzon-Sirignano~\cite{abramzon1989droplet} model predictions. Note that, in calculating the analytical models, the Frossling's correlation (Eq.~\eqref{eqn:Frossling}) \cite{frossling1963evaporation} is used to estimate $Sh_0$. The A--S model agrees reasonably well for the less deformable cases, i.e., $We\le 3$ but deviates significantly for the largely deforming cases of $We=6$ and $We=9$. The classical model substantially overpredicts the evaporation rate even for the most deformable case. Finally a close correlation between evaporation rate and deformation is demonstrated in Fig.~\ref{Fig19_combined}d where  average Sherwood number and surface area normalized by the respective values obtained for a nearly spherical droplet case are plotted together. As seen, the normalized Sherwood number correlates very well with the normalized droplet surface area for all the Weber numbers considered here. \citet{HAYWOOD19941401} investigated variation of Sherwood number for a deforming evaporating droplet and reported that  oscillations observed in Sherwood number were in-phase with fluctuations in droplet's aspect ratio. These results suggest that effects of droplet deformation on evaporation rate can be incorporated into the low order models provided that deformation is predicted reliably in terms of the known flow parameters such as $We$ and $Re$. However, the inherent complexity of intermediate Reynolds number flows, combined with challenges posed by an evolving droplet interface, has significantly restricted development of accurate analytical solutions for predicting droplet shape \cite{R.J.Haywood}.

\begin{figure}[t]
\centering
\includegraphics[width=\textwidth]{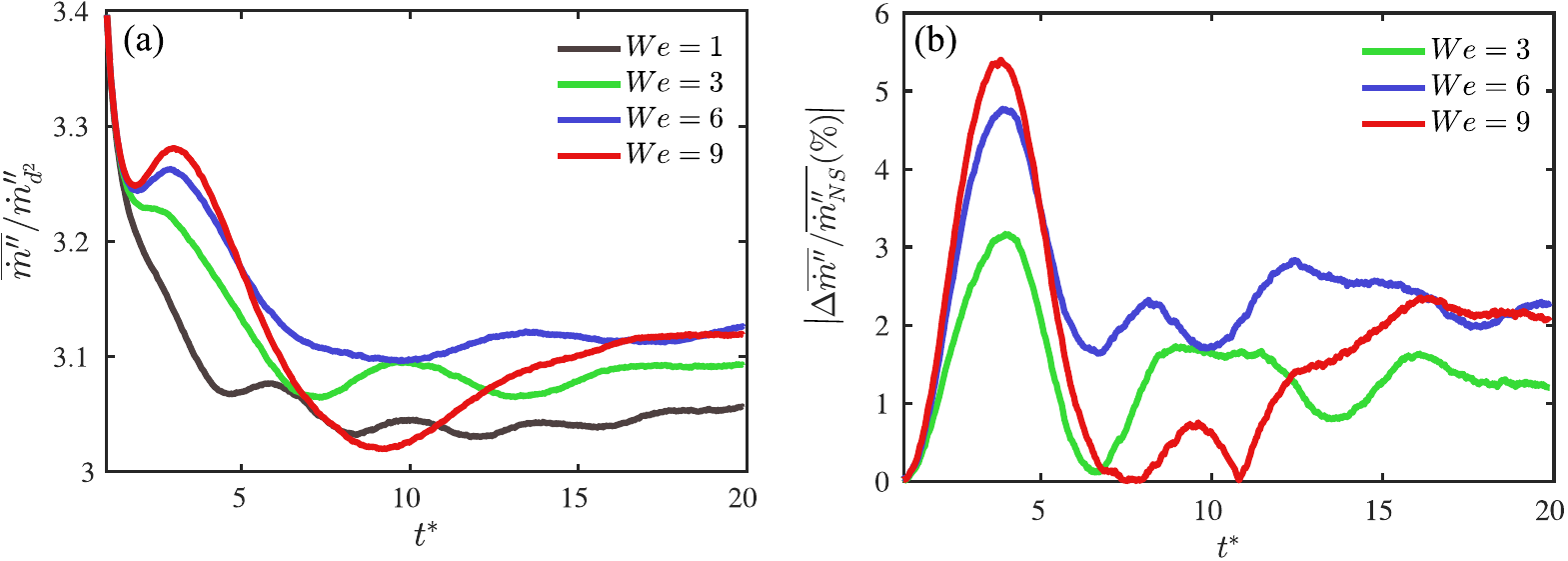}
\caption{(a) Surface-averaged mass flux of a deformable evaporating droplets for $We\in[1,\;3.\;6,\;9]$ at $Re=100$ and $B_M=5.$ (b) Percentage difference in surface-averaged mass flux between a deforming droplet in the range $We\in[1,\;3,\;9]$ and a nearly spherical droplet of $We=1$. }
\label{Fig:We1369-surfflux}
\end{figure}

Average mass flux is defined as $\overline{\dot{m}^{\prime\prime}} = \frac{1}{A} \int_A \dot{m}^{\prime\prime} \, \mathrm{d}A$ where $\dot{m}^{\prime\prime}=\frac{d\dot{m}}{dA}$ (so, $\dot{m} = \overline{\dot{m}^{\prime\prime}}A$).  Figure~\ref{Fig:We1369-surfflux}a shows average mass flux of deforming droplets normalized by $\dot{m}_{d^2}^{\prime\prime}$ where $\dot{m}_{d^2}^{\prime\prime}$ is computed from the $d^2$-law. As seen, the average mass flux is not influenced significantly by droplet deformation confirming the close correlation between evaporation rate and deformation as already seen in Fig.~\ref{Fig19_combined}d. Figure~\ref{Fig:We1369-surfflux}b shows the maximum deviation of surface averaged mass flux for a deforming droplet  from that of the nearly spherical (NS) one for the range of $We\in[3,\;6,\;9]$. Although the maximum deviation becomes more than $5\%$ in the transient regime, it reduces to about $2\%$ in the quasi-steady state.

\begin{figure}[t]
\centering
\includegraphics[width=\textwidth]{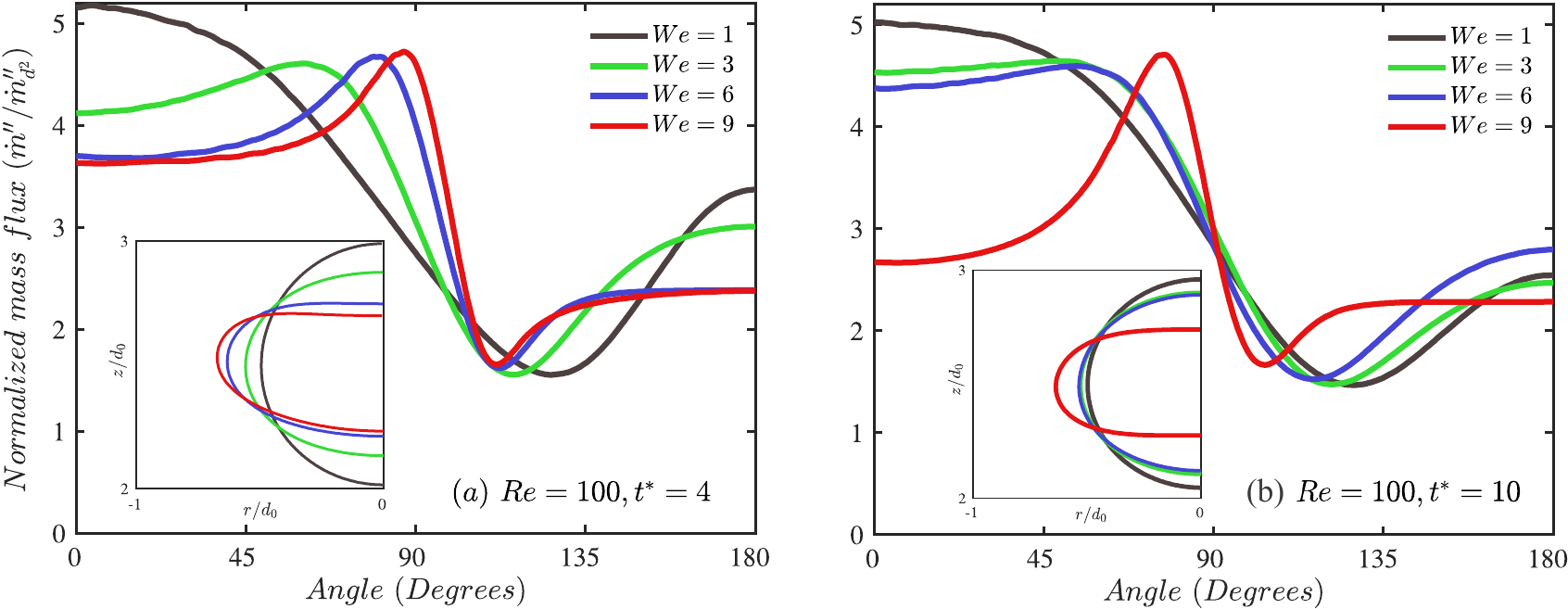} \\
\includegraphics[width=\textwidth]{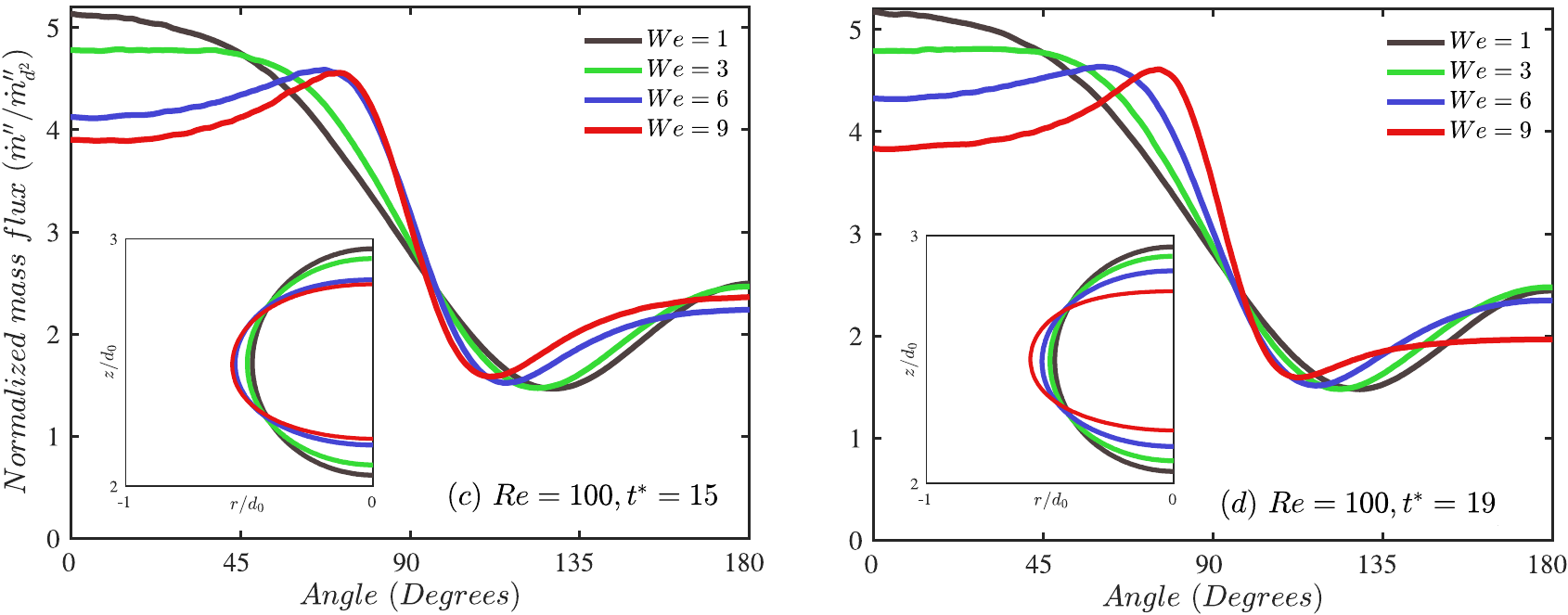}
\caption{Evolution of the normalized local mass flux distribution over deforming droplets as a function of angle for $Re=100$ and $B_M=5$ at $t^*\in [4,\; 10,\; 15,\; 19]$. Insets show the instantaneous droplet shapes. }
\label{Fig:We1369-flux}
\end{figure}

Figure~\ref{Fig:We1369-flux} shows evolution of normalized local mass flux distribution, $\dot{m}^{\prime\prime} / \dot{m}_{d^2}^{\prime\prime}$, over deforming droplets at various times. The interfaces are also indicated in the insets in this figure. It is seen that mass flux in upstream region of the droplet is considerably higher than that in the downstream region due to a stronger convective flow there, as also reported by \citet{SETIYA2023104455}. Difference between  front and rear edges reduces but a steeper transition occurs on the shoulder (i.e., around $\theta \approx 90^0$) as deformation increases.  Variation of mass flux over a nearly spherical droplet ($We=1$) is smoother compared to the deformed cases and the maximum value of mass flux is always located at the front stagnation point. As Weber number increases and droplet deforms into an oblate spheroid shape, mass flux in the downstream generally increases. Temporal evolution of the surface-averaged normalized mass flux from the rear portion of the droplet ($\theta \ge 90$) is summarized in Table~\ref{tab:flxsum} for $1\le We\le 9$. Clearly, the most strongly deformed droplet ($We = 9$) consistently exhibits the highest mass flux values at all time instances investigated. The deformed cases of $We \ge 3$ generally exhibit higher mass flux compared to the nearly spherical case ($We = 1$), which highlights the important role of droplet deformation in enhancing convective mass transfer from the leeward side of an evaporating droplet. This finding is consistent with findings reported by \citet{SETIYA2023104455} regarding the increase of mass flux in the rear side of the deformed droplets. Although, in their study, local variation of mass flux was not presented explicitly, this phenomenon was attributed to lower average concentration of vapor in the wake of deformed droplets. In their study, \citet{SETIYA2023104455} suggested that lower values of mass flux on windward side of deformed droplets are compensated for by larger mass flux in downstream; therefore, surface averaged mass flux remains similar regardless of the Weber number. However, as seen in Fig.~\ref{Fig:We1369-flux}, while local normalized mass flux of a nearly spherical droplet starts to decrease starting from 45 degrees until 135 degrees, $\dot{m}^{\prime\prime} / \dot{m}_{d^2}^{\prime\prime}$ of the deformed droplets increase from approximately $\theta=45^0$ until $\theta=90^0$ beyond that of the nearly spherical droplet. Considering that this region constitutes a larger portion of the droplet surface area, the close values of surface-averaged mass flux as seen in Fig.~\ref{Fig:We1369-flux}b are partly attributed to the increase of normalized mass flux of deformed droplet between $45^0 \le \theta \le 90^0$.

\begin{table}[t]
\centering
\caption{Temporal variation of the surface-averaged normalized mass flux from rear side of deforming droplets, i.e., $90^0 \le \theta \le 180^0$.}
\begin{tabular}{l|cccc}
\toprule
Time & We = 1 & We = 3 & We = 6 & We = 9 \\
\midrule
$t^* = 4$  & 2.0707 & 2.1640 & 2.7751 & 3.3841 \\
$t^* = 10$  & 1.9311 & 1.9665 & 1.9661 & 2.2970 \\
$t^* = 15$ & 1.9182 & 1.9090 & 2.0397 & 2.0313 \\
$t^* = 19$  & 1.9166 & 1.9247 & 1.9838 & 2.1886 \\
\bottomrule
\end{tabular}
\label{tab:flxsum}
\end{table}

\begin{figure}[t]
\centering
\includegraphics[scale=0.4]{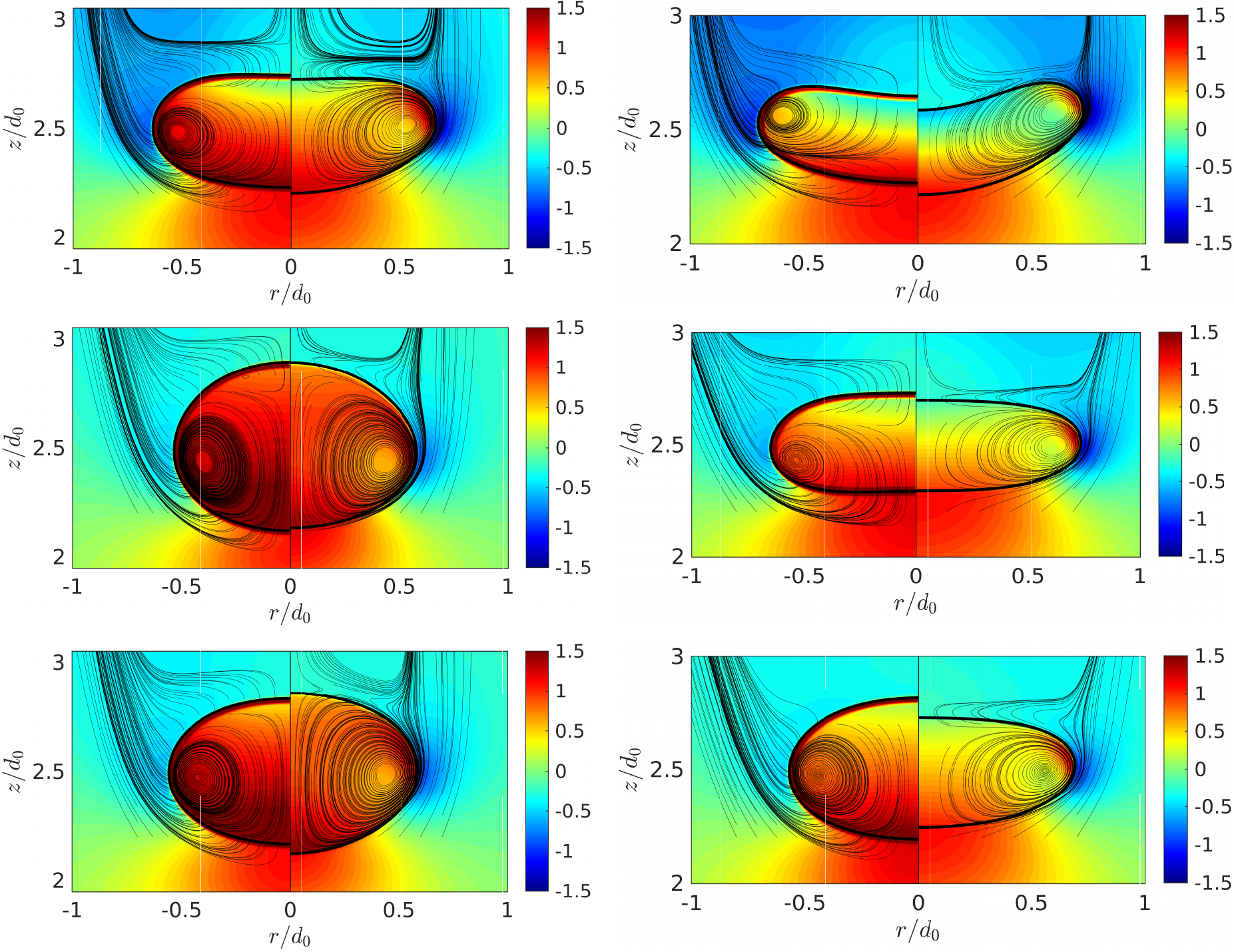}
\caption{Effect of Stefan flow on droplet deformation for Weber numbers of $We=6$ (left column) $We=9$ (right column) at times $t^*=5,\; 10,\; 15$ (from top to bottom). Right and left portions in each sub-plot show the cases in the presence and absence of the Stefan flow, respectively. The contours indicate the pressure field. ($Re=100$, $B_M=5$).}
\label{Fig:StefanonshapeWe9}
\end{figure}

We next examine influence of Stefan flow on droplet deformation. As discussed before, Stefan flow can substantially alter flow field around an evaporating droplet, which, in turn, affects its deformation~\cite{salimnezhad2024hybrid}. \citet{HAYWOOD19941401} reported that normal stresses induced on droplet surface due to surface blowing are typically 3 orders of magnitude smaller than aerodynamic stresses. The authors suggested that evaporation, even in highly volatile fuels, is not likely to significantly affect droplet shape. \citet{MASHAYEK20011517} also suggested that deformation of a droplet is not influenced significantly by evaporation. To shed further light on this problem, we plot evolution of moderately ($We=6$) and highly deformable ($We=9$) droplets in the presence and absence of Stefan flow in Fig.~\ref{Fig:StefanonshapeWe9}. The other flow parameters are fixed at $Re=100$ and $B_M=5$.  As seen, Stefan flow has a considerable impact on evolution and final shape of the droplet. In addition, while stream lines are tangential to the droplet surface in the absence of Stefan flow, they become more normal to the interface due to blowing effect of evaporation. This also influences flow inside the droplet. As time progresses, the droplet in the presence of Stefan flow exhibits a relatively moderate deformation with a less pronounced oblate shape. In contrast, when Stefan flow is switched off, the droplet undergoes a greater deformation, elongates in the radial direction, and gets more flattened. Figure~\ref{Fig:StefanonshapeWe9} also shows shape oscillations in the transient regime as also manifested in the mass transfer rates in Fig.~\ref{Fig19_combined}.

\section{Conclusions}
\label{conlusions} 
Extensive interface-resolved simulations are performed to study evaporation of a deformable droplet under convection using the immersed-boundary/front-fracking (IB/FT) method. Effects of convection, evaporation intensity and droplet deformability are investigated. In particular, performance of low-order the classical and Abramzon-Sirignano (A-S)  evaporation models is investigated against the interface-resolved simulation for the range of flow parameters relevant to spray combustion applications (i.e., $20 \leq Re \leq 200$, $0.65\le We \le 9$, and $1 \leq B_M \leq 15$). Simulations are also performed to demonstrate effects of Stefan flow on flow and mass fraction fields around a droplet, and resulting effects on evaporation characteristics. 

It is found that a flow separation occurs leading to a recirculation wake behind the droplet at high Reynolds numbers. The wake gets intensified as either the Reynolds number, Weber number, or mass transfer number is increased. This, in turn, can lead to reduction of  average vapor mass faction distribution in the wake region. The combined effects of these phenomena result in an enhanced evaporation rate from leeward side of the droplet. The results reveal that Stefan flow significantly alters local and global evaporation characteristics, particularly in the wake region, where it results in an earlier flow separation and enlargement of the recirculation zone. The stagnation points are pushed further away from the droplet surface as evaporation intensity is increased. Stefan flow also considerably alters flow field and impacts distribution of  vapor concentration in the wake region. Vapor mass fraction field in the wake region is also strongly influenced by Reynolds number, i.e., vapor mass fraction profiles are qualitatively different at low and high Reynolds number regimes. It is also shown that  flow and vapor mass fraction fields in near droplet region are dominated by the Stefan flow. 

The A-S model generally outperforms the classical model in the presence of significant convection. Both models fail to predict the local Sherwood number in the wake region behind the droplet, leading to significant discrepancies compared to the interface-resolved simulations. The discrepancy increases as the convection (Reynolds number) increases. 

It is demonstrated that droplet deformability significantly enhances the evaporation rate. Performance of the low-order evaporation models deteriorates as deformation rate increases and they fail to predict the evaporation rate in the highly deforming cases.  In particular, while the A--S model predicts the evaporation rates of a nearly spherical droplets reasonably accurately, it underpredicts the evaporation rate for the deformed cases. It is found that change in evaporation rate is closely correlated with change in surface area, which implies that effects of droplet deformation on evaporation rate can be incorporated into the evaporation models provided that deformation is predicted reliably in terms of the known flow parameters such as $Re$ and $We$. However, accurate prediction of the droplet shape evolution using analytical models still remains a challenging task.

Interface-resolved simulations reveal that Stefan flow also has a significant influence on droplet deformation at high Weber numbers and could act to reduce droplet deformation, which may further decrease evaporation rate.


\section*{Acknowledgment}
We acknowledge financial support from the Scientific and Technical Research Council of Turkey (TUBITAK) [Grant Number 124M335].

\section*{CRediT authorship contribution statement}
\textbf{F. Salimezhad:} Conceptualization, methodology, investigation, software, formal analysis,  visualization, writing – original draft, writing – review \& editing.
\textbf{M. Muradoglu:} Conceptualization, resources, investigation, writing – review \& editing, supervision, project administration, funding acquisition.

\section*{declaration of competing interest}
The authors declare that they have no known competing financial interests or personal relationships that could have appeared to influence the work reported in this paper.

  \bibliographystyle{elsarticle-num-names} 
  \bibliography{References}

\begin{thebibliography}{53}
\expandafter\ifx\csname natexlab\endcsname\relax\def\natexlab#1{#1}\fi
\providecommand{\url}[1]{\texttt{#1}}
\providecommand{\href}[2]{#2}
\providecommand{\path}[1]{#1}
\providecommand{\DOIprefix}{doi:}
\providecommand{\ArXivprefix}{arXiv:}
\providecommand{\URLprefix}{URL: }
\providecommand{\Pubmedprefix}{pmid:}
\providecommand{\doi}[1]{\href{http://dx.doi.org/#1}{\path{#1}}}
\providecommand{\Pubmed}[1]{\href{pmid:#1}{\path{#1}}}
\providecommand{\bibinfo}[2]{#2}
\ifx\xfnm\relax \def\xfnm[#1]{\unskip,\space#1}\fi
\bibitem[{Law(1982)}]{LAW1982171}
\bibinfo{author}{C.~Law},
\newblock \bibinfo{title}{Recent advances in droplet vaporization and combustion},
\newblock \bibinfo{journal}{Progress in Energy and Combustion Science} \bibinfo{volume}{8} (\bibinfo{year}{1982}) \bibinfo{pages}{171--201}. \URLprefix \url{https://www.sciencedirect.com/science/article/pii/0360128582900119}. \DOIprefix\doi{https://doi.org/10.1016/0360-1285(82)90011-9}.
\bibitem[{Faeth(1987)}]{FAETH1987293}
\bibinfo{author}{G.~Faeth},
\newblock \bibinfo{title}{Mixing, transport and combustion in sprays},
\newblock \bibinfo{journal}{Progress in Energy and Combustion Science} \bibinfo{volume}{13} (\bibinfo{year}{1987}) \bibinfo{pages}{293--345}. \URLprefix \url{https://www.sciencedirect.com/science/article/pii/0360128587900025}. \DOIprefix\doi{https://doi.org/10.1016/0360-1285(87)90002-5}.
\bibitem[{Sirignano(1983)}]{SIRIGNANO1983291}
\bibinfo{author}{W.~A. Sirignano},
\newblock \bibinfo{title}{Fuel droplet vaporization and spray combustion theory},
\newblock \bibinfo{journal}{Progress in Energy and Combustion Science} \bibinfo{volume}{9} (\bibinfo{year}{1983}) \bibinfo{pages}{291--322}. \URLprefix \url{https://www.sciencedirect.com/science/article/pii/0360128583900114}. \DOIprefix\doi{https://doi.org/10.1016/0360-1285(83)90011-4}.
\bibitem[{Gökalp et~al.(1992)Gökalp, Chauveau, Simon, and Chesneau}]{GOKALP1992286}
\bibinfo{author}{I.~Gökalp}, \bibinfo{author}{C.~Chauveau}, \bibinfo{author}{O.~Simon}, \bibinfo{author}{X.~Chesneau},
\newblock \bibinfo{title}{Mass transfer from liquid fuel droplets in turbulent flow},
\newblock \bibinfo{journal}{Combustion and Flame} \bibinfo{volume}{89} (\bibinfo{year}{1992}) \bibinfo{pages}{286--298}. \URLprefix \url{https://www.sciencedirect.com/science/article/pii/001021809290016I}. \DOIprefix\doi{https://doi.org/10.1016/0010-2180(92)90016-I}.
\bibitem[{Dauxois et~al.(2021)Dauxois, Peacock, Bauer, Caulfield, Cenedese, Gorl\'e, Haller, Ivey, Linden, Meiburg, Pinardi, Vriend, and Woods}]{PhysRevFluids.6.020501}
\bibinfo{author}{T.~Dauxois}, \bibinfo{author}{T.~Peacock}, \bibinfo{author}{P.~Bauer}, \bibinfo{author}{C.~P. Caulfield}, \bibinfo{author}{C.~Cenedese}, \bibinfo{author}{C.~Gorl\'e}, \bibinfo{author}{G.~Haller}, \bibinfo{author}{G.~N. Ivey}, \bibinfo{author}{P.~F. Linden}, \bibinfo{author}{E.~Meiburg}, \bibinfo{author}{N.~Pinardi}, \bibinfo{author}{N.~M. Vriend}, \bibinfo{author}{A.~W. Woods},
\newblock \bibinfo{title}{Confronting grand challenges in environmental fluid mechanics},
\newblock \bibinfo{journal}{Phys. Rev. Fluids} \bibinfo{volume}{6} (\bibinfo{year}{2021}) \bibinfo{pages}{020501}. \URLprefix \url{https://link.aps.org/doi/10.1103/PhysRevFluids.6.020501}. \DOIprefix\doi{10.1103/PhysRevFluids.6.020501}.
\bibitem[{Mittal et~al.(2020)Mittal, Ni, and Seo}]{mittal_ni_seo_2020}
\bibinfo{author}{R.~Mittal}, \bibinfo{author}{R.~Ni}, \bibinfo{author}{J.-H. Seo},
\newblock \bibinfo{title}{The flow physics of covid-19},
\newblock \bibinfo{journal}{Journal of Fluid Mechanics} \bibinfo{volume}{894} (\bibinfo{year}{2020}) \bibinfo{pages}{F2}. \DOIprefix\doi{10.1017/jfm.2020.330}.
\bibitem[{Sazhin(2006)}]{sazhin2006advanced}
\bibinfo{author}{S.~S. Sazhin},
\newblock \bibinfo{title}{Advanced models of fuel droplet heating and evaporation},
\newblock \bibinfo{journal}{Progress in energy and combustion science} \bibinfo{volume}{32} (\bibinfo{year}{2006}) \bibinfo{pages}{162--214}.
\bibitem[{Birouk and Gökalp(2006)}]{BIROUK2006408}
\bibinfo{author}{M.~Birouk}, \bibinfo{author}{I.~Gökalp},
\newblock \bibinfo{title}{Current status of droplet evaporation in turbulent flows},
\newblock \bibinfo{journal}{Progress in Energy and Combustion Science} \bibinfo{volume}{32} (\bibinfo{year}{2006}) \bibinfo{pages}{408--423}. \URLprefix \url{https://www.sciencedirect.com/science/article/pii/S0360128506000220}. \DOIprefix\doi{https://doi.org/10.1016/j.pecs.2006.05.001}.
\bibitem[{Wilhelmsen et~al.(2016)Wilhelmsen, Trinh, Lervik, Badam, Kjelstrup, and Bedeaux}]{PhysRevE.93.032801}
\bibinfo{author}{O.~Wilhelmsen}, \bibinfo{author}{T.~T. Trinh}, \bibinfo{author}{A.~Lervik}, \bibinfo{author}{V.~K. Badam}, \bibinfo{author}{S.~Kjelstrup}, \bibinfo{author}{D.~Bedeaux},
\newblock \bibinfo{title}{Coherent description of transport across the water interface: From nanodroplets to climate models},
\newblock \bibinfo{journal}{Phys. Rev. E} \bibinfo{volume}{93} (\bibinfo{year}{2016}) \bibinfo{pages}{032801}. \URLprefix \url{https://link.aps.org/doi/10.1103/PhysRevE.93.032801}. \DOIprefix\doi{10.1103/PhysRevE.93.032801}.
\bibitem[{Jenny et~al.(2012)Jenny, Roekaerts, and Beishuizen}]{JENNY2012846}
\bibinfo{author}{P.~Jenny}, \bibinfo{author}{D.~Roekaerts}, \bibinfo{author}{N.~Beishuizen},
\newblock \bibinfo{title}{Modeling of turbulent dilute spray combustion},
\newblock \bibinfo{journal}{Progress in Energy and Combustion Science} \bibinfo{volume}{38} (\bibinfo{year}{2012}) \bibinfo{pages}{846--887}. \URLprefix \url{https://www.sciencedirect.com/science/article/pii/S0360128512000445}. \DOIprefix\doi{https://doi.org/10.1016/j.pecs.2012.07.001}.
\bibitem[{Fuchs(2013)}]{fuchs2013evaporation}
\bibinfo{author}{N.~A. Fuchs}, \bibinfo{title}{Evaporation and droplet growth in gaseous media}, \bibinfo{publisher}{Elsevier}, \bibinfo{year}{2013}.
\bibitem[{Spalding(1953)}]{SPALDING1953847}
\bibinfo{author}{D.~Spalding},
\newblock \bibinfo{title}{The combustion of liquid fuels},
\newblock \bibinfo{journal}{Symposium (International) on Combustion} \bibinfo{volume}{4} (\bibinfo{year}{1953}) \bibinfo{pages}{847--864}. \URLprefix \url{https://www.sciencedirect.com/science/article/pii/S0082078453801104}. \DOIprefix\doi{https://doi.org/10.1016/S0082-0784(53)80110-4}, \bibinfo{note}{fourth Symposium (International) on Combustion}.
\bibitem[{Godsave(1953)}]{GODSAVE1953818}
\bibinfo{author}{G.~Godsave},
\newblock \bibinfo{title}{Studies of the combustion of drops in a fuel spray—the burning of single drops of fuel},
\newblock \bibinfo{journal}{Symposium (International) on Combustion} \bibinfo{volume}{4} (\bibinfo{year}{1953}) \bibinfo{pages}{818--830}. \URLprefix \url{https://www.sciencedirect.com/science/article/pii/S0082078453801074}. \DOIprefix\doi{https://doi.org/10.1016/S0082-0784(53)80107-4}, \bibinfo{note}{fourth Symposium (International) on Combustion}.
\bibitem[{Renksizbulut and Haywood(1988)}]{RENKSIZBULUT1988189}
\bibinfo{author}{M.~Renksizbulut}, \bibinfo{author}{R.~Haywood},
\newblock \bibinfo{title}{Transient droplet evaporation with variable properties and internal circulation at intermediate reynolds numbers},
\newblock \bibinfo{journal}{International Journal of Multiphase Flow} \bibinfo{volume}{14} (\bibinfo{year}{1988}) \bibinfo{pages}{189--202}. \URLprefix \url{https://www.sciencedirect.com/science/article/pii/0301932288900055}. \DOIprefix\doi{https://doi.org/10.1016/0301-9322(88)90005-5}.
\bibitem[{Williams(1973)}]{WILLIAMS19731}
\bibinfo{author}{A.~Williams},
\newblock \bibinfo{title}{Combustion of droplets of liquid fuels: A review},
\newblock \bibinfo{journal}{Combustion and Flame} \bibinfo{volume}{21} (\bibinfo{year}{1973}) \bibinfo{pages}{1--31}. \URLprefix \url{https://www.sciencedirect.com/science/article/pii/0010218073900023}. \DOIprefix\doi{https://doi.org/10.1016/0010-2180(73)90002-3}.
\bibitem[{Faeth(1977)}]{FAETH1977191}
\bibinfo{author}{G.~Faeth},
\newblock \bibinfo{title}{Current status of droplet and liquid combustion},
\newblock \bibinfo{journal}{Progress in Energy and Combustion Science} \bibinfo{volume}{3} (\bibinfo{year}{1977}) \bibinfo{pages}{191--224}. \URLprefix \url{https://www.sciencedirect.com/science/article/pii/0360128577900120}. \DOIprefix\doi{https://doi.org/10.1016/0360-1285(77)90012-0}.
\bibitem[{Abramzon and Sirignano(1989)}]{abramzon1989droplet}
\bibinfo{author}{B.~Abramzon}, \bibinfo{author}{W.~A. Sirignano},
\newblock \bibinfo{title}{Droplet vaporization model for spray combustion calculations},
\newblock \bibinfo{journal}{International journal of heat and mass transfer} \bibinfo{volume}{32} (\bibinfo{year}{1989}) \bibinfo{pages}{1605--1618}.
\bibitem[{Sirignano(2010)}]{sirignano2010fluid}
\bibinfo{author}{W.~A. Sirignano}, \bibinfo{title}{Fluid dynamics and transport of droplets and sprays}, \bibinfo{publisher}{Cambridge university press}, \bibinfo{year}{2010}.
\bibitem[{Sazhin(2017)}]{SAZHIN201769}
\bibinfo{author}{S.~S. Sazhin},
\newblock \bibinfo{title}{Modelling of fuel droplet heating and evaporation: Recent results and unsolved problems},
\newblock \bibinfo{journal}{Fuel} \bibinfo{volume}{196} (\bibinfo{year}{2017}) \bibinfo{pages}{69--101}. \URLprefix \url{https://www.sciencedirect.com/science/article/pii/S0016236117300583}. \DOIprefix\doi{https://doi.org/10.1016/j.fuel.2017.01.048}.
\bibitem[{Renksizbulut and Yuen(1983)}]{10.1115/1.3245591}
\bibinfo{author}{M.~Renksizbulut}, \bibinfo{author}{M.~C. Yuen},
\newblock \bibinfo{title}{{Numerical Study of Droplet Evaporation in a High-Temperature Stream}},
\newblock \bibinfo{journal}{Journal of Heat Transfer} \bibinfo{volume}{105} (\bibinfo{year}{1983}) \bibinfo{pages}{389--397}. \DOIprefix\doi{10.1115/1.3245591}.
\bibitem[{Haywood et~al.(1989)Haywood, Nafziger, and Renksizbulut}]{haywood1989detailed}
\bibinfo{author}{R.~J. Haywood}, \bibinfo{author}{R.~Nafziger}, \bibinfo{author}{M.~Renksizbulut},
\newblock \bibinfo{title}{A detailed examination of gas and liquid phase transient processes in convective droplet evaporation},
\newblock \bibinfo{journal}{Journal of Heat Transfer} \bibinfo{volume}{111} (\bibinfo{year}{1989}) \bibinfo{pages}{495--502}. \DOIprefix\doi{10.1115/1.3250704}.
\bibitem[{R.~J.~Haywood and Raithby(1994)}]{R.J.Haywood}
\bibinfo{author}{M.~R. R.~J.~Haywood}, \bibinfo{author}{G.~D. Raithby},
\newblock \bibinfo{title}{Numerical solution of deforming evaporating droplets at intermediate reynolds numbers},
\newblock \bibinfo{journal}{Numerical Heat Transfer, Part A: Applications} \bibinfo{volume}{26} (\bibinfo{year}{1994}) \bibinfo{pages}{253--272}. \URLprefix \url{https://doi.org/10.1080/10407789408955991}. \DOIprefix\doi{10.1080/10407789408955991}. \href{http://arxiv.org/abs/https://doi.org/10.1080/10407789408955991}{{\tt arXiv:https://doi.org/10.1080/10407789408955991}}.
\bibitem[{Haywood et~al.(1994)Haywood, Renksizbulut, and Raithby}]{HAYWOOD19941401}
\bibinfo{author}{R.~Haywood}, \bibinfo{author}{M.~Renksizbulut}, \bibinfo{author}{G.~Raithby},
\newblock \bibinfo{title}{Transient deformation and evaporation of droplets at intermediate reynolds numbers},
\newblock \bibinfo{journal}{International Journal of Heat and Mass Transfer} \bibinfo{volume}{37} (\bibinfo{year}{1994}) \bibinfo{pages}{1401--1409}. \URLprefix \url{https://www.sciencedirect.com/science/article/pii/0017931094901864}. \DOIprefix\doi{https://doi.org/10.1016/0017-9310(94)90186-4}.
\bibitem[{Mashayek(2001)}]{MASHAYEK20011517}
\bibinfo{author}{F.~Mashayek},
\newblock \bibinfo{title}{Dynamics of evaporating drops. part i: formulation and evaporation model},
\newblock \bibinfo{journal}{International Journal of Heat and Mass Transfer} \bibinfo{volume}{44} (\bibinfo{year}{2001}) \bibinfo{pages}{1517--1526}. \URLprefix \url{https://www.sciencedirect.com/science/article/pii/S001793100000199X}. \DOIprefix\doi{https://doi.org/10.1016/S0017-9310(00)00199-X}.
\bibitem[{Chiang et~al.(1992)Chiang, Raju, and Sirignano}]{CHIANG19921307}
\bibinfo{author}{C.~Chiang}, \bibinfo{author}{M.~Raju}, \bibinfo{author}{W.~Sirignano},
\newblock \bibinfo{title}{Numerical analysis of convecting, vaporizing fuel droplet with variable properties},
\newblock \bibinfo{journal}{International Journal of Heat and Mass Transfer} \bibinfo{volume}{35} (\bibinfo{year}{1992}) \bibinfo{pages}{1307--1324}. \URLprefix \url{https://www.sciencedirect.com/science/article/pii/001793109290186V}. \DOIprefix\doi{https://doi.org/10.1016/0017-9310(92)90186-V}.
\bibitem[{Schlottke and Weigand(2008)}]{SCHLOTTKE20085215}
\bibinfo{author}{J.~Schlottke}, \bibinfo{author}{B.~Weigand},
\newblock \bibinfo{title}{Direct numerical simulation of evaporating droplets},
\newblock \bibinfo{journal}{Journal of Computational Physics} \bibinfo{volume}{227} (\bibinfo{year}{2008}) \bibinfo{pages}{5215--5237}. \URLprefix \url{https://www.sciencedirect.com/science/article/pii/S0021999108000740}. \DOIprefix\doi{https://doi.org/10.1016/j.jcp.2008.01.042}.
\bibitem[{Palmore and Desjardins(2019)}]{PALMORE2019108954}
\bibinfo{author}{J.~Palmore}, \bibinfo{author}{O.~Desjardins},
\newblock \bibinfo{title}{A volume of fluid framework for interface-resolved simulations of vaporizing liquid-gas flows},
\newblock \bibinfo{journal}{Journal of Computational Physics} \bibinfo{volume}{399} (\bibinfo{year}{2019}) \bibinfo{pages}{108954}. \URLprefix \url{https://www.sciencedirect.com/science/article/pii/S002199911930659X}. \DOIprefix\doi{https://doi.org/10.1016/j.jcp.2019.108954}.
\bibitem[{Ni et~al.(2021)Ni, Hespel, Han, and Foucher}]{NI2021120736}
\bibinfo{author}{Z.~Ni}, \bibinfo{author}{C.~Hespel}, \bibinfo{author}{K.~Han}, \bibinfo{author}{F.~Foucher},
\newblock \bibinfo{title}{Numerical simulation of heat and mass transient behavior of single hexadecane droplet under forced convective conditions},
\newblock \bibinfo{journal}{International Journal of Heat and Mass Transfer} \bibinfo{volume}{167} (\bibinfo{year}{2021}) \bibinfo{pages}{120736}. \URLprefix \url{https://www.sciencedirect.com/science/article/pii/S0017931020336723}. \DOIprefix\doi{https://doi.org/10.1016/j.ijheatmasstransfer.2020.120736}.
\bibitem[{Ranz(1952)}]{ranz1952evaporation}
\bibinfo{author}{W.~E. Ranz},
\newblock \bibinfo{title}{Evaporation from drops-i and-ii},
\newblock \bibinfo{journal}{Chem. Eng. Progr} \bibinfo{volume}{48} (\bibinfo{year}{1952}) \bibinfo{pages}{141--146}.
\bibitem[{Clift et~al.(2005)Clift, Grace, and Weber}]{clift2005bubbles}
\bibinfo{author}{R.~Clift}, \bibinfo{author}{J.~R. Grace}, \bibinfo{author}{M.~E. Weber},
\newblock \bibinfo{title}{Bubbles, drops, and particles}  (\bibinfo{year}{2005}).
\bibitem[{Dodd et~al.(2021)Dodd, Mohaddes, Ferrante, and Ihme}]{DODD2021121157}
\bibinfo{author}{M.~S. Dodd}, \bibinfo{author}{D.~Mohaddes}, \bibinfo{author}{A.~Ferrante}, \bibinfo{author}{M.~Ihme},
\newblock \bibinfo{title}{Analysis of droplet evaporation in isotropic turbulence through droplet-resolved dns},
\newblock \bibinfo{journal}{International Journal of Heat and Mass Transfer} \bibinfo{volume}{172} (\bibinfo{year}{2021}) \bibinfo{pages}{121157}. \URLprefix \url{https://www.sciencedirect.com/science/article/pii/S001793102100260X}. \DOIprefix\doi{https://doi.org/10.1016/j.ijheatmasstransfer.2021.121157}.
\bibitem[{Verwey and Birouk(2018)}]{VERWEY201833}
\bibinfo{author}{C.~Verwey}, \bibinfo{author}{M.~Birouk},
\newblock \bibinfo{title}{Fuel vaporization: Effect of droplet size and turbulence at elevated temperature and pressure},
\newblock \bibinfo{journal}{Combustion and Flame} \bibinfo{volume}{189} (\bibinfo{year}{2018}) \bibinfo{pages}{33--45}. \URLprefix \url{https://www.sciencedirect.com/science/article/pii/S0010218017303991}. \DOIprefix\doi{https://doi.org/10.1016/j.combustflame.2017.10.010}.
\bibitem[{Scapin et~al.(2022)Scapin, Dalla~Barba, Lupo, Rosti, Duwig, and Brandt}]{scapin2022finite}
\bibinfo{author}{N.~Scapin}, \bibinfo{author}{F.~Dalla~Barba}, \bibinfo{author}{G.~Lupo}, \bibinfo{author}{M.~E. Rosti}, \bibinfo{author}{C.~Duwig}, \bibinfo{author}{L.~Brandt},
\newblock \bibinfo{title}{Finite-size evaporating droplets in weakly compressible homogeneous shear turbulence},
\newblock \bibinfo{journal}{Journal of Fluid Mechanics} \bibinfo{volume}{934} (\bibinfo{year}{2022}) \bibinfo{pages}{A15}. \DOIprefix\doi{https://doi.org/10.1017/jfm.2021.1140}.
\bibitem[{Setiya and Palmore(2023)}]{SETIYA2023104455}
\bibinfo{author}{M.~Setiya}, \bibinfo{author}{J.~Palmore},
\newblock \bibinfo{title}{Quasi-steady evaporation of deformable liquid fuel droplets},
\newblock \bibinfo{journal}{International Journal of Multiphase Flow} \bibinfo{volume}{164} (\bibinfo{year}{2023}) \bibinfo{pages}{104455}. \URLprefix \url{https://www.sciencedirect.com/science/article/pii/S0301932223000769}. \DOIprefix\doi{https://doi.org/10.1016/j.ijmultiphaseflow.2023.104455}.
\bibitem[{Salimnezhad et~al.(2025)Salimnezhad, Turkeri, Gokalp, and Muradoglu}]{salimnezhad2024hybrid}
\bibinfo{author}{F.~Salimnezhad}, \bibinfo{author}{H.~Turkeri}, \bibinfo{author}{I.~Gokalp}, \bibinfo{author}{M.~Muradoglu},
\newblock \bibinfo{title}{A hybrid immersed-boundary/front-tracking method for interface-resolved simulation of droplet evaporation},
\newblock \bibinfo{journal}{Computers and Fluids} \bibinfo{volume}{291} (\bibinfo{year}{2025}) \bibinfo{pages}{1065703}.
\bibitem[{Rusche(2003)}]{rusche2003computational}
\bibinfo{author}{H.~Rusche}, \bibinfo{title}{Computational fluid dynamics of dispersed two-phase flows at high phase fractions}, Ph.D. thesis, Imperial College London (University of London), \bibinfo{year}{2003}.
\bibitem[{Sato and Ni{\v{c}}eno(2013)}]{sato2013sharp}
\bibinfo{author}{Y.~Sato}, \bibinfo{author}{B.~Ni{\v{c}}eno},
\newblock \bibinfo{title}{A sharp-interface phase change model for a mass-conservative interface tracking method},
\newblock \bibinfo{journal}{Journal of Computational Physics} \bibinfo{volume}{249} (\bibinfo{year}{2013}) \bibinfo{pages}{127--161}.
\bibitem[{Bhuvankar and Dabiri(2020)}]{BHUVANKAR2020115919}
\bibinfo{author}{P.~Bhuvankar}, \bibinfo{author}{S.~Dabiri},
\newblock \bibinfo{title}{Numerical simulation of sliding bubbles in saturated flow boiling},
\newblock \bibinfo{journal}{Chemical Engineering Science} \bibinfo{volume}{228} (\bibinfo{year}{2020}) \bibinfo{pages}{115919}. \URLprefix \url{https://www.sciencedirect.com/science/article/pii/S0009250920304516}. \DOIprefix\doi{https://doi.org/10.1016/j.ces.2020.115919}.
\bibitem[{Tryggvason et~al.(2001)Tryggvason, Bunner, Esmaeeli, Juric, Al-Rawahi, Tauber, Han, Nas, and Jan}]{tryggvason2001front}
\bibinfo{author}{G.~Tryggvason}, \bibinfo{author}{B.~Bunner}, \bibinfo{author}{A.~Esmaeeli}, \bibinfo{author}{D.~Juric}, \bibinfo{author}{N.~Al-Rawahi}, \bibinfo{author}{W.~Tauber}, \bibinfo{author}{J.~Han}, \bibinfo{author}{S.~Nas}, \bibinfo{author}{Y.-J. Jan},
\newblock \bibinfo{title}{A front-tracking method for the computations of multiphase flow},
\newblock \bibinfo{journal}{Journal of computational physics} \bibinfo{volume}{169} (\bibinfo{year}{2001}) \bibinfo{pages}{708--759}.
\bibitem[{Esmaeeli and Tryggvason(2004{\natexlab{a}})}]{esmaeeli2004front}
\bibinfo{author}{A.~Esmaeeli}, \bibinfo{author}{G.~Tryggvason},
\newblock \bibinfo{title}{A front tracking method for computations of boiling in complex geometries},
\newblock \bibinfo{journal}{International Journal of Multiphase Flow} \bibinfo{volume}{30} (\bibinfo{year}{2004}{\natexlab{a}}) \bibinfo{pages}{1037--1050}.
\bibitem[{Esmaeeli and Tryggvason(2004{\natexlab{b}})}]{esmaeeli2004computations}
\bibinfo{author}{A.~Esmaeeli}, \bibinfo{author}{G.~Tryggvason},
\newblock \bibinfo{title}{Computations of film boiling. part ii: multi-mode film boiling},
\newblock \bibinfo{journal}{International journal of heat and mass transfer} \bibinfo{volume}{47} (\bibinfo{year}{2004}{\natexlab{b}}) \bibinfo{pages}{5463--5476}.
\bibitem[{Irfan and Muradoglu(2017)}]{irfan2017front}
\bibinfo{author}{M.~Irfan}, \bibinfo{author}{M.~Muradoglu},
\newblock \bibinfo{title}{A front tracking method for direct numerical simulation of evaporation process in a multiphase system},
\newblock \bibinfo{journal}{Journal of Computational Physics} \bibinfo{volume}{337} (\bibinfo{year}{2017}) \bibinfo{pages}{132--153}.
\bibitem[{Irfan and Muradoglu(2018)}]{irfan2018front}
\bibinfo{author}{M.~Irfan}, \bibinfo{author}{M.~Muradoglu},
\newblock \bibinfo{title}{A front tracking method for particle-resolved simulation of evaporation and combustion of a fuel droplet},
\newblock \bibinfo{journal}{Computers \& Fluids} \bibinfo{volume}{174} (\bibinfo{year}{2018}) \bibinfo{pages}{283--299}.
\bibitem[{Mittal et~al.(2008)Mittal, Dong, Bozkurttas, Najjar, Vargas, and Von~Loebbecke}]{mittal2008versatile}
\bibinfo{author}{R.~Mittal}, \bibinfo{author}{H.~Dong}, \bibinfo{author}{M.~Bozkurttas}, \bibinfo{author}{F.~Najjar}, \bibinfo{author}{A.~Vargas}, \bibinfo{author}{A.~Von~Loebbecke},
\newblock \bibinfo{title}{A versatile sharp interface immersed boundary method for incompressible flows with complex boundaries},
\newblock \bibinfo{journal}{Journal of computational physics} \bibinfo{volume}{227} (\bibinfo{year}{2008}) \bibinfo{pages}{4825--4852}.
\bibitem[{Unverdi and Tryggvason(1992)}]{unverdi1992front}
\bibinfo{author}{S.~O. Unverdi}, \bibinfo{author}{G.~Tryggvason},
\newblock \bibinfo{title}{A front-tracking method for viscous, incompressible, multi-fluid flows},
\newblock \bibinfo{journal}{Journal of computational physics} \bibinfo{volume}{100} (\bibinfo{year}{1992}) \bibinfo{pages}{25--37}.
\bibitem[{Juric and Tryggvason(1998)}]{juric1998computations}
\bibinfo{author}{D.~Juric}, \bibinfo{author}{G.~Tryggvason},
\newblock \bibinfo{title}{Computations of boiling flows},
\newblock \bibinfo{journal}{International journal of multiphase flow} \bibinfo{volume}{24} (\bibinfo{year}{1998}) \bibinfo{pages}{387--410}.
\bibitem[{Borges et~al.(2008)Borges, Carmona, Costa, and Don}]{borges2008improved}
\bibinfo{author}{R.~Borges}, \bibinfo{author}{M.~Carmona}, \bibinfo{author}{B.~Costa}, \bibinfo{author}{W.~S. Don},
\newblock \bibinfo{title}{An improved weighted essentially non-oscillatory scheme for hyperbolic conservation laws},
\newblock \bibinfo{journal}{Journal of computational physics} \bibinfo{volume}{227} (\bibinfo{year}{2008}) \bibinfo{pages}{3191--3211}.
\bibitem[{Leonard(1979)}]{LEONARD197959}
\bibinfo{author}{B.~Leonard},
\newblock \bibinfo{title}{A stable and accurate convective modelling procedure based on quadratic upstream interpolation},
\newblock \bibinfo{journal}{Computer Methods in Applied Mechanics and Engineering} \bibinfo{volume}{19} (\bibinfo{year}{1979}) \bibinfo{pages}{59--98}. \URLprefix \url{https://www.sciencedirect.com/science/article/pii/0045782579900343}. \DOIprefix\doi{https://doi.org/10.1016/0045-7825(79)90034-3}.
\bibitem[{Chorin(1968)}]{chorin1968numerical}
\bibinfo{author}{A.~J. Chorin},
\newblock \bibinfo{title}{Numerical solution of the navier-stokes equations},
\newblock \bibinfo{journal}{Mathematics of computation} \bibinfo{volume}{22} (\bibinfo{year}{1968}) \bibinfo{pages}{745--762}.
\bibitem[{Tryggvason et~al.(2011)Tryggvason, Scardovelli, and Zaleski}]{tryggvason2011direct}
\bibinfo{author}{G.~Tryggvason}, \bibinfo{author}{R.~Scardovelli}, \bibinfo{author}{S.~Zaleski}, \bibinfo{title}{Direct numerical simulations of gas--liquid multiphase flows}, \bibinfo{publisher}{Cambridge university press}, \bibinfo{year}{2011}.
\bibitem[{Frossling(1963)}]{frossling1963evaporation}
\bibinfo{author}{N.~Frossling}, \bibinfo{title}{The evaporation of falling drops}, \bibinfo{publisher}{UKAEA Research Group, Atomic Energy Research Establishment}, \bibinfo{year}{1963}.
\bibitem[{Bird(1960)}]{bird1960stewart}
\bibinfo{author}{R.~B. Bird},
\newblock \bibinfo{title}{„stewart, we, and lightfoot, en},
\newblock \bibinfo{journal}{Transport phenomena} \bibinfo{volume}{2} (\bibinfo{year}{1960}).
\bibitem[{Hase and Weigand(2004)}]{hase2004transient}
\bibinfo{author}{M.~Hase}, \bibinfo{author}{B.~Weigand},
\newblock \bibinfo{title}{Transient heat transfer of deforming droplets at high reynolds numbers},
\newblock \bibinfo{journal}{International Journal of Numerical Methods for Heat \& Fluid Flow} \bibinfo{volume}{14} (\bibinfo{year}{2004}) \bibinfo{pages}{85--97}.

\end{thebibliography}





\end{document}